\documentclass[fleqn,10pt]{wlscirep}
\usepackage[utf8]{inputenc}
\usepackage[T1]{fontenc}
\usepackage{subcaption}
\usepackage[symbol]{footmisc}
\usepackage[autostyle]{csquotes}
\newcommand{\cbrk}{\protect\\}

\usepackage{tocloft}
\usepackage{geometry}
\usepackage{lipsum}

\usepackage{makecell} 
\usepackage{siunitx}

\title{Tracking Patterns in Toxicity and Antisocial Behavior Over User Lifetimes on Large Social Media Platforms}

\author[1,*]{Katy Blumer}
\author[1]{Jon Kleinberg}
\affil[1]{Cornell University, Department of Computer Science, Ithaca, NY, USA}
\affil[*]{kblumer@cs.cornell.edu}

\begin{abstract}

An increasing amount of attention has been devoted to the problem of \enquote{toxic} or antisocial behavior on social media. 
In this paper we analyze such behavior at very large scales: over a 14-year time span on nearly 500 million comments from Reddit and Wikipedia, grounded in two different proxies for toxicity.

At the individual level, we analyze users' toxicity levels over the course of their time on the site, and find a striking reversal in trends: both Reddit and Wikipedia users tended to become less toxic over their life cycles on the site in the early (pre-2013) history of the site, but more toxic over their life cycles in the later (post-2013) history of the site. 
We also find that toxicity on Reddit and Wikipedia differ in a key way, with the most toxic behavior on Reddit exhibited in aggregate by the most active users, and the most toxic behavior on Wikipedia exhibited in aggregate by the least active users.
Finally, we consider the toxicity of discussion around widely-shared pieces of content, and find that the trends for toxicity in discussion about content bear interesting similarities with the trends for toxicity in discussion by users.

\end{abstract}
\begin{document}

\flushbottom
\maketitle

%
\thispagestyle{empty}

\section*{Introduction}


Antisocial behavior on social media is recognized as an increasingly important problem for the experience of users on many platforms. Often referred to with the umbrella term \enquote{toxicity},\cite{gillespie2020content} such behavior has been implicated in patterns of online harassment,\cite{castano-pulgarin2021internet,gagliardone2015countering} outcomes of civic processes,\cite{tucker2018social,haidt2022yes,wike2022social,gonzalez-bailon2023social} and physical violence.\cite{gallacher2021online,bhavnani2009rumor} 
However, its underlying mechanisms are still poorly understood.
One reason for this challenge is that the culture and dynamics of social media are also constantly changing, both within and across platforms.\cite{2012eternal,kiene2016surviving}
Users also influence and are influenced by the communities they interact with. \cite{danescu-niculescu-mizil2013no,cheng2014how} 
%


We note at the outset that it’s worth being critical of the term \enquote{toxicity}. The term is commonly assumed to need no definition,\cite{gillespie2020content,arouh2020toxic} and often used without explanation - but some think it’s too vague to be meaningful, or even that it’s used to avoid more uncomfortable topics such as radicalization or identity-based hatred.\cite{gillespie2020content, matias2024canWITHLINK}
It is therefore important to discuss the sense in which we will use the term in our work.

Google's Jigsaw, the makers of the Perspective API,\cite{lees2022new, wulczyn2017ex,rieder2021fabrics} give a practical reason for choosing the term \enquote{toxicity}: when teaching humans to annotate their training data for generally problematic text, they found that \enquote{toxic} was the easiest concept for annotators to understand\cite{jigsawCurrent} That is, they tried various terms for various forms of problematic behavior, and the annotators were most likely to agree on which comments met the definition of \enquote{toxic} than for any of the other terms. For the API metric used in this work, \enquote{severe toxicity}, the specific definition they used was: \enquote{a hateful, aggressive, disrespectful comment or otherwise very likely to make a user leave a discussion or give up on sharing their perspective}.\cite{perspective_api_attribute_defs} Whether or not we agree with their definition,\cite{sheth2022defining, arora2023detecting, fortuna2020toxic} the Perspective API model is clearly measuring \textit{some} signal.\cite{lees2022new} And since we’re interested in patterns at large scale over long time periods, we think it can give us useful information. Also, the fact that the API is commonly used in research\cite{fortuna2020toxic} makes it easier to put results in context, and its use in real-world applications\cite{lee2019alphabetmade,guo2023evaluating} means that it’s worth studying.

Another way to address the subtlety of the underlying concepts is to quantify them with multiple independent measures.  Accordingly, we supplement the Perspective API score with an independent metric: a \enquote{downvote score} that also measures some kinds of antisocial behavior. Reddit’s content ranking depends on user voting, where any user can \enquote{upvote} or \enquote{downvote} a post or comment. Reddit asks users to vote based on whether a comment \enquote{contributes to conversation} (though in practice, users sometimes vote based on other factors, such as whether they agree with the comment.).\cite{reddiquette} 
A comment's score reflects community judgment, with negative scores (more downvotes than upvotes) often indicating toxic content that users found objectionable. While comments might also be downvoted for being unpopular, off-topic, or misleading, a negative score generally signals content the community deemed inappropriate or undesirable for wider visibility.
As we will see, the broad conclusions we draw hold under both numerical metrics, despite their reliance on disjoint sources of information in the underlying data.



A large body of research addresses online toxicity and antisocial behavior, many specifically using the Perspective API toxicity score.\cite{xia2020exploring,avalle2024persistent,cambo2022model,mamakos2023social} 
Xia et al\cite{xia2020exploring} studied various predictors of comment toxicity on Reddit, including toxicity of the author's previous comments and of the previous comments in the discussion, while Avalle et al\cite{avalle2024persistent} found that comments increase in toxicity the longer a conversation lasts.

Other metrics of \enquote{toxicity} or \enquote{antisocial behavior} include banned users,\cite{cheng2015antisocial} hand-coded \enquote{civility},\cite{coe2014online,papacharissi2004democracy} and custom-designed toxicity or hate speech classifiers.\cite{almerekhi2022investigating,cinelli2021dynamics}
However, most existing work focuses on conversation-level\cite{avalle2024persistent,xia2020exploring,cinelli2021dynamics,saveski2021structure,coe2014online,youngreusser2024responding,fariello2021does} or community-level\cite{waller2021quantifying,lupu2023offline,johnson2019hidden} patterns. Mamakos and Finkel\cite{mamakos2023social} do track individual users between subreddits (communities within Reddit), finding that users who comment in politically partisan contexts are more toxic than users who don't. Meanwhile, Chandrasekharan et al\cite{chandrasekharan2022quarantined} follow changes in user population and overall toxicity over time for subreddits that were affected by quarantining. Sun et al\cite{sun2021overtime} find that mean \enquote{incivility} across Reddit is fairly stable from 2006-2019, which accords with our findings on toxicity (Fig. \ref{fig_supp:SF3a_mean_scores_all_reddit}).

Few works follow individual user trajectories over time, as this work does. Of those that do, Cheng et al\cite{cheng2015antisocial} analyze users who will go on to be banned from a community. Similar to this work, they find that \enquote{future banned users} are different from \enquote{never-banned users} from their very first post, and that these users' behavior changes over time according to their main metrics: post deletion rate and text quality. They also measure slopes in users' post deletion rates over time, though they focus on whether the slopes change between the first and second half of a user's lifetime.
Mekacher et al\cite{mekacher2023systemic} track user toxicity over the first 150 days of users' lifetimes on Gettr, a Twitter alternative popular with the far right, and find that lifetime toxicity is flat; however, age in days is a different measure of lifetime than the number-of-posts measure we use. 
Almerekhi et al\cite{almerekhi2022investigating} follow user change at a more local level, using a binary definition of \enquote{toxicity} and counting events where a user made a \enquote{toxic} comment after a \enquote{non-toxic} comment or vice versa.
Waller and Anderson\cite{waller2021quantifying} also find a temporal shift in other characteristics at a similar time to the shift found in this work. They found a sharp change in user polarization, measured by the partisanship of communities that users commented in. However, this polarization event wasn't relative to user lifetimes; existing users all polarized at approximately the same time, though most of the polarization was driven by new users.

The present work is interested in whether patterns of toxicity change over time, not just globally across a platform, but within individual users.
We analyze trajectories across four dimensions: activity level (total lifetime number of comments), two different metrics of antisocial behavior (Perspective API toxicity score and \enquote{downvote score}, discussed below), two different social media sites (Reddit and Wikipedia), and two types of histories (people/users and information/URLs). 

In nearly all cases, we observe a striking reversal: users do change over their lifetimes, but the direction of that change hasn't been stable over time. Before the mid-2010s, the longer a user spent on a social media site, the less toxic they became. After the mid-2010s, the longer they spent on the site, the more toxic they became. 

We also observe another pattern: First, low-activity users are generally separable from higher-activity users in their toxicity behavior. On Reddit, higher-activity users have higher toxicity scores; on Wikipedia, the reverse is true. This raises the question of how this difference comes about: one possibility is that high-activity users start out looking exactly like low-activity users and gradually reach their final level of toxicity; a second possibility is that high and low-activity users have different toxicity from the beginning. We find that the data is more consistent with the latter principle: Low- and high-activity users have different average toxicity metrics from their very first post. That is, it doesn't seem that users start out the same, then only become more or less toxic the more they comment. Instead, they’re distinct from each other from the beginning.

\section*{Results}

\subsection*{Definitions}

We use two metrics: toxicity and downvote scores. We also use three separate datasets. See the Methods section for a more detailed description. 

\paragraph{Toxicity score} The toxicity score is from the Perspective API,\cite{perspectiveapi} a machine learning model that assigns a piece of text a score between 0 and 1, where 1 is most toxic. The API defines \enquote{toxicity} as \enquote{a very hateful, aggressive, disrespectful comment or otherwise very likely to make a user leave a discussion or give up on sharing their perspective}.\cite{perspective_api_attribute_defs} Scores for comments not in English were excluded, to prevent bias between models for different languages.\cite{nogara2023toxic}
\paragraph{Downvote score}  The downvote score comes from the Reddit voting system, where any user can \enquote{upvote} or \enquote{downvote} a comment based on whether it \enquote{contributes to conversation}.\cite{reddiquette} Here \enquote{downvote score} for a group of comments denotes the percentage of comments that were downvoted, meaning comments that received more downvotes than upvotes.

\paragraph{Datasets} See Methods and Table \ref{tab:dataset_sizes} for more details. The Reddit user dataset contains all comments made by a random sample of 1 million users ($\sim1\%$), for a total of 123 million comments. The Wikipedia dataset contains all edits made to English Wikipedia talk pages, after filtering out bot accounts, for a total of 122 million edits from 2.8 million accounts, on 21.8 million talk pages. Supplementary Section \ref{section_supp:G_size_plots} shows the number of accounts created in different years.

The Reddit URL dataset consists of all Reddit posts of YouTube or Imgur URLs. Scores for each post were calculated as the mean of scores for all comments on the post. Downvote score for each post was defined as the percentage of comments on that post that were downvoted.

\begin{figure}[h!]
    \centering
    
    \includegraphics[width=0.95\columnwidth]{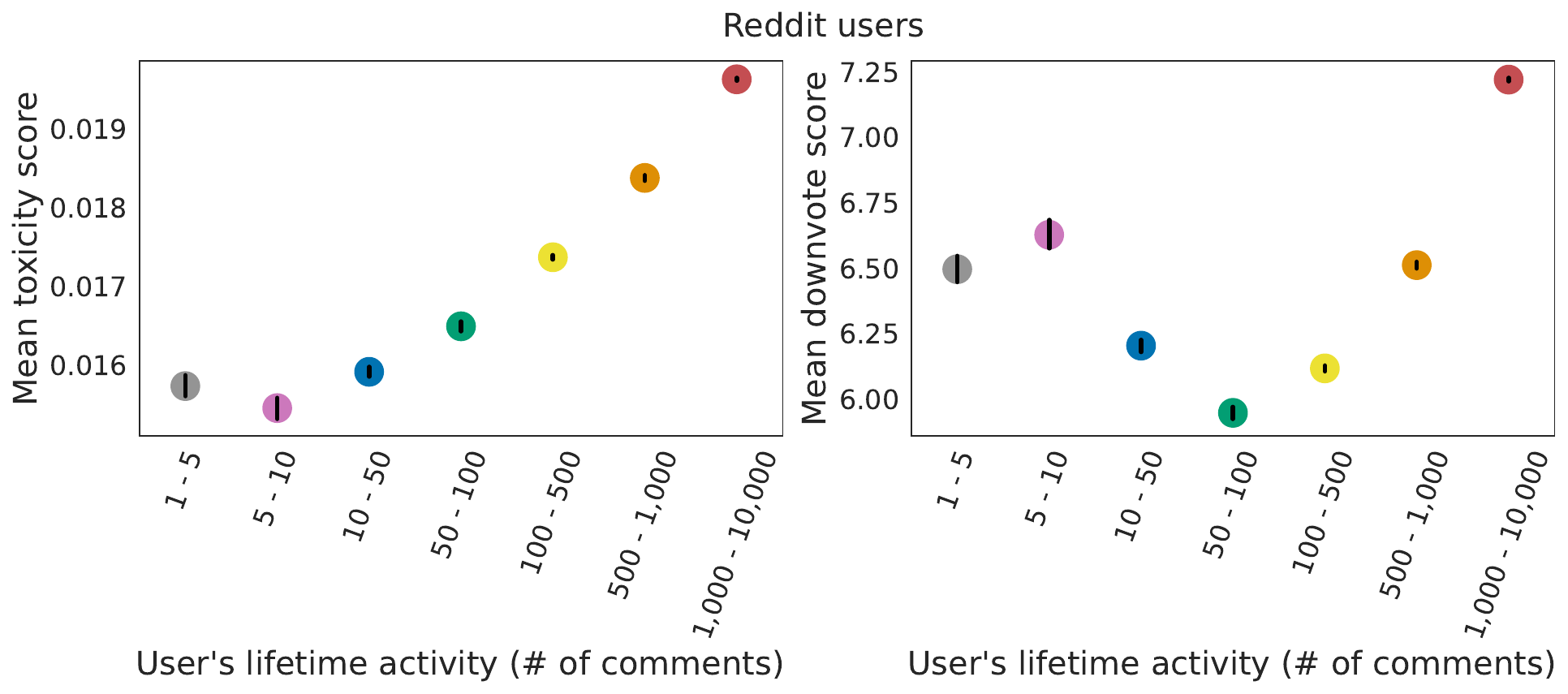} 
    
    \caption{Mean toxicity and downvote scores for users with different lifetime comment volumes.
    \cbrk
    Black bars represent 95\% confidence intervals over the mean, from 1,000 bootstrap runs.}
    
    \label{fig:A_means}
\end{figure}

\subsection*{Reddit users}

As a first question, we consider whether users who comment often are different from more infrequent users in the metrics we study (Fig \ref{fig:A_means}). On average, these groups are indeed different, with the most prolific users showing the highest average for both toxicity and downvote scores. For toxicity scores, there is a clear upward trend as we move from less-active to more-active users. 

For the downvote score, the most active users still have the highest average score, but here the trend is more U-shaped: moderate-activity users have the lowest scores, with the most- and least-active users both having high scores.  An intuitive explanation for the U-shape might be that downvotes capture a wider range of factors than toxicity, such as off-topic or unpopular opinions, and that inexperienced commenters are more likely to break community norms and thus be downvoted for reasons other than toxicity\cite{danescu-niculescu-mizil2013no}. Or there might be a selection effect: users whose comments are downvoted might be more likely to stop commenting, and thus have a lower lifetime number of comments.\cite{dror2012churn} However, our current analysis doesn't allow us to determine if either explanation is accurate.

(To note, there are a number of options for stratifying users by activity.  We use the lifetime number of comments, but we could also have used time between first and last post, or mean time between posts. We find that any of the three options show similar results, with more-active users more toxic on average (see Supplementary Section \ref{section_supp:E_alt_buckets}). The number-of-comments method also has the advantage that each user contributes a comparable number of comments to their bucket's average.)

\begin{figure}[h!]
    \centering
    \includegraphics[width=0.95\columnwidth]{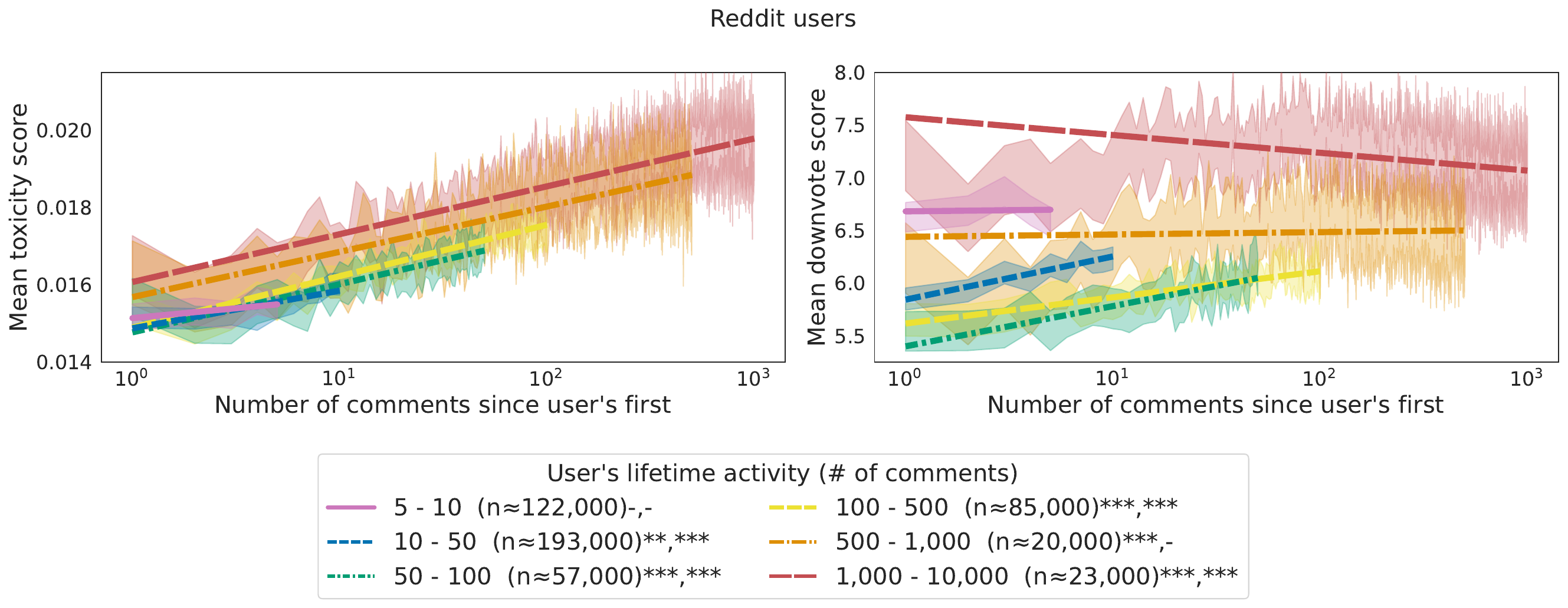} 
    \caption{Trends in each metric over users’ lifetimes, from their first comment to their thousandth. Users are grouped by their \enquote{activity} (lifetime comment count), and users with higher activity tend to have higher average toxicity scores.
    \cbrk
    Trend lines are fit to the means over all users in an activity bucket. Shaded bands are 95\% confidence intervals for the mean. 
    \cbrk
    The x-axis is logarithmic, for visibility of shorter activity buckets. For a linear x-axis scale and fit line, see Supplementary Section \ref{section_supp:C_linear_scale}.
    \cbrk
    Asterisks represent confidence that the relevant slope is positive or negative, for toxicity score and downvote score respectively. (See Supplementary Section \ref{section_supp:SB2_pval_tables} for \textit{p}-values.)
    }
    \label{fig:B_rainbow}
\end{figure}

\begin{figure}[h!!]
    \centering
    
    \begin{subfigure}[b]{0.95\columnwidth}
        \centering
        \includegraphics[width=1\columnwidth]{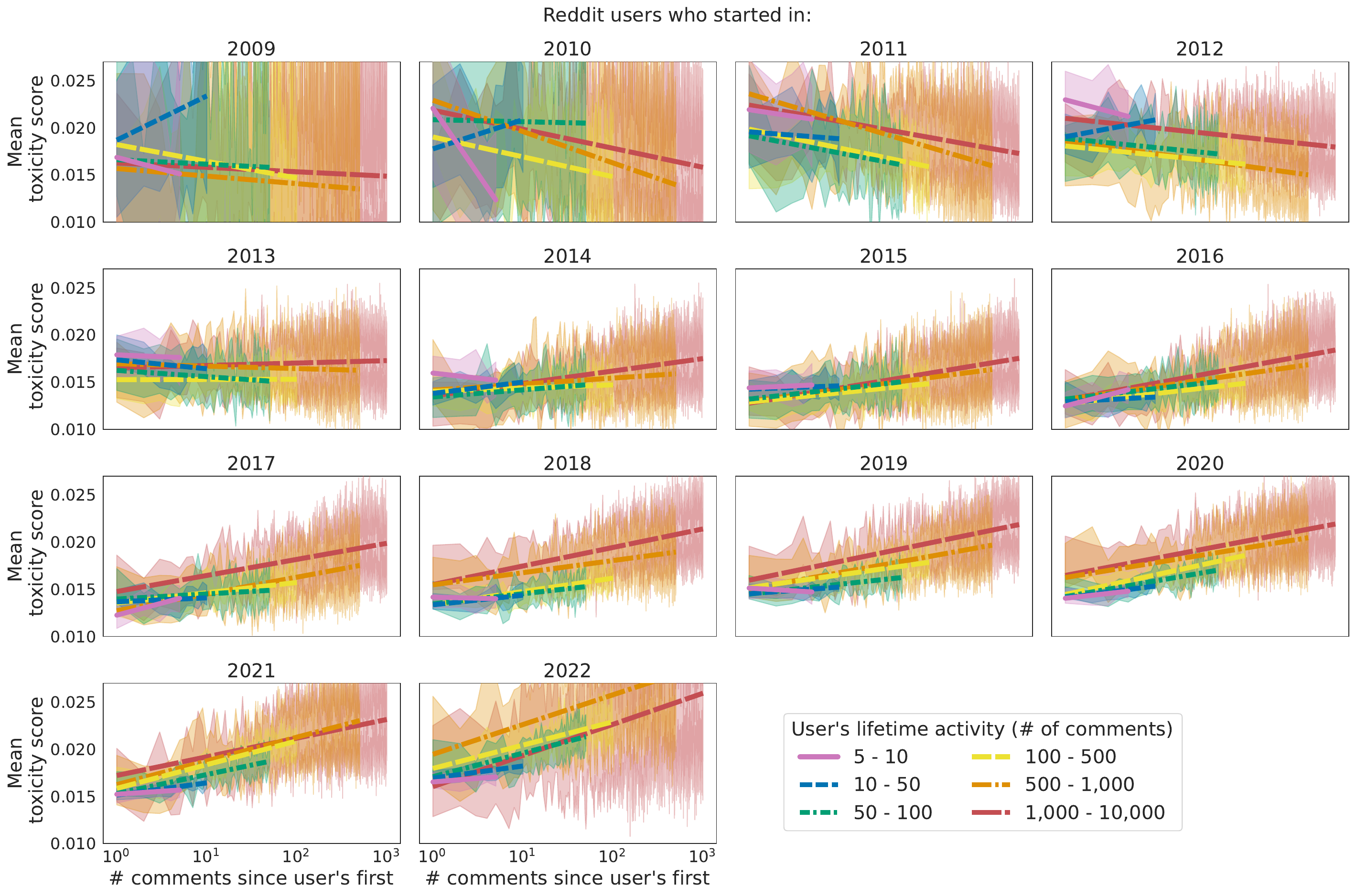} 
        \caption{}
        \label{fig:C1_rainbow_grid_tox}
    \end{subfigure}

    \begin{subfigure}[b]{0.95\columnwidth}
        \centering
        \includegraphics[width=1\columnwidth]{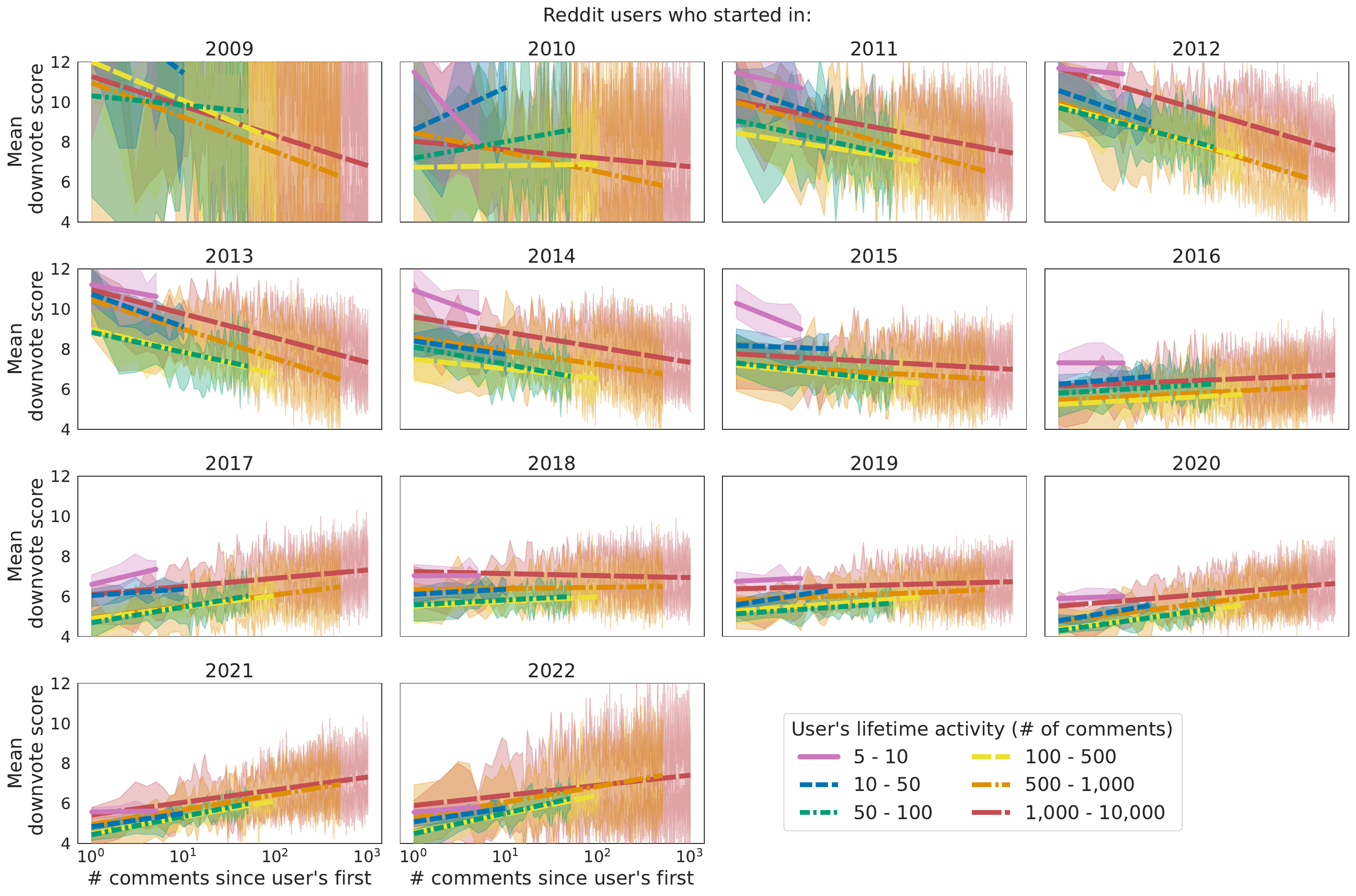} 
        \caption{}
        \label{fig:C2_rainbow_grid_votes}
    \end{subfigure}

    \caption{Trends over user lifetimes in \subref{fig:C1_rainbow_grid_tox}) toxicity score and \subref{fig:C2_rainbow_grid_votes}) downvote score, for users who began in different years.
    \cbrk
    Individual plots have the same structure as Fig. \ref{fig:B_rainbow}, with 95\% confidence intervals over bucket means and trend lines fitted to the means.
    }
    \label{fig:C_rainbow_grid}
\end{figure}


A natural next question is whether the most-active users made more toxic comments from the beginning, or if they became more toxic over their lifetime on the site.
In fact, in Fig. \ref{fig:B_rainbow} we see a consistent increase in toxicity score over users' lifetimes (less so for downvote score). But knowing that our user sample spans over a decade of time, we might wonder if this effect was consistent over Reddit's lifespan as well. In Figure \ref{fig:C_rainbow_grid}, we separate users into cohorts based on the year in which they began commenting. 
This separation highlights one of the key findings from our analysis: in the earlier years, there is a \textit{downward} trend for both toxicity and downvote scores, meaning that users become less toxic over time; but the trend gradually reverses over the years until, by 2022, there's a clear \textit{upward} slope, meaning that users become \textit{more} toxic over time. This appears for both toxicity and downvote score, though the reversal happens a few years later for toxicity score.

\begin{figure}[h!]
    \centering
    \includegraphics[width=0.95\columnwidth]{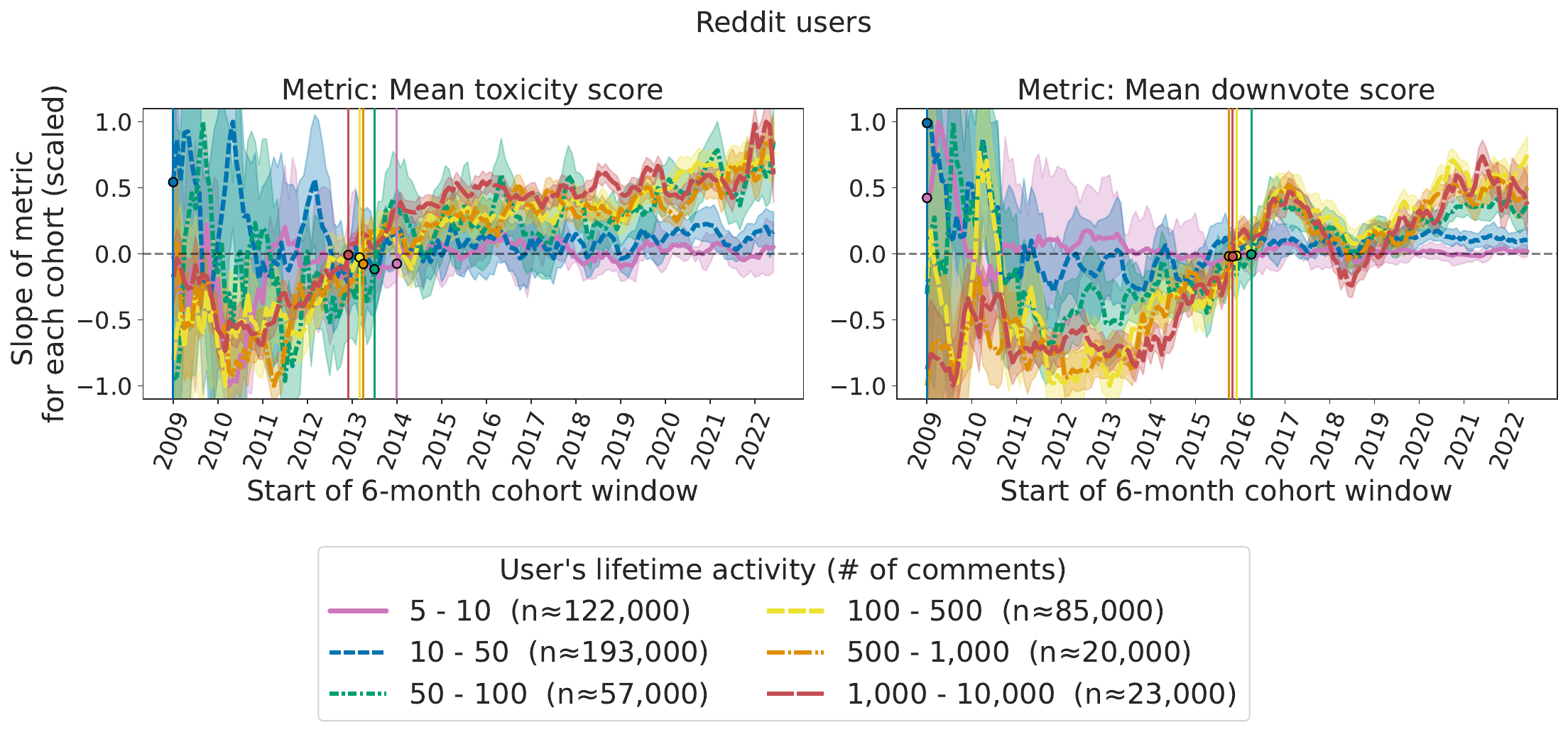} 
    \caption{Changes in trends over time for toxicity and downvote scores. Each point on a line in this plot represents the slope of an entire line from a plot such as those in Figure \ref{fig:C_rainbow_grid}.
    \cbrk
    The approximate point where each line crosses zero is marked with a vertical line. This is the time point where slopes go from mostly-negative (users become less toxic over their lifetime) to mostly-positive (users become more toxic over their lifetime). (Details in Supplementary Section \ref{section_supp:SH_integral_plots}.)
    \cbrk
    Shaded areas represent 95\% confidence intervals around the fitted slope. For visibility, the values are scaled so that the maximum absolute value is 1 for each lifetime activity bucket.}
    \label{fig:D_slope}
\end{figure}

We calculate statistical significance of slopes using a two-sided \textit{t}-test, where significance means that the true slope of the data is significantly different from zero (Section \ref{section_supp:SB2_pval_tables}). Similarly, 95\% confidence intervals such as those in Fig. \ref{fig:D_slope} are calculated using the standard error of the fitted slope.
(Patterns appear similar whether the x-axis is logarithmic or linear - see Section \ref{section_supp:C_linear_scale}).

A more granular view of these trends can be seen in Figure \ref{fig:D_slope}, where each point represents the slope of an entire lifetime trend line such as those in Figure \ref{fig:C_rainbow_grid}.  
Here the trend is even more clear: for both toxicity and downvote score, lifetime toxicity slopes are negative in earlier years and become positive in later years. When plotting them, we see that the lines all cross 0 at around the same time point (estimated using cumulative sums - see Supplementary Section \ref{section_supp:SH_integral_plots}).

We can thus see a clear shift from negative to positive lifetime slope in the mid-2010s. For the toxicity score, this occurs around 2013. For the downvote score, the shift happens later, in 2015. (This time estimate is also robust to bootstrapping - see Fig. \ref{fig_supp:SB_bootstrap}).

\begin{figure}[h]
    \centering
    
    \begin{subfigure}[b]{0.45\columnwidth}
        \centering
        \includegraphics[width=1\columnwidth]{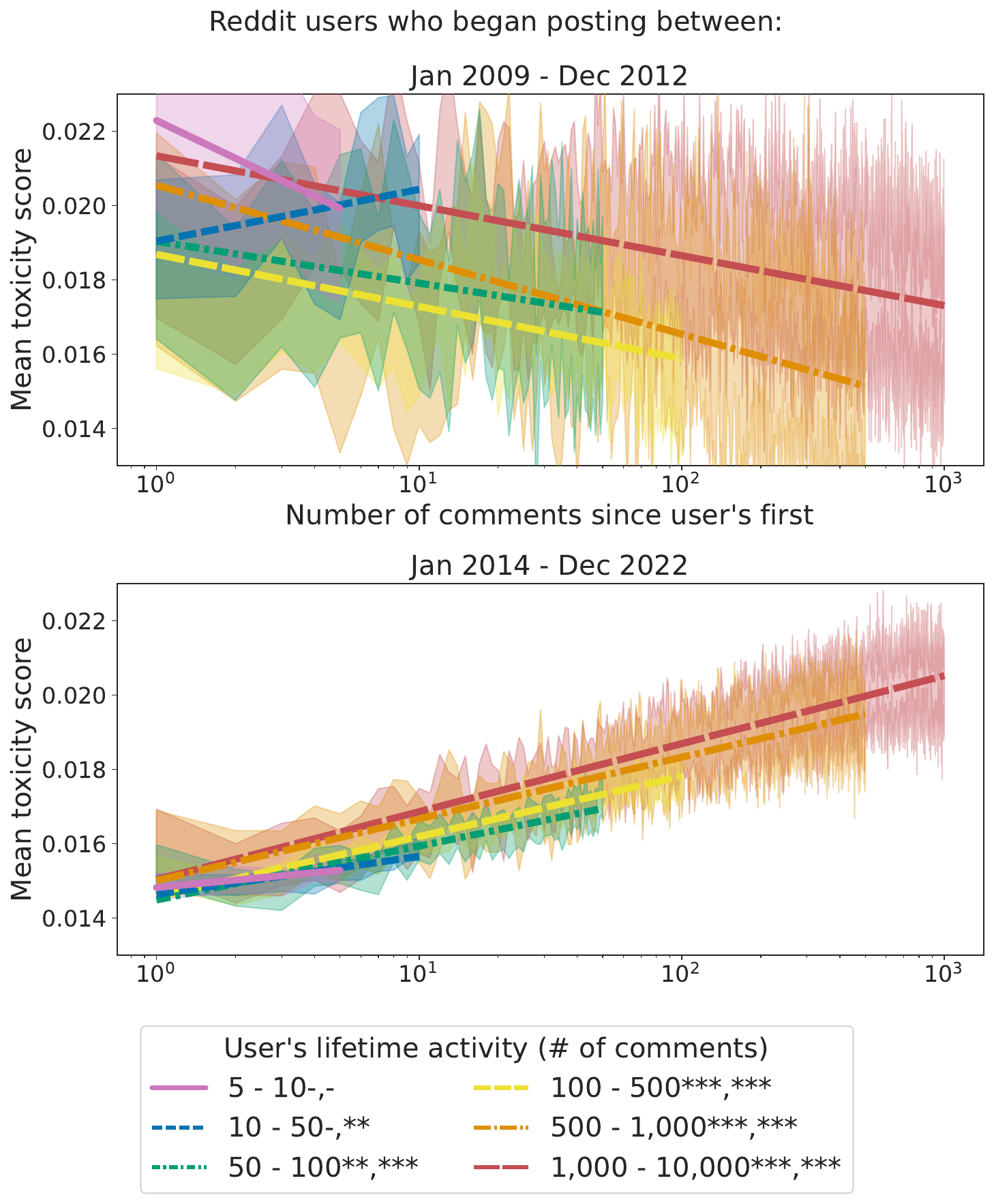} 
        \caption{}
        \label{fig:E1_rainbow_cut_tox}
    \end{subfigure}
    \begin{subfigure}[b]{0.45\columnwidth}
        \centering
        \includegraphics[width=1\columnwidth]{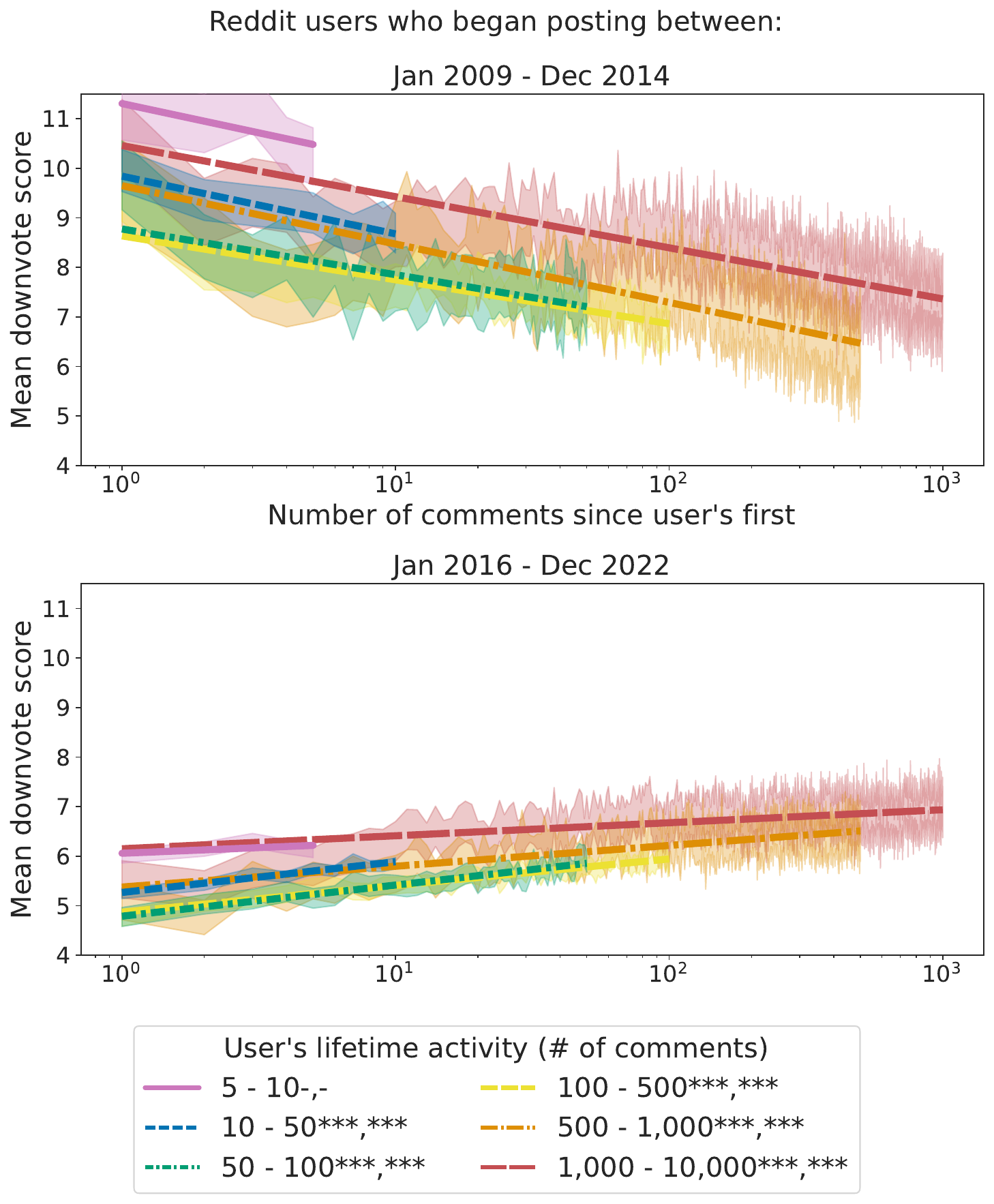} 
        \caption{}
        \label{fig:E2_rainbow_cut_vote}
    \end{subfigure}
    
    \caption{Trends in each metric over users’ lifetimes, split into cohorts \subref{fig:E1_rainbow_cut_tox}) from before and after 2013 for the toxicity score, and \subref{fig:E2_rainbow_cut_vote}) before and after 2015 for the downvote score.
    \cbrk
    The structure is the same as Fig. \ref{fig:B_rainbow}, with 95\% confidence intervals around bucket means and trend lines fitted to the means.
    \cbrk
    Asterisks represent confidence levels for each slope fit, for the earlier and later cohort respectively.
    }
    \label{fig:E_rainbow_cut}
\end{figure}

Figure \ref{fig:E_rainbow_cut} sums up what we've learned so far. First, before the mid-2010s, users' comments, on average, become less toxic the longer they've been on the site. Users who begin commenting after this point, however, make more toxic comments the longer they've been commenting. 
Second, more prolific users, on average, make more toxic comments than lower-activity users, though this effect is less strong when limiting to users from the same time period. 
Third, more prolific users score higher than others from their very first comment, rather than becoming toxic later in their lifetimes.

\subsection*{Wikipedia: a second set of users}

\begin{figure}[h]
    \centering
    
    \begin{subfigure}[b]{0.45\columnwidth}
        \centering
        \includegraphics[width=1\columnwidth]{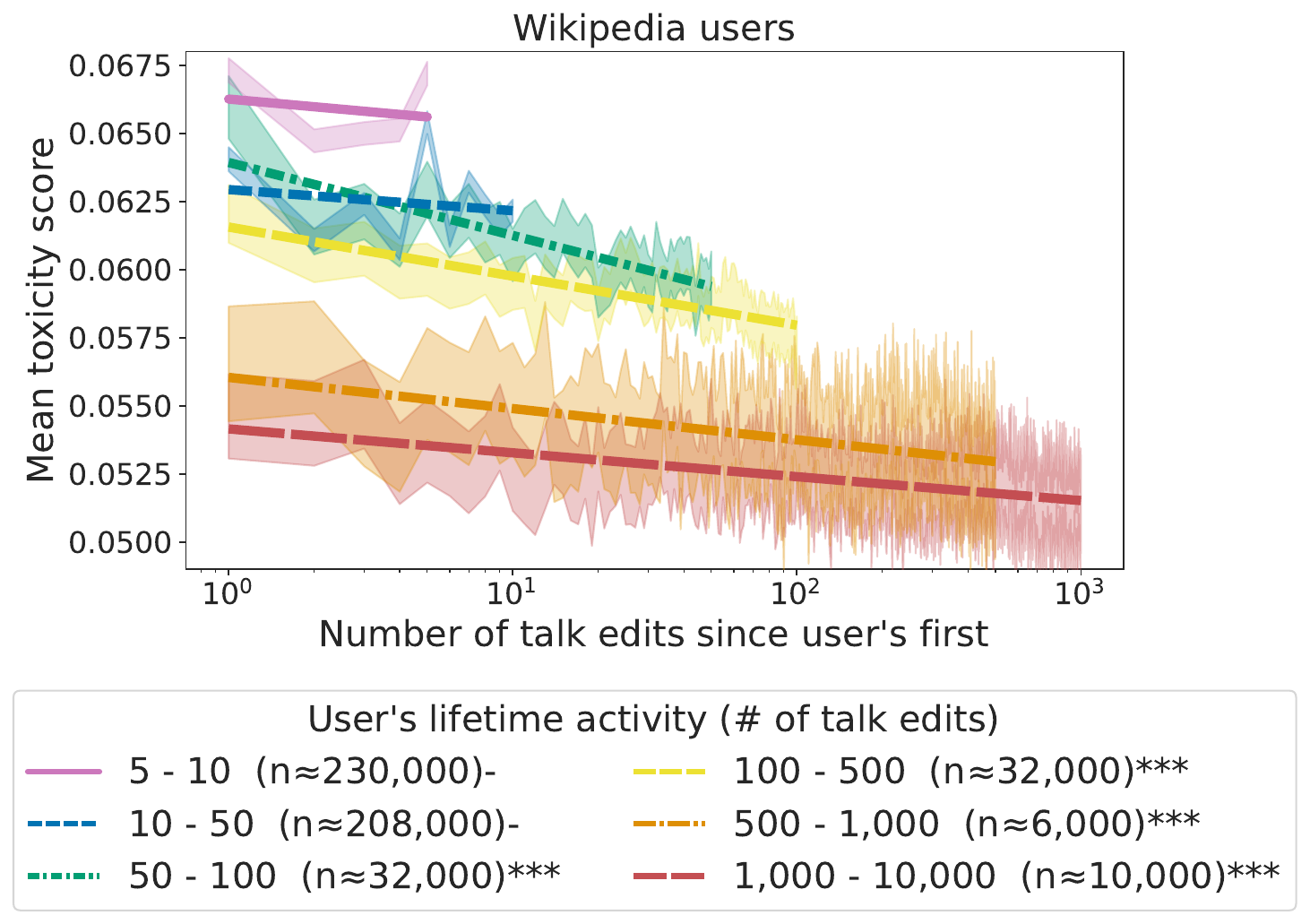} 
        \caption{}
        \label{fig:WA1_rainbow}
    \end{subfigure}
    \begin{subfigure}[b]{0.45\columnwidth}
        \centering
        \includegraphics[width=1\columnwidth]{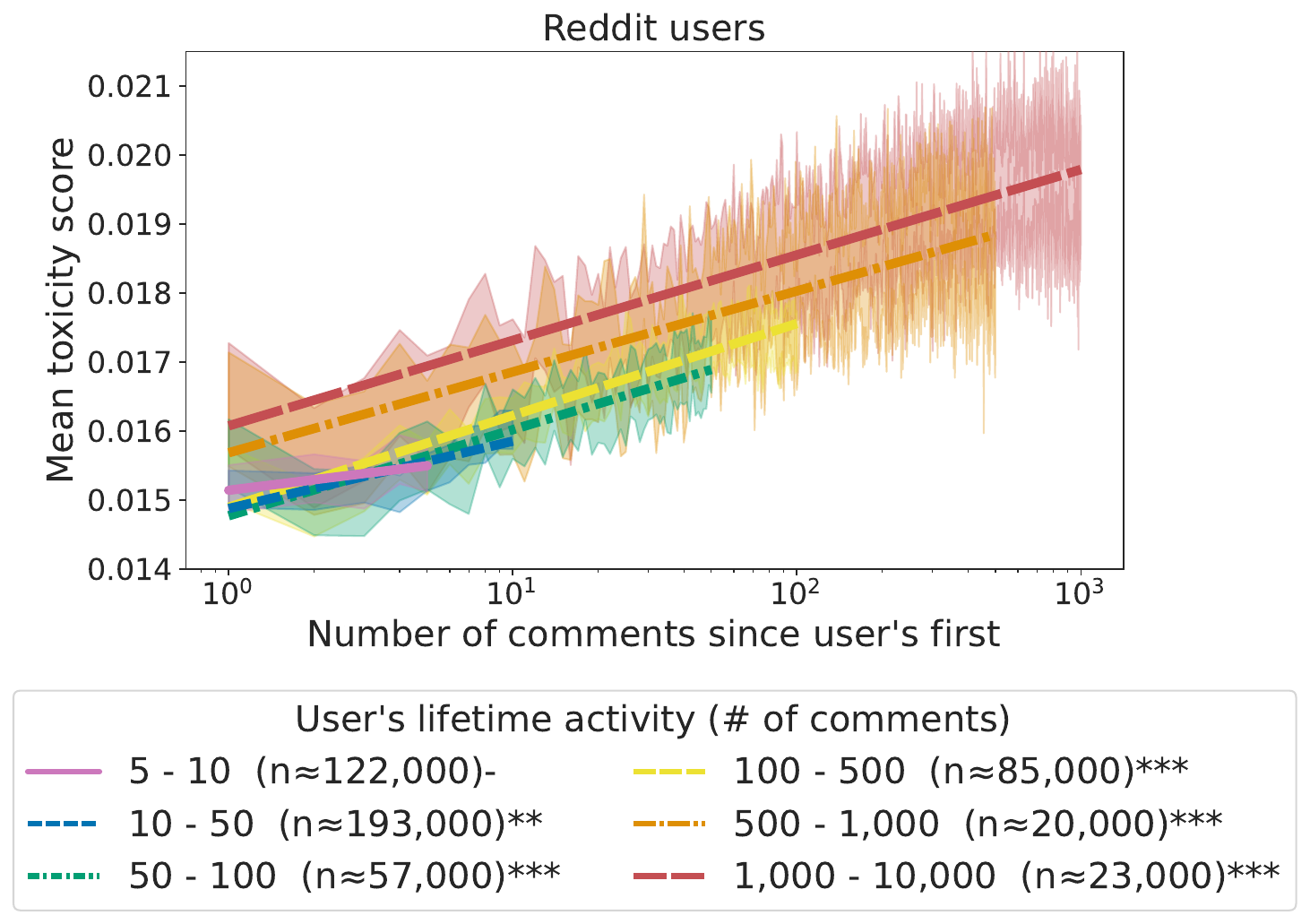} 
        \caption{}
        \label{fig:WA2_compare_reddit}
    \end{subfigure}
    \caption{Trends in toxicity over \subref{fig:WA1_rainbow}) Wikipedia and \subref{fig:WA2_compare_reddit}) Reddit users’ lifetimes, from their first comment to their thousandth. Unlike Reddit users, Wikipedia users with longer comment histories make comments that are \textit{less} toxic, on average, than less-active users' comments.
    \cbrk
    Panel \subref{fig:WA2_compare_reddit}) is a reproduction of Fig. \ref{fig:B_rainbow} for ease of comparison.
    }
    \label{fig:WA_rainbow}
\end{figure}

\begin{figure}[h]
    \centering
    
    \includegraphics[width=0.45\columnwidth]{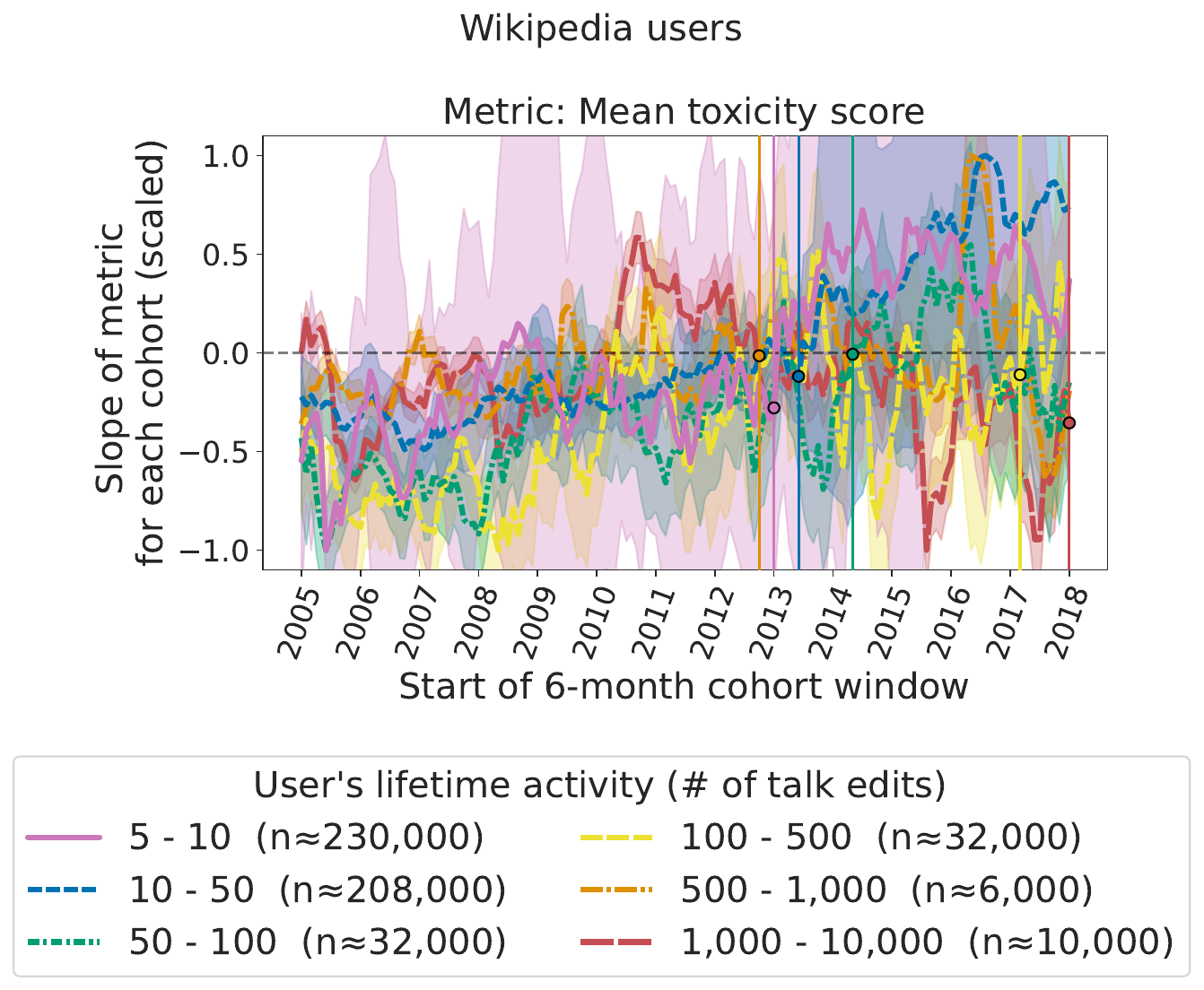} 
    
    \caption{Changes in trends over time for toxicity scores. Each point on a line in this plot represents the slope of an entire line from a plot such as those in Figure \ref{fig:WA_rainbow}.
    \cbrk
    The structure is the same as Fig. \ref{fig:D_slope}.  Trends are much noisier than in Fig. \ref{fig:D_slope} (note that most user-activity buckets have much smaller user counts than in the Reddit dataset.) The approximate point where each line crosses zero is marked with a vertical line. This is the point where slopes go from mostly-negative to mostly-positive. (Crossing points for this figure can be seen more clearly in Fig. \ref{fig_supp:SHb_integral_figWB_1mo}.)
}
    \label{fig:WB_slope}
\end{figure}

Reddit, like most social media sites, has undergone many changes since its inception. To see if this was simply an artifact of culture change on a particular site, we can apply the same analysis to a second site: Wikipedia. 
While Wikipedia isn't primarily a social media site, it hosts vigorous discussions between editors and is known for strongly enforced cultural norms.\cite{cooke2020wikipedia}
As an analogy to Reddit comments, we use edits to Wikipedia \enquote{talk pages}, which are pages attached to Wikipedia articles or user accounts for discussion of edits to the relevant page.  In Figure \ref{fig:WA_rainbow}, we compare lifetime toxicity scores for Wikipedia users and Reddit users. The two are similar in many ways: there's a clear ordering of users with different lifetime comment counts, and different groups on the same site show similar lifetime trends.
But there's also a striking difference: on Wikipedia, high-activity users are \textit{less} toxic than less-active ones. 

Figure \ref{fig:WB_slope} shows that there's also a similar change from negative to positive slope in 2013.
That is, as on Reddit, pre-2013 Wikipedia users became less toxic over their life cycles on the site, but post-2013 users became more toxic over their life cycles. (This is only apparent in the shortest-lived user buckets, but note how much smaller the other buckets are; the difference between bucket sizes is larger than on Reddit, especially when considering bucket sizes in the earlier history of the site, as in Supplementary Section \ref{section_supp:G_size_plots}.)

\subsection*{Toxicity involving links and pages: does information behave like people?}

We can follow other entities over time, not just users. Discourse around a particular idea can change over time, just as people do. Many pieces of content are commented on many times in the same dataset. 

\begin{figure}[h]
    \centering
    \includegraphics[width=0.95\columnwidth]{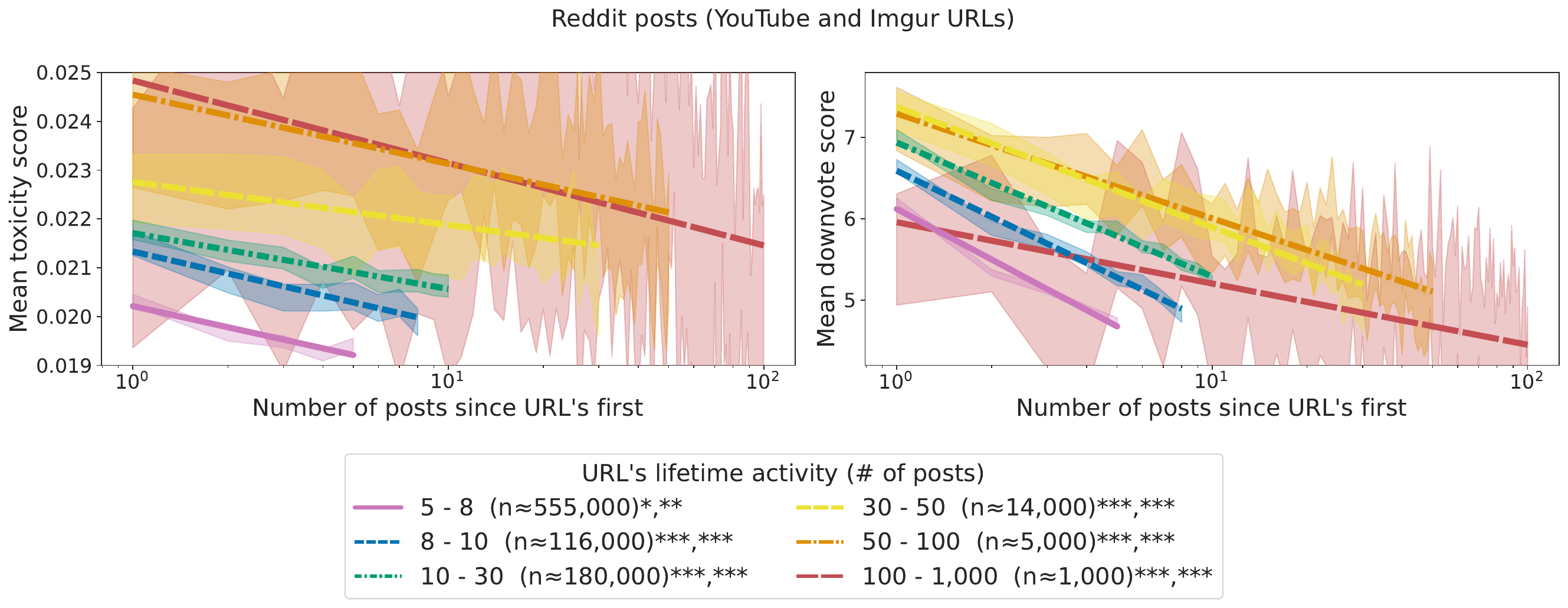} 
    
    \caption{Trends in each metric over the lifetime of a piece of content (URL), from the first time it was posted on Reddit to the hundredth time.
    \cbrk
    Shaded bands are 95\% confidence intervals over the post scores for each time a URL was posted. Post scores are calculated as the mean score for all comments on that post. Trend lines are fit to the mean post score over all URLs in a bucket. 
    \cbrk
    The structure is similar to Fig. \ref{fig:B_rainbow}, except that a URL's score at each time point is now the mean of all comment scores on a given \textit{post}, rather than the score of a single comment.
    \cbrk
    Asterisks represent confidence that the relevant slope is positive or negative, for toxicity score and downvote score respectively.
    }
    \label{fig:LA1_rainbow_reddit_links}
\end{figure}


\begin{figure}[h]
    \centering
    \includegraphics[width=0.45\columnwidth]{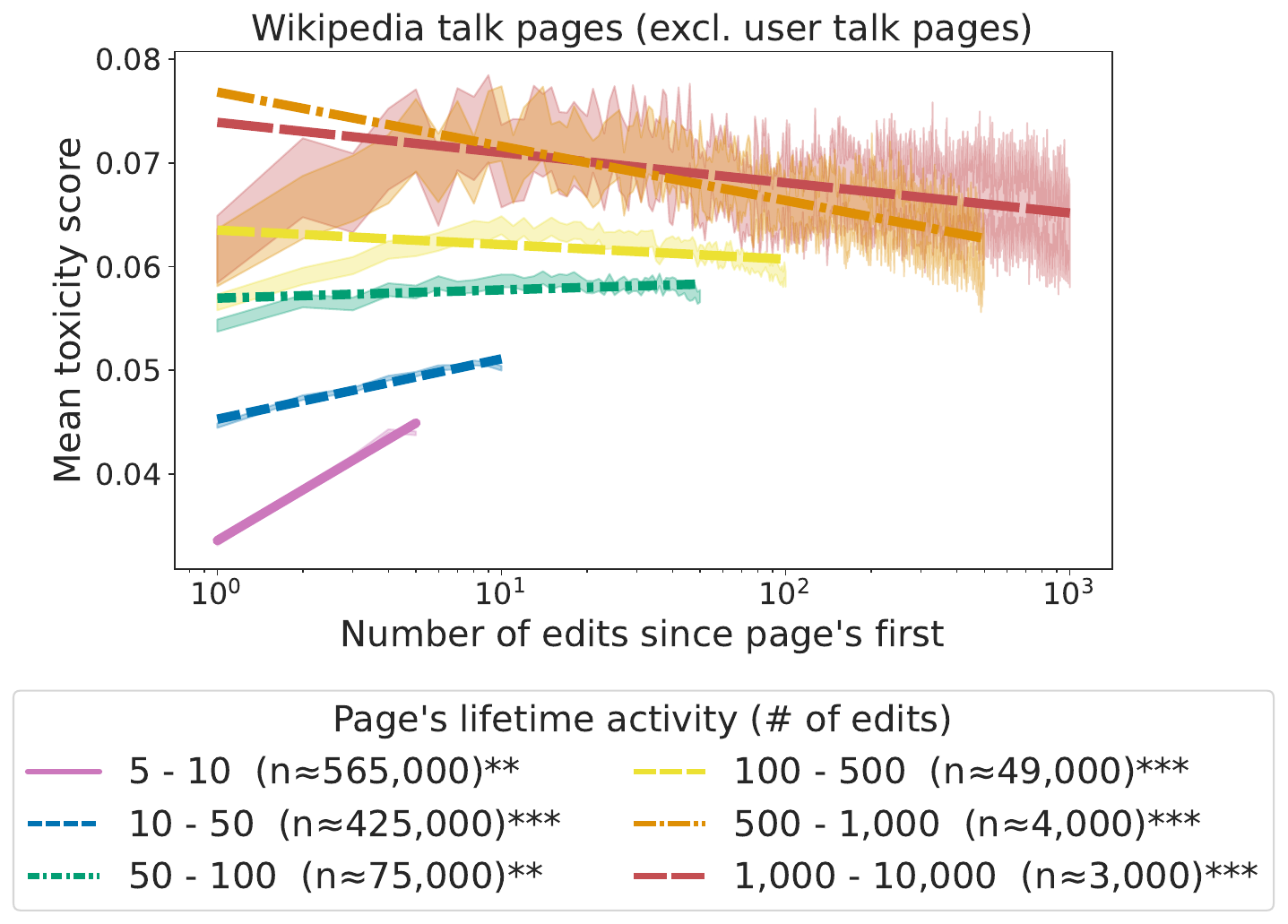} 
    
    \caption{Trends in toxicity score over the lifetime of a talk page, from the first \enquote{comment} (edit) to the hundredth.
    \cbrk
    The structure is the same as Fig. \ref{fig:B_rainbow}. Shaded bands are 95\% confidence intervals over the mean score for the \textit{n}th comment on each talk page. Trend lines are fit to the means over all pages in a bucket.
    }
    \label{fig:LA2_rainbow_wiki_page}
\end{figure}


In the case of Reddit, we can follow specific URLs that are posted many times as top-level link posts. We focus on YouTube and Imgur URLs, since each points to a fixed video or image, and both sites make it easy to detect different URLs that point to the same item. As with users, we order posts of the same URL in chronological order, from the first time the URL was posted to the last. When we apply a score intended to quantify toxicity, the score for each post is the mean score over all comments on the post.

For Wikipedia, we use article \enquote{talk pages}: pages attached to each article used for editor discussion. Here a \enquote{comment} is an edit to a talk page, and we again arrange comments in chronological order.
Using these definitions, Figure \ref{fig:LA1_rainbow_reddit_links} and \ref{fig:LA2_rainbow_wiki_page} look strikingly similar to Figure \ref{fig:B_rainbow}.
As with users, more-active discussions tend to be more toxic, and Figure \ref{fig:LA1_rainbow_reddit_links} shows a general downward trend over time. (However, Figure \ref{fig:LA2_rainbow_wiki_page} shows less consistent trends). 

Interestingly, Wikipedia pages look more similar to Reddit users than Wikipedia users: Wikipedia users are the only set where higher activity corresponds to \textit{lower} toxicity. Reddit users, Reddit content, and Wikipedia content all follow the opposite pattern.
Trajectories are also roughly similar between buckets, at least for Reddit URLs. However, there's not as much of a clear pattern over time for URLs, unlike for users as in Figures \ref{fig:D_slope} and \ref{fig:WB_slope}. The corresponding figures for URLs are in the supplement, Figures \ref{fig_supp:SAc2a_slope_yt_img}-\ref{fig_supp:SAc2b-c_wiki_page_slopes}.


\subsection*{Other metrics: Are these patterns universal?}
We've seen that similar patterns hold across two different metrics and two different datasets. Are these patterns specific to antisocial behavior, or would any metric of user behavior look similar?
In fact, every metric we tried, except for those related to antisocial behavior, does look different from the patterns we've observed. In Supplementary Section \ref{section_supp:D_alt_metrics_controls}, we examine two kinds of scores for Reddit users: various binary thresholds on a comment's upvote score, and \enquote{controversiality}, a binary measure defined by Reddit as comments with upvote/downvote ratios close to 1.\cite{justachetan2019whatWITHLINK} 

Recall the three main findings: users' mean toxicity is correlated with their activity level; users change over their lifetimes; and the direction of change reverses in the mid-2010s.
For positive vote thresholds, there's correlation between mean score and activity level, though the shape of user trajectories is clearly different (Fig \ref{fig_supp:SDa_rainbow_metric_pos_thresholds}). Lifetime slopes also don't change much over time, though it's interesting to note that lower-activity users have consistently negative trends, and higher-activity users have consistently positive trends (Fig \ref{fig_supp:SDb_slope_metric_pos_thresholds}).
For negative vote thresholds, there's still some correlation between mean score and activity level, except for the least active users (Fig 
\ref{fig_supp:SDa_rainbow_metric_neg_thresholds}). However, different activity levels no longer have similar slopes, and for the more extreme thresholds, slopes don't change over time (Fig \ref{fig_supp:SDb_slope_metric_neg_thresholds}).
Controversiality, the metric that's most orthogonal to our toxicity and downvote scores, also looks the most different from the patterns we found. Controversiality doesn't show much of a difference between mean score for different activity levels (Fig \ref{fig_supp:SDa_rainbow_metric_controversiality}). Slopes also show little difference between activity buckets and over time, and are generally positive (Fig \ref{fig_supp:SDb_slope_metric_controversiality}).





\section*{Discussion}
In summary, we've seen that users, on average, show different amounts of antisocial behavior depending on how long they've been posting. However, the direction of change over their lifetime depends on when they began posting: before 2013-2015, users get less toxic over their lifetimes, while after that time, users got more toxic over their lifetimes. Also, users who would go on to post more were different from their very first post. On Reddit, high-activity users are \textit{more} toxic than lower-activity users, whereas on Wikipedia, higher-activity users were \textit{less} toxic. 
When tracking lifetimes of information instead of users, URLs that would go on to draw more discussion (higher-activity information) generally had more toxic discussion from the beginning, similar to the pattern for Reddit users.

It is interesting to ask how these findings about the contrast in toxicity by activity level --- first for Reddit, then for Wikipedia --- accord with intuition about social media.
Earlier, it may have seemed \enquote{obvious} that the most active users were the most toxic, given social media's reputation for toxic behavior. When Figure \ref{fig:WA1_rainbow} shows the opposite, it might seem equally \enquote{obvious} that people with toxic behavior by definition don't follow norms and can't last long. 
But the point is that both of these \enquote{obvious} outcomes are observed, in different settings.
To quote Duncan Watts in \textit{Everything is Obvious}, \enquote{When every answer and its opposite appears equally obvious \textelp{}  something is wrong with the entire argument of \enquote{obviousness}.}\cite{watts2011everything, lazarsfeld1949american} 
The question is thus not which tendency is true of social media in general - whether high- or low-activity users are more toxic - but which, if any, is characteristic of a particular community. Viewed this way, as a comparison across different sites, it accords with some underlying intuition that high-activity Wikipedia users are less toxic than low-activity ones.\cite{rawat2019automatic}
Indeed, Wikipedia is known for its strongly enforced social norms, and is often favorably compared to other social media giants; Wikipedia has been described as \enquote{one of the few remaining places that retains the faintly utopian glow of the early World Wide Web}.\cite{cooke2020wikipedia}

Given that our work describes a change happening in 2013-2015, it's useful to note other works that have observed changes around the mid-2010s. Waller and Anderson\cite{waller2021quantifying} find a shift in polarization of Reddit communities in 2016, and Jonathan Haidt's new book\cite{haidt2024anxious} discusses a \enquote{tidal wave} of adolescent mental illness around 2013\cite{blanchflower2024declining}, which he suggests is due to the rise of smartphones.\cite{remnick2024jonathan} While the methods used here aren't well suited to establishing causal mechanisms, connections between these and other potential factors may be an interesting avenue for future research.

Using the methodology in this paper, we have surfaced several related patterns in social media data: that users stratify by activity with respect to toxicity level, that this toxicity level changes over user lifetimes, and that these trends have not been stable over the history of the platform. This methodology may also be useful to form the basis for other investigations of how social media can affect user behavior.
Finally, several limitations should be acknowledged. Most importantly, the analysis can't determine the cause of the patterns we've found. While our findings are consistent across metrics and platforms, the multi-decade time period we analyze encompasses many changes to social norms, user demographics, technology, political polarization, platform moderation, external manipulation, and more - any of which could affect our results. 
The datasets also have limitations. On Reddit, 10\% of comments are deleted and unavailable for analysis, which could confound results, especially if toxic comments were more likely to be deleted. Neither metric used can perfectly capture antisocial behavior - as previously discussed, the Perspective API toxicity score has known flaws and biases, and the downvote metric captures comments that are downvoted for reasons other than antisocial behavior. 
Our analysis is also very high-level, aggregating across large groups of users.
Future work could analyze user trajectories at a more granular level, distilling patterns in the behavior of individual users and perhaps providing more insight into causality.






 \section*{Methods}

\subsection*{Metric: Perspective API \enquote{toxicity} score}
We used the Perspective API\cite{perspectiveapi}  \enquote{SEVERE\_TOXICITY} score as our toxicity metric.  The API defines \enquote{SEVERE\_TOXICITY} as \enquote{A very hateful, aggressive, disrespectful comment or otherwise very likely to make a user leave a discussion or give up on sharing their perspective},\cite{ perspective_api_attribute_defs} and uses machine learning to assign text a score between 0 and 1 based on how well the text fits that definition.
Scores were generated for the text of each comment in the dataset, or in the case of Wikipedia, for the text of each talk page edit. The Perspective model is regularly updated, so scores for the same piece of text will change over time. Scores for the Reddit data were generated between August 2023 and February 2024.  Scores for the Wikipedia data were included with the Wikiconv dataset,\cite{hua2018wikiconv} and generated sometime before 2018.

Toxicity scores for non-English text were ignored for most datasets, due to the potential for between-language bias.\cite{nogara2023toxic} For the Reddit user dataset, English text was determined using Perspective API's language detection feature. For Wikipedia, this was done by only generating scores for the English section of Wikipedia. Comments were not filtered by language for the Reddit post URL dataset, due to the additional cost of language detection for such a large number of comments.

\subsection*{Metric: Downvote score}
Reddit’s content ranking depends on user voting, where any user can \enquote{upvote} or \enquote{downvote} a post or comment. Reddit asks users to vote based on whether a comment \enquote{contributes to conversation}, and is on-topic to the community it’s posted in.\cite{reddiquette} To prevent manipulation, Reddit only publishes the total score (upvotes minus downvotes), and some \enquote{fuzzing} is applied.
By \enquote{downvote score}, we mean the percentage of comments that are downvoted in a given set of comments. The downvote score is a binary value for each Reddit comment: a comment is considered \enquote{downvoted} if it got more downvotes than upvotes, meaning its vote score is 0 or lower (since all comments start with a vote score of 1).  Vote score thresholds of -1 and -2 were qualitatively similar; see Supplementary Section \ref{section_supp:D_alt_metrics_controls}. (Positive thresholds, however, were different.)

\subsection*{Dataset: Reddit user sample}
Reddit data was obtained from the Pushshift archive of all comments and posts on Reddit.\cite{baumgartner2020pushshift} The Reddit user sample consists of all comments made by a sample of 1 million user accounts between January 2005 and December 2022. This represents roughly 1\% of the 98.6 million users who commented on Reddit during that period. Users were randomly sampled from all users who had made at least one, but fewer than 50,000 comments. 

The 50,000-comment threshold was chosen to avoid bots and other unusual accounts after a rough analysis of high-volume users: 27\% of accounts with > 100,000 comments have \enquote{bot} in their name, while only 0.25\% of accounts with < 10,000 comments do. In our sample, only 1,200 users (0.12\%) have more than 10,000 comments. Of those, seven have \enquote{bot} in the name, and on inspection, six of those seven are actually bots.

Initial analysis of the dataset showed that data from 2005-2009 was noisy, due to the sparseness of data from when the Reddit user base was very small. We removed all users who had begun commenting before 2009, roughly 1,500 users (0.15\%).

After filtering, our dataset had 123 million comments (Table \ref{tab:dataset_sizes}).

92.7\% of comments were entirely in English, and 95.5\% contained at least some English (Supplementary Section \ref{section_supp:SJ_languages}). Toxicity scores for comments that were not entirely in English were ignored, as scoring accuracy could be compromised.\cite{nogara2023toxic} Results remained the same after this change.

Roughly 10\% of comments on Reddit are deleted (Supplementary Section \ref{section_supp:SK_deleted_data}), When comments are deleted, author information is also removed, so it's impossible to tell which comments belong to authors in our sample. Readers should be aware that some data are missing.

\subsection*{Dataset: Wikipedia talk page edits}
Wikipedia analysis used the Wikiconv dataset\cite{hua2018wikiconv} of all talk page edits on English Wikipedia. Bot accounts were roughly filtered out by removing any account with a name ending with \enquote{bot} (case-insensitive), as Wikipedia policy recommends for bot accounts.\cite{wiki_bot_account_policy}  Accounts containing \enquote{MediaWiki} in the name were also removed. 
After filtering, the dataset had 122 million edits from 2.8 million authors (Table \ref{tab:dataset_sizes}). 
When analyzing by pages, we ignore edits to user talk pages, which generally contain interpersonal conversations rather than discussions of a specific topic. User talk pages comprise $\sim60\%$ of edits in the dataset; versions of Figures \ref{fig:B_rainbow} and \ref{fig:D_slope} for user talk pages can be found in Supplementary Section  \ref{section_supp:A3_url_rainbow_slopes}. 

\subsection*{Dataset: Reddit post URLs}
Reddit posts can either be text, images, videos, or URLs. To compare conversations about the same topic over time, we chose a subset of URL posts: YouTube and Imgur links. This choice was made since these URLs are easily canonicalized, meaning that links that point to the same page can be identified. Canonicalization means we can be more confident that we've collected \textit{all} discussions of the same URL into one group.
The post dataset consists of all posts of YouTube and Imgur URLs that were posted at least 5 times; only posts with at least one comment were counted. The dataset has 104 million comments on 11 million posts (Table \ref{tab:dataset_sizes}). For each post, an average score for each metric was calculated using all comments on that post.

\begin{table}[]
\centering
\begin{tabular}{lllll}
\hline
\textbf{Dataset}          & \textbf{Comments} & \textbf{Authors} & \textbf{Pages / Posts}  &\textbf{Date tox scores collected}\\ \hline
Reddit users              & 123,564,988          & 998,472                 & \multicolumn{1}{c}{-} &     Aug - Oct 2023\\
Reddit URLs               & 239,797,094          & \multicolumn{1}{c}{-}   & 10,973,235                 &Jan - Feb 2024\\
Wikipedia users and pages & 122,890,496          & 2,881,627               & 21,801,067                 &2018\\ \hline
\end{tabular}
\caption{Datasets used in this paper.}
\label{tab:dataset_sizes}
\end{table}





\section*{Additional information}

\textbf{Author contributions statement} K.B. and J.K. conceived the experiment(s),  K.B. analyzed the data.  All authors reviewed the manuscript. 
\\
\noindent\textbf{Competing interests} The authors declare no competing interests.
\\
\noindent\textbf{Data availability} The Wikipedia dataset analysed during the current study is available in the Wikiconv repository, \url{https://github.com/conversationai/wikidetox/tree/main/wikiconv}. The Reddit dataset analysed during the current study was publicly available from Pushshift at the time of analysis, but Pushshift data is currently available only under the request procedure outlined at \url{https://web.archive.org/web/20250327162604/https://www.reddit.com/r/pushshift/comments/14ei799/pushshift_live_again_and_how_moderators_can/}.

\let\oldaddcontentsline\addcontentsline
\renewcommand{\addcontentsline}[3]{}
\bibliography{zotero,zzmanual}
\let\addcontentsline\oldaddcontentsline

\clearpage


\appendix



\renewcommand{\thesection}{\Alph{section}}

\renewcommand{\thepage}{S\arabic{page}}
\renewcommand{\thesection}{S.\Alph{section}}
\renewcommand{\thetable}{S\arabic{table}}
\renewcommand{\thefigure}{S\arabic{figure}}
\renewcommand{\figurename}{Supplementary Figure}
\renewcommand{\contentsname}{Supplementary Material}

\clearpage
\tableofcontents
\clearpage

\section{Redundant figures from the main text: datasets, metrics that were excluded for space}
\label{section_supp:A_many_rainbow_slopes}


\subsection{Wikipedia users}

\begin{figure}[h!]
    \centering
    \includegraphics[width=0.95\columnwidth]{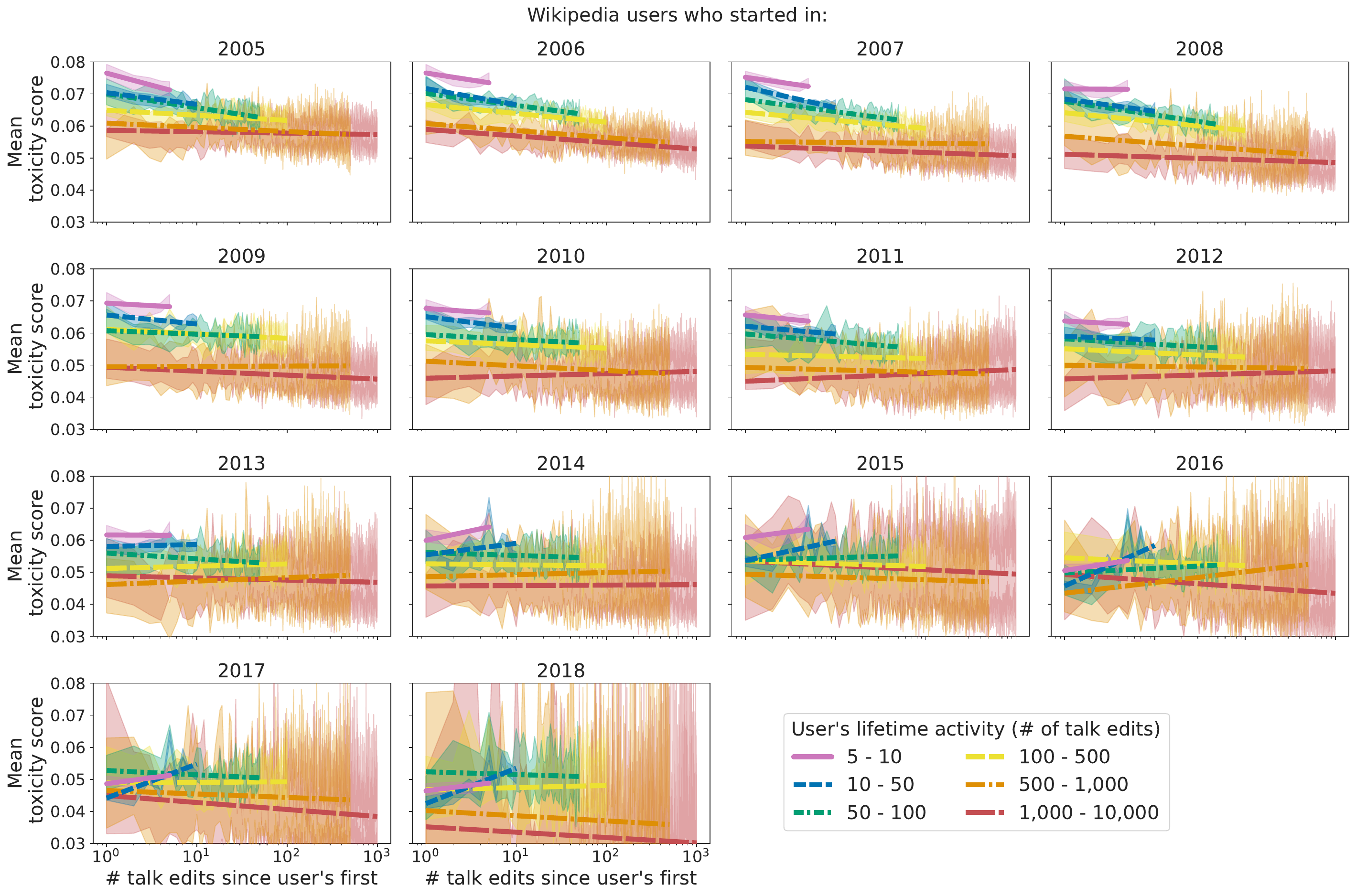} 
    \caption{Trends in toxicity score over user lifetimes for users who began in different years.
    \cbrk
    Structure is the same as Fig. \ref{fig:C_rainbow_grid}; individual plots have the same structure as Fig. \ref{fig:B_rainbow}, showing 95\% confidence intervals and trend lines fitted to bucket means.
    }
    \label{fig_supp:SAb_rainbow_grid_W_AUTH0_tox}
\end{figure}

\clearpage
\subsection{URLs} \label{section_supp:A3_url_rainbow_slopes}

\begin{figure}[h!]
    \centering
    \includegraphics[width=0.45\columnwidth]{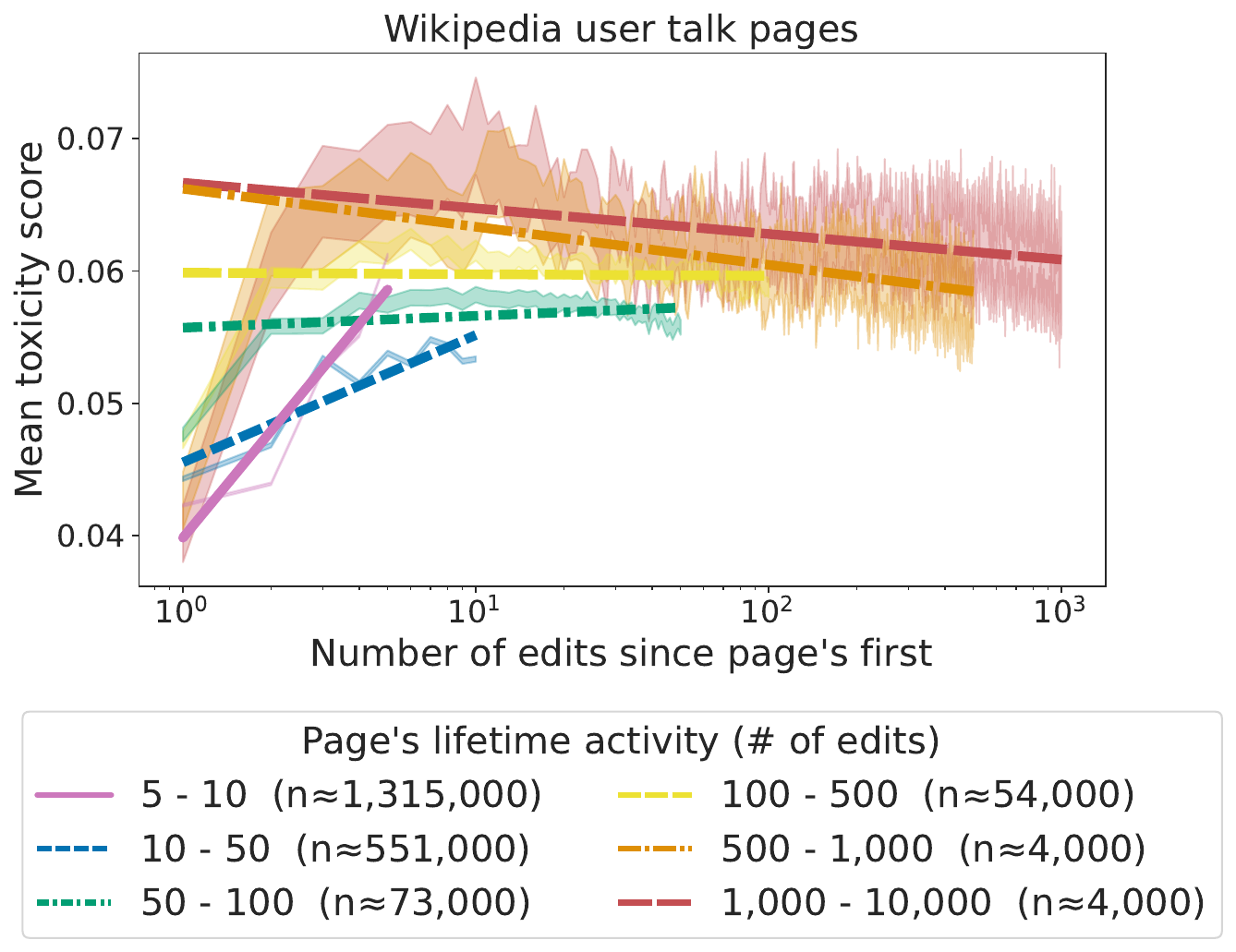} 
    \caption{Trends in toxicity over the lifetime of a Wikipedia talk page, from the first edit (\enquote{comment}) to the thousandth.
    \cbrk
    The structure is the same as Fig. \ref{fig:B_rainbow}, showing 95\% confidence intervals and trend lines fitted to bucket means. Asterisks represent confidence levels for each slope.}
    \label{fig_supp:SAc1_rainbow_wiki_page_usertalk}
\end{figure}

\begin{figure}[h!]
    \centering
    \includegraphics[width=0.95\columnwidth]{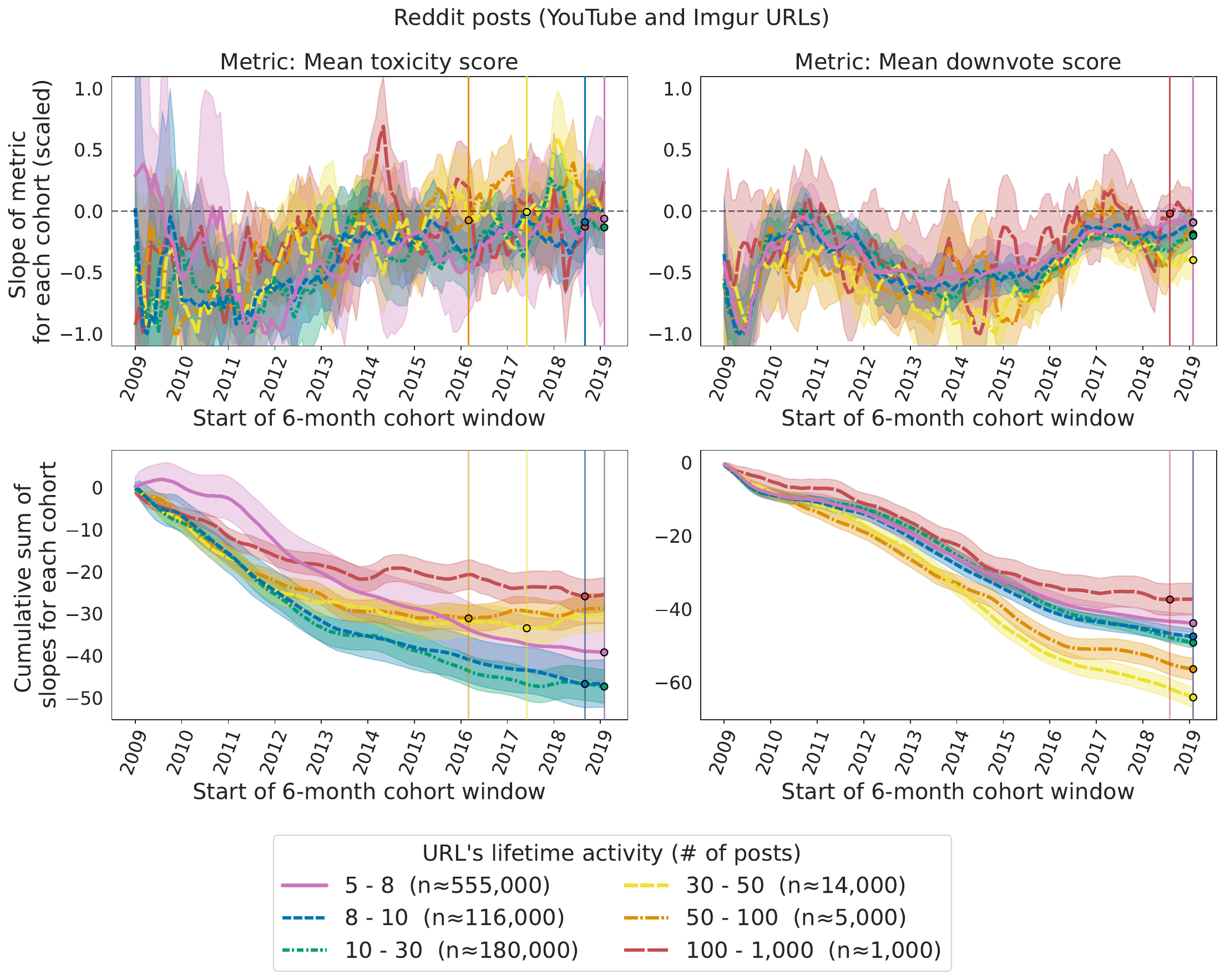} 
    \caption{Changes in trends over time for toxicity and downvote scores. Each point on a line in this plot represents the slope of an entire line from a plot such as those in Figure \ref{fig:LA1_rainbow_reddit_links}.
    \cbrk
    The structure is the same as Fig. \ref{fig:D_slope}/Fig.\ref{fig_supp:SHa_integral_1mo}. The lower plots show the cumulative sum of the upper plots. The minimum value for each lifetime activity bucket is marked by a vertical line. If there were an internal minimum, that would be the point where the slopes could be said to cross zero. For visibility, the slopes are scaled so that the maximum absolute value is 1 for each lifetime activity bucket.
    }
    \label{fig_supp:SAc2a_slope_yt_img}
\end{figure}

\begin{figure}[h]
    \centering
    
    \begin{subfigure}[b]{0.45\columnwidth}
        \centering
        \includegraphics[width=1\columnwidth]{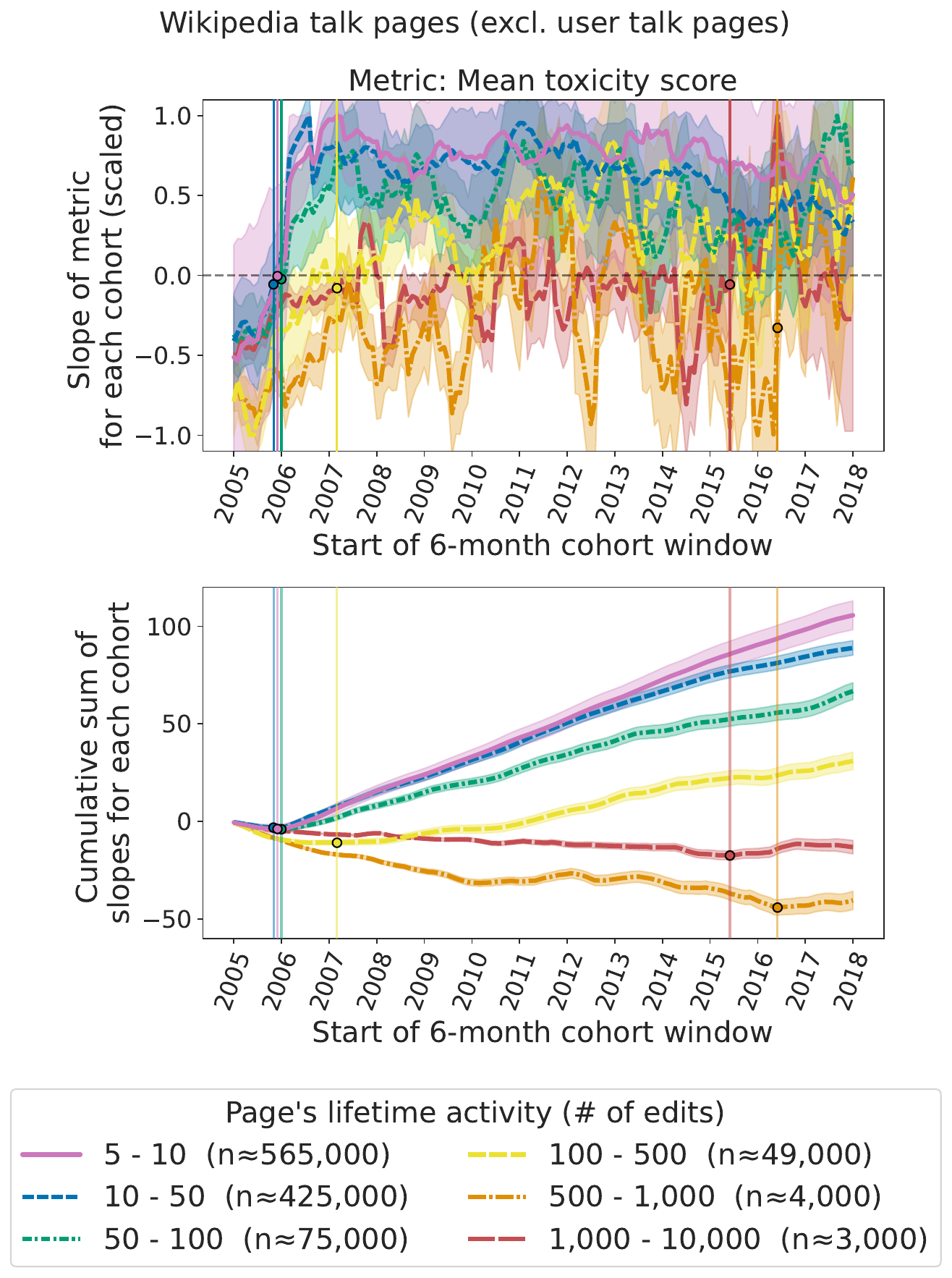} 
        \caption{}
        \label{fig_supp:SAc2b_wiki_page_slopes_other}
    \end{subfigure}
    \begin{subfigure}[b]{0.45\columnwidth}
        \centering
        \includegraphics[width=1\columnwidth]{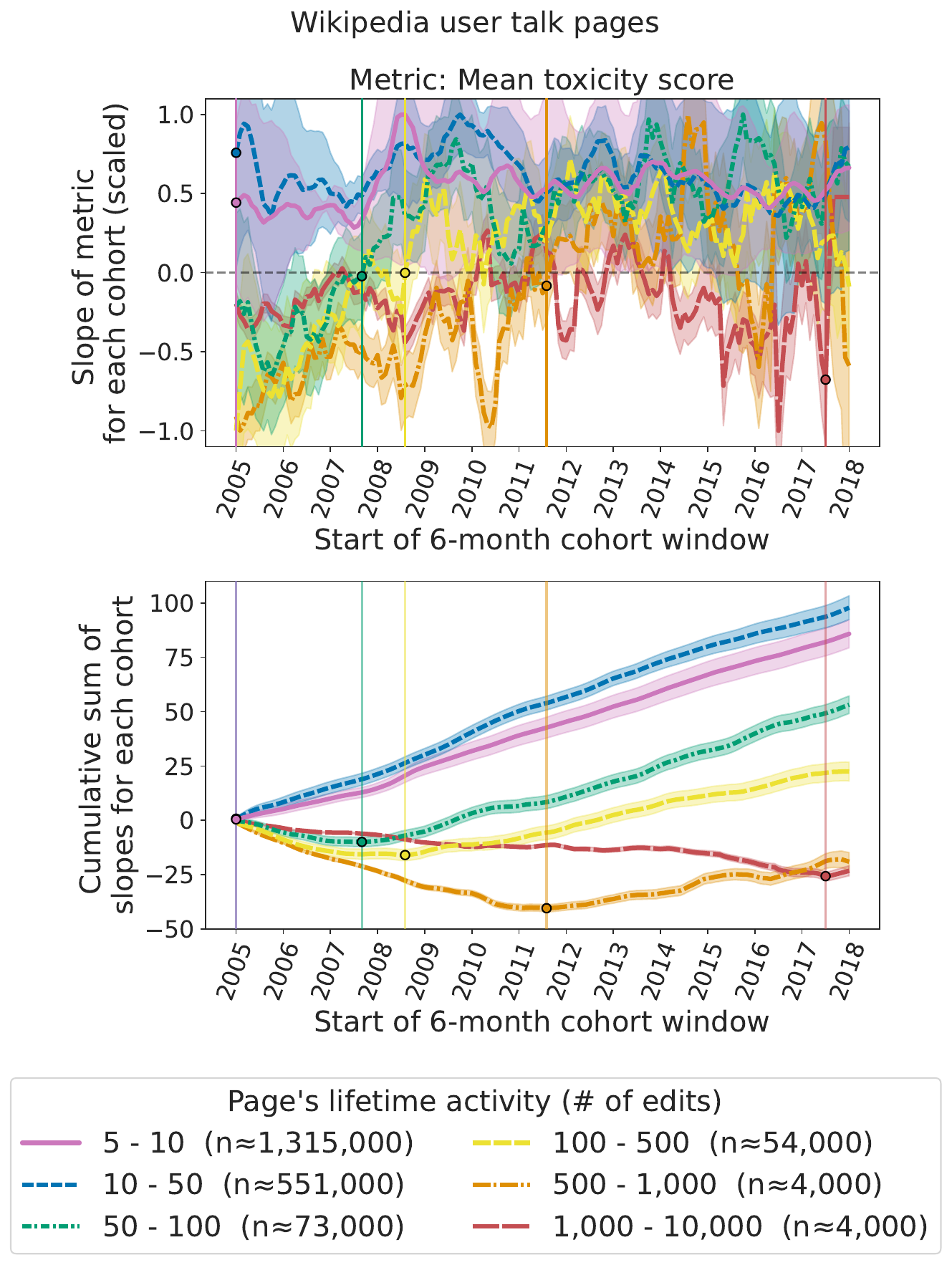} 
        \caption{}
        \label{fig_supp:SAc2c_wiki_page_slopes_user}
    \end{subfigure}
    
    \caption{Changes in toxicity trends over time for \subref{fig_supp:SAc2b_wiki_page_slopes_other}) non-user and \subref{fig_supp:SAc2c_wiki_page_slopes_user}) user talk pages. Each point on a line in this plot represents the slope of an entire line from a plot such as those in Figure \ref{fig:LA2_rainbow_wiki_page}.
    \cbrk
    The structure is the same as Fig. \ref{fig:WB_slope}/Fig.\ref{fig_supp:SHb_integral_figWB_1mo}. The lower plots show the cumulative sum of the upper plots. The minimum value for each lifetime activity bucket is marked by a vertical line. If there were an internal minimum, that would be the point where the slopes could be said to cross zero. For visibility, the slopes are scaled so that the maximum absolute value is 1 for each lifetime activity bucket.}
    \label{fig_supp:SAc2b-c_wiki_page_slopes}
\end{figure}

\clearpage
\section{Statistical validation}

\subsection{Bootstrapping: robustness check for date of trend reversal}
\label{section_supp:SB1_bootstrap}

\begin{figure}[h!]
    \centering
    \includegraphics[width=0.95\columnwidth]{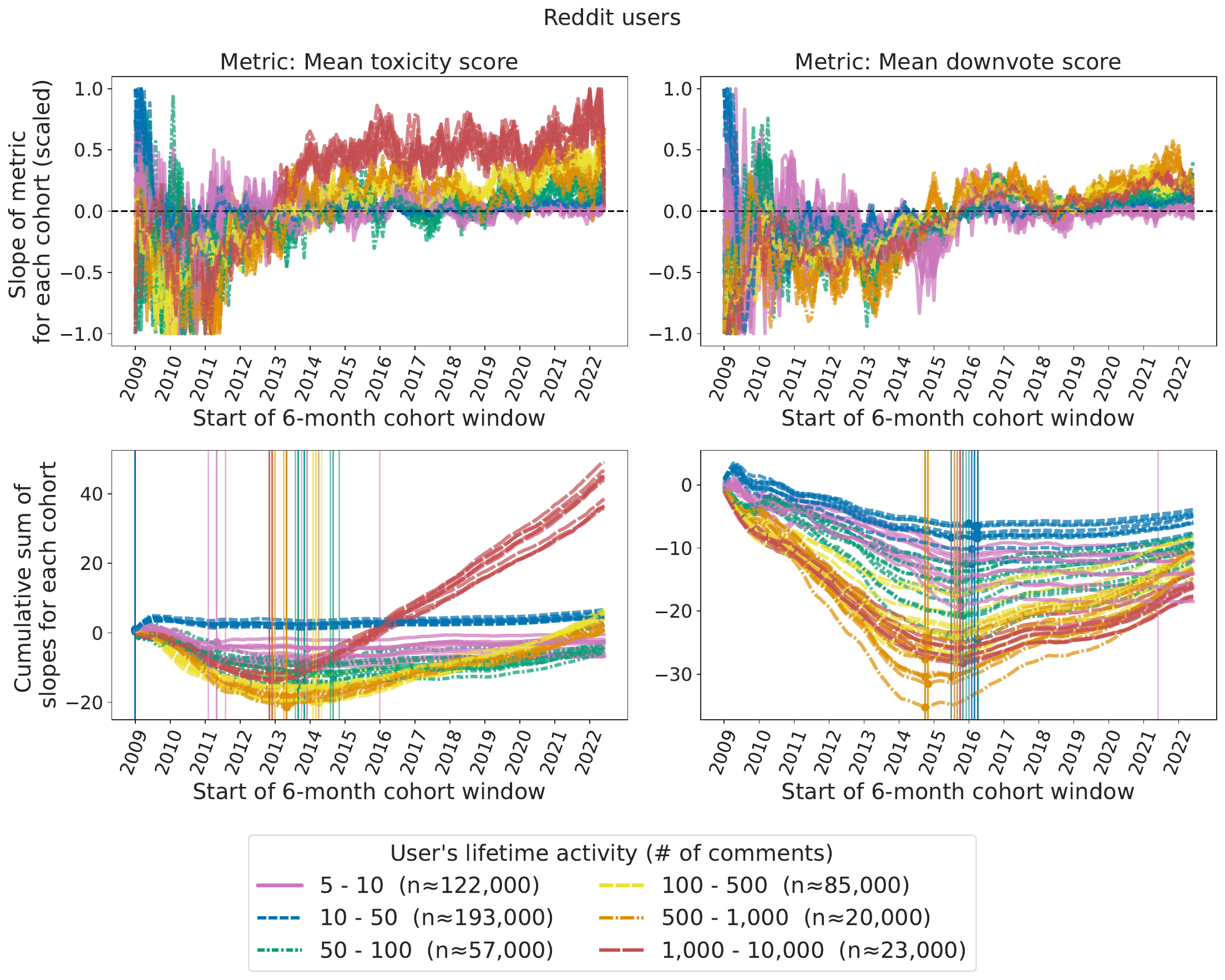} 
    \caption{Results of 10 bootstrapping runs for Figure \ref{fig:D_slope}/Fig. \ref{fig_supp:SHa_integral_1mo}. For each run, we sampled with replacement from our dataset to generate a dataset of the same size (123 million comments). All 10 runs are plotted on top of each other, as are the vertical lines marking the minimum (zero-crossing) point, where cohort lifetime slopes change from negative to positive. 
    \cbrk
    Note: This figure was generated before filtering out the 7\% of comments that were not fully in English.}
    \label{fig_supp:SB_bootstrap}
\end{figure}

\sisetup{
    round-mode=places,
    round-precision=2,
    group-digits=true,  
} 

\clearpage 
\subsection{Tables of \textit{t}-statistics and \textit{p}-values for slopes}
\label{section_supp:SB2_pval_tables}

All \textit{p}-values in this section are for a two-sided \textit{t}-test whose null hypothesis is that the slope is zero.\cite{webb2021mostly}

Note that the slopes are fit to values that are averages over a group of users. Hierarchical linear modeling might be a useful direction for future work, but is not something we currently have experience with. We did try using scipy.stats.ODR.RealData to do the fitting, which takes into account a standard error for each data point. (We set fit\_type=2, which uses regular least squares instead of ODR, for consistency.) However, since slopes obtained from this method were very similar, and p-values were usually slightly better, we felt comfortable sticking with simple linear regression. 

N.B. In this case, we're not \textit{technically} using linear regression, since our x-axis is log-transformed (reasoning below). However, doing linear regression on log-transformed data gives slopes and p-values that are identical to several decimal places. Since p-values for linear regression slopes can be obtained directly from scipy.stats.linregress, this also gives us confidence in our hand-calculated p-values.

Fitting of slopes was done with nonlinear least squares regression, fitting to the function $y = m \; log(x) + b$. This is equivalent to a linear regression on data with a log-transformed x-axis. This was done because when the \textit{y-axis} is log-transformed, linear regression will be off due to asymmetric error.\cite{holbert2018problems,clauset2009powerlaw} In hindsight, I \textit{think} this should only be a problem when the y-axis is transformed, not when the x-axis is transformed. However, we experimented with several different ways to transform and fit the data, so it made more sense to be consistent in the code. And again, in practice slopes and p-values were the same either way.


\begin{table}[h!]
\centering
\begin{tabular}{r | S[scientific-notation=true] | S[scientific-notation=true] | S[table-format=1.2e1]@{\extracolsep{-25pt}} | S[scientific-notation=true]@{\extracolsep{1pt}} | r}\toprule
\multicolumn{1}{c|}{\textbf{\makecell{User's lifetime \\ activity  \\ (\# of comments)}}} & 
\multicolumn{1}{c|}{\textbf{\makecell{Slope over \\ lifetime for \\ mean toxicity score}}} & 
\multicolumn{1}{c|}{\textbf{\makecell{Standard error \\ of slope}}} & 
\multicolumn{1}{c|}{\textbf{\makecell{t-statistic}}} & 
\multicolumn{1}{c|}{\textbf{\makecell{\textit{p}-value}}} & 
\multicolumn{1}{c|}{\textbf{\makecell{Significance}}} \\ \midrule
5 - 10 & 0.00022066376926530243 & 9.895447946631003e-05 & 2.2299523018604677 & 0.11200152437733198 & n.s. \\ 
10 - 50 & 0.0004239188528676792 & 9.565344954865701e-05 & 4.431819812750612 & 0.0021916880466482324 & ** \\ 
50 - 100 & 0.0005418265334115765 & 3.866824661751371e-05 & 14.01218262547677 & 1.3877691751140363e-18 & *** \\ 
100 - 500 & 0.0005782908566555609 & 2.3215013014082288e-05 & 24.91021031540099 & 3.4753065123033514e-44 & *** \\ 
500 - 1,000 & 0.0005090724303207039 & 2.1074118080626052e-05 & 24.156286321120437 & 6.625869730111254e-86 & *** \\ 
1,000 - 10,000 & 0.0005377163835489373 & 1.4325370120613761e-05 & 37.535950486555336 & 5.401764235384197e-193 & *** \\ 
\end{tabular}
\caption{Slopes for mean toxicity score in Figure \ref{fig:B_rainbow}.}
\end{table}

\begin{table}[h!]
\centering
\begin{tabular}{r | S[scientific-notation=true] | S[scientific-notation=true] | S[table-format=1.2e1]@{\extracolsep{-25pt}} | S[scientific-notation=true]@{\extracolsep{1pt}} | r}\toprule
\multicolumn{1}{c|}{\textbf{\makecell{User's lifetime \\ activity  \\ (\# of comments)}}} & 
\multicolumn{1}{c|}{\textbf{\makecell{Slope over \\ lifetime for \\ mean downvote score}}} & 
\multicolumn{1}{c|}{\textbf{\makecell{Standard error \\ of slope}}} & 
\multicolumn{1}{c|}{\textbf{\makecell{t-statistic}}} & 
\multicolumn{1}{c|}{\textbf{\makecell{\textit{p}-value}}} & 
\multicolumn{1}{c|}{\textbf{\makecell{Significance}}} \\ \midrule
5 - 10 & 0.00962657956396136 & 0.0995258704162882 & 0.09672439460912159 & 0.9290448055098682 & n.s. \\ 
10 - 50 & 0.17803311087389564 & 0.025231709352398 & 7.055927459665967 & 0.00010654082464232175 & *** \\ 
50 - 100 & 0.16587136452723586 & 0.015873304005257074 & 10.449706278686591 & 5.883065707738837e-14 & *** \\ 
100 - 500 & 0.10775922700357654 & 0.008201016737091234 & 13.139739919833035 & 2.4114237248210818e-23 & *** \\ 
500 - 1,000 & 0.009697594890538412 & 0.008062616748914335 & 1.2027850501320474 & 0.22963104026586972 & n.s. \\ 
1,000 - 10,000 & -0.07335679815680914 & 0.0057814490774774395 & -12.688306542832365 & 2.6184388021406513e-34 & *** \\ 
\end{tabular}
\caption{Slopes for mean downvote score in Figure \ref{fig:B_rainbow}.}
\end{table}

\begin{table}[h!]
\centering
\begin{tabular}{r | S[scientific-notation=true] | S[scientific-notation=true] | S[table-format=1.2e1]@{\extracolsep{-25pt}} | S[scientific-notation=true]@{\extracolsep{1pt}} | r}\toprule
\multicolumn{1}{c|}{\textbf{\makecell{User's lifetime \\ activity  \\ (\# of comments)}}} & 
\multicolumn{1}{c|}{\textbf{\makecell{Slope over \\ lifetime for \\ mean toxicity score}}} & 
\multicolumn{1}{c|}{\textbf{\makecell{Standard error \\ of slope}}} & 
\multicolumn{1}{c|}{\textbf{\makecell{t-statistic}}} & 
\multicolumn{1}{c|}{\textbf{\makecell{\textit{p}-value}}} & 
\multicolumn{1}{c|}{\textbf{\makecell{Significance}}} \\ \midrule
5 - 10 & -0.0014698631869512124 & 0.00046852074473213446 & -3.137242488145475 & 0.05177398527800015 & n.s. \\ 
10 - 50 & 0.000604564099517461 & 0.0004334094176840061 & 1.3949030059107803 & 0.20055458858058595 & n.s. \\ 
50 - 100 & -0.0004884379911469767 & 0.0001735611852888153 & -2.814212119686721 & 0.0070686308740315115 & ** \\ 
100 - 500 & -0.0006092022268307855 & 9.03855903476219e-05 & -6.740037040061375 & 1.0988900323185627e-09 & *** \\ 
500 - 1,000 & -0.0008700504800731039 & 6.10495993403914e-05 & -14.251534645166206 & 6.76029168287745e-39 & *** \\ 
1,000 - 10,000 & -0.0005834687062314786 & 3.422630636063379e-05 & -17.047375784100655 & 2.1837139459556828e-57 & *** \\ 
\end{tabular}
\caption{Slopes for 2009-2012 in Figure \ref{fig:E1_rainbow_cut_tox}.}
\end{table}

\begin{table}[h!]
\centering
\begin{tabular}{r | S[scientific-notation=true] | S[scientific-notation=true] | S[table-format=1.2e1]@{\extracolsep{-25pt}} | S[scientific-notation=true]@{\extracolsep{1pt}} | r}\toprule
\multicolumn{1}{c|}{\textbf{\makecell{User's lifetime \\ activity  \\ (\# of comments)}}} & 
\multicolumn{1}{c|}{\textbf{\makecell{Slope over \\ lifetime for \\ mean toxicity score}}} & 
\multicolumn{1}{c|}{\textbf{\makecell{Standard error \\ of slope}}} & 
\multicolumn{1}{c|}{\textbf{\makecell{t-statistic}}} & 
\multicolumn{1}{c|}{\textbf{\makecell{\textit{p}-value}}} & 
\multicolumn{1}{c|}{\textbf{\makecell{Significance}}} \\ \midrule
5 - 10 & 0.00028886448131260967 & 0.00010691785055541796 & 2.701742317227792 & 0.07367727225825789 & n.s. \\ 
10 - 50 & 0.000443753149271154 & 8.97875781906097e-05 & 4.942255468001511 & 0.001131848455035649 & ** \\ 
50 - 100 & 0.0006298312429574438 & 4.262770058059352e-05 & 14.775163435490999 & 1.741780525988563e-19 & *** \\ 
100 - 500 & 0.0007064477473798933 & 2.4153557010888264e-05 & 29.2481868016968 & 3.255445483633394e-50 & *** \\ 
500 - 1,000 & 0.0007219960777577376 & 2.2781976205249392e-05 & 31.691547355377196 & 1.7136304664996922e-121 & *** \\ 
1,000 - 10,000 & 0.0007950883736326253 & 1.5664459480968984e-05 & 50.75747264682778 & 9.90262247642098e-279 & *** \\ 
\end{tabular}
\caption{Slopes for 2014-2022 in Figure \ref{fig:E1_rainbow_cut_tox}.}
\end{table}

\begin{table}[h!]
\centering
\begin{tabular}{r | S[scientific-notation=true] | S[scientific-notation=true] | S[table-format=1.2e1]@{\extracolsep{-25pt}} | S[scientific-notation=true]@{\extracolsep{1pt}} | r}\toprule
\multicolumn{1}{c|}{\textbf{\makecell{User's lifetime \\ activity  \\ (\# of comments)}}} & 
\multicolumn{1}{c|}{\textbf{\makecell{Slope over \\ lifetime for \\ mean downvote score}}} & 
\multicolumn{1}{c|}{\textbf{\makecell{Standard error \\ of slope}}} & 
\multicolumn{1}{c|}{\textbf{\makecell{t-statistic}}} & 
\multicolumn{1}{c|}{\textbf{\makecell{\textit{p}-value}}} & 
\multicolumn{1}{c|}{\textbf{\makecell{Significance}}} \\ \midrule
5 - 10 & 0.4964871477844504 & 0.7395955863888852 & 0.6712954443232623 & 0.5500977512373594 & n.s. \\ 
10 - 50 & -0.5055848554925814 & 0.05642716745817646 & -8.959954544365857 & 1.9148566498147525e-05 & *** \\ 
50 - 100 & -0.4007650897729617 & 0.05259880831940391 & -7.619280789392293 & 8.321158336380817e-10 & *** \\ 
100 - 500 & -0.3848675243229472 & 0.024297279144827114 & -15.839943313360065 & 8.966772253258958e-29 & *** \\ 
500 - 1,000 & -0.5116569313977508 & 0.01811904089236967 & -28.238632190140972 & 1.8956874584363983e-105 & *** \\ 
1,000 - 10,000 & -0.3724983933507882 & 0.011288513235103463 & -32.99800297814632 & 4.800131573010445e-162 & *** \\ 
\end{tabular}
\caption{Slopes for 2009-2014 in Figure \ref{fig:E2_rainbow_cut_vote}.}
\end{table}

\begin{table}[h!]
\centering
\begin{tabular}{r | S[scientific-notation=true] | S[scientific-notation=true] | S[table-format=1.2e1]@{\extracolsep{-25pt}} | S[scientific-notation=true]@{\extracolsep{1pt}} | r}\toprule
\multicolumn{1}{c|}{\textbf{\makecell{User's lifetime \\ activity  \\ (\# of comments)}}} & 
\multicolumn{1}{c|}{\textbf{\makecell{Slope over \\ lifetime for \\ mean downvote score}}} & 
\multicolumn{1}{c|}{\textbf{\makecell{Standard error \\ of slope}}} & 
\multicolumn{1}{c|}{\textbf{\makecell{t-statistic}}} & 
\multicolumn{1}{c|}{\textbf{\makecell{\textit{p}-value}}} & 
\multicolumn{1}{c|}{\textbf{\makecell{Significance}}} \\ \midrule
5 - 10 & 0.09887149434717947 & 0.08495664886119998 & 1.163787598410493 & 0.328655246107501 & n.s. \\ 
10 - 50 & 0.2694943368253844 & 0.033099711352355024 & 8.14189386597808 & 3.8468734352320935e-05 & *** \\ 
50 - 100 & 0.2729215938336545 & 0.017335739119528078 & 15.743291471559946 & 1.379941428202079e-20 & *** \\ 
100 - 500 & 0.23336812124229211 & 0.009148318675381099 & 25.50940009012864 & 4.604530262770917e-45 & *** \\ 
500 - 1,000 & 0.18227526510639921 & 0.00898320915955046 & 20.29066248698151 & 3.6929091647537666e-67 & *** \\ 
1,000 - 10,000 & 0.11400568931499282 & 0.006795472708406716 & 16.77671211510507 & 7.756806340792174e-56 & *** \\ 
\end{tabular}
\caption{Slopes for 2016-2022 in Figure \ref{fig:E2_rainbow_cut_vote}.}
\end{table}

\begin{table}[h!]
\centering
\begin{tabular}{r | S[scientific-notation=true] | S[scientific-notation=true] | S[table-format=1.2e1]@{\extracolsep{-25pt}} | S[scientific-notation=true]@{\extracolsep{1pt}} | r}\toprule
\multicolumn{1}{c|}{\textbf{\makecell{User's lifetime \\ activity  \\ (\# of talk edits)}}} & 
\multicolumn{1}{c|}{\textbf{\makecell{Slope over \\ lifetime for \\ mean toxicity score}}} & 
\multicolumn{1}{c|}{\textbf{\makecell{Standard error \\ of slope}}} & 
\multicolumn{1}{c|}{\textbf{\makecell{t-statistic}}} & 
\multicolumn{1}{c|}{\textbf{\makecell{\textit{p}-value}}} & 
\multicolumn{1}{c|}{\textbf{\makecell{Significance}}} \\ \midrule
5 - 10 & -0.00040686480082083575 & 0.001130752871624086 & -0.3598176144681915 & 0.7428222781541292 & n.s. \\ 
10 - 50 & -0.0003369574011015926 & 0.0006886988994348992 & -0.4892666466841715 & 0.6377848405694821 & n.s. \\ 
50 - 100 & -0.0011636444461129581 & 0.00010357804055963797 & -11.234470548252524 & 4.8936958850275744e-15 & *** \\ 
100 - 500 & -0.0007826506009080401 & 5.638979195932375e-05 & -13.879295768152467 & 7.2174505528477405e-25 & *** \\ 
500 - 1,000 & -0.0004971729634435642 & 4.426096852000126e-05 & -11.232762862360204 & 2.9904564654935667e-26 & *** \\ 
1,000 - 10,000 & -0.0003809192202332764 & 2.317638840972658e-05 & -16.435659150129435 & 6.663985872535434e-54 & *** \\ 
\end{tabular}
\caption{Slopes for mean toxicity score in Figure \ref{fig:WA1_rainbow}.}
\end{table}

\begin{table}[h!]
\centering
\begin{tabular}{r | S[scientific-notation=true] | S[scientific-notation=true] | S[table-format=1.2e1]@{\extracolsep{-25pt}} | S[scientific-notation=true]@{\extracolsep{1pt}} | r}\toprule
\multicolumn{1}{c|}{\textbf{\makecell{URL's lifetime \\ activity  \\ (\# of posts)}}} & 
\multicolumn{1}{c|}{\textbf{\makecell{Slope over \\ lifetime for \\ mean toxicity score}}} & 
\multicolumn{1}{c|}{\textbf{\makecell{Standard error \\ of slope}}} & 
\multicolumn{1}{c|}{\textbf{\makecell{t-statistic}}} & 
\multicolumn{1}{c|}{\textbf{\makecell{\textit{p}-value}}} & 
\multicolumn{1}{c|}{\textbf{\makecell{Significance}}} \\ \midrule
5 - 8 & -0.0006222782647420603 & 0.00015145815901611593 & -4.108581992442194 & 0.026106248482295164 & * \\ 
8 - 10 & -0.0006487009970742791 & 9.112769037055368e-05 & -7.118593639720903 & 0.00038653412714685187 & *** \\ 
10 - 30 & -0.0004969828474427207 & 4.641560690461959e-05 & -10.70723578954771 & 5.084851096757777e-06 & *** \\ 
30 - 50 & -0.00038315122112559774 & 7.841069232496521e-05 & -4.886466497932019 & 3.7829122599719974e-05 & *** \\ 
50 - 100 & -0.0006163413556423373 & 0.0001258903661576216 & -4.895857994968768 & 1.1483236235307187e-05 & *** \\ 
100 - 1,000 & -0.0007338510662325882 & 0.00016933249852096787 & -4.333787504716455 & 3.551724496503646e-05 & *** \\ 
\end{tabular}
\caption{Slopes for mean toxicity score in Figure \ref{fig:LA1_rainbow_reddit_links}.}
\end{table}

\begin{table}[h!]
\centering
\begin{tabular}{r | S[scientific-notation=true] | S[scientific-notation=true] | S[table-format=1.2e1]@{\extracolsep{-25pt}} | S[scientific-notation=true]@{\extracolsep{1pt}} | r}\toprule
\multicolumn{1}{c|}{\textbf{\makecell{URL's lifetime \\ activity  \\ (\# of posts)}}} & 
\multicolumn{1}{c|}{\textbf{\makecell{Slope over \\ lifetime for \\ mean downvote score}}} & 
\multicolumn{1}{c|}{\textbf{\makecell{Standard error \\ of slope}}} & 
\multicolumn{1}{c|}{\textbf{\makecell{t-statistic}}} & 
\multicolumn{1}{c|}{\textbf{\makecell{\textit{p}-value}}} & 
\multicolumn{1}{c|}{\textbf{\makecell{Significance}}} \\ \midrule
5 - 8 & -0.8973601336205956 & 0.09276110220052901 & -9.673883905354005 & 0.0023453572347835573 & ** \\ 
8 - 10 & -0.8157412081488229 & 0.04504956726440586 & -18.10763693602333 & 1.8258104445961635e-06 & *** \\ 
10 - 30 & -0.7153210519074649 & 0.036902430582762775 & -19.384117539444496 & 5.2084151428401766e-08 & *** \\ 
30 - 50 & -0.6416191974003946 & 0.04012124514859633 & -15.9920061060927 & 1.3022858574013777e-15 & *** \\ 
50 - 100 & -0.5597865894445473 & 0.04034118948180641 & -13.87630351596368 & 2.0231643528439063e-18 & *** \\ 
100 - 1,000 & -0.32717673331684644 & 0.04735678983719915 & -6.90876080159991 & 4.955902283526293e-10 & *** \\ 
\end{tabular}
\caption{Slopes for mean downvote score in Figure \ref{fig:LA1_rainbow_reddit_links}.}
\end{table}

\begin{table}[h!]
\centering
\begin{tabular}{r | S[scientific-notation=true] | S[scientific-notation=true] | S[table-format=1.2e1]@{\extracolsep{-25pt}} | S[scientific-notation=true]@{\extracolsep{1pt}} | r}\toprule
\multicolumn{1}{c|}{\textbf{\makecell{Page's lifetime \\ activity  \\ (\# of edits)}}} & 
\multicolumn{1}{c|}{\textbf{\makecell{Slope over \\ lifetime for \\ mean toxicity score}}} & 
\multicolumn{1}{c|}{\textbf{\makecell{Standard error \\ of slope}}} & 
\multicolumn{1}{c|}{\textbf{\makecell{t-statistic}}} & 
\multicolumn{1}{c|}{\textbf{\makecell{\textit{p}-value}}} & 
\multicolumn{1}{c|}{\textbf{\makecell{Significance}}} \\ \midrule
5 - 10 & 0.007046421696020495 & 0.0006228127347843739 & 11.313869005038988 & 0.001481003414916036 & ** \\ 
10 - 50 & 0.0025287541032074415 & 0.00022054796540948028 & 11.465778423811035 & 3.030993144612311e-06 & *** \\ 
50 - 100 & 0.000347872867812894 & 0.00011220095809085758 & 3.100444717514757 & 0.0032301918407520243 & ** \\ 
100 - 500 & -0.0006084513899718075 & 0.00013958799354373122 & -4.358909205046974 & 3.226933435562591e-05 & *** \\ 
500 - 1,000 & -0.002272361858669181 & 8.283667242082068e-05 & -27.431834141348627 & 1.2549045341649557e-101 & *** \\ 
1,000 - 10,000 & -0.0012668758188631316 & 5.8183825269990124e-05 & -21.77367701392017 & 2.593237920634398e-86 & *** \\ 
\end{tabular}
\caption{Slopes for mean toxicity score in Figure \ref{fig:LA2_rainbow_wiki_page}.}
\end{table}


\clearpage
\section{Calculation of zero-crossing point for slopes: Integral plots}
\label{section_supp:SH_integral_plots}

Figures such as Figure \ref{fig:D_slope} are graphs of values that might appear to have a \enquote{zero-crossing} point - that is, on the left of each plot, the values are mostly below 0, and on the right of the plot, the values are mostly above 0. 

Since these values are slopes, zero is an important threshold. It represents the point where the slope goes from negative (representing, for example, toxicity scores that are decreasing over a user's lifetime) to positive (toxicity scores increasing over a lifetime).

\begin{figure}[h!]
    \centering
    \includegraphics[width=0.8\columnwidth]{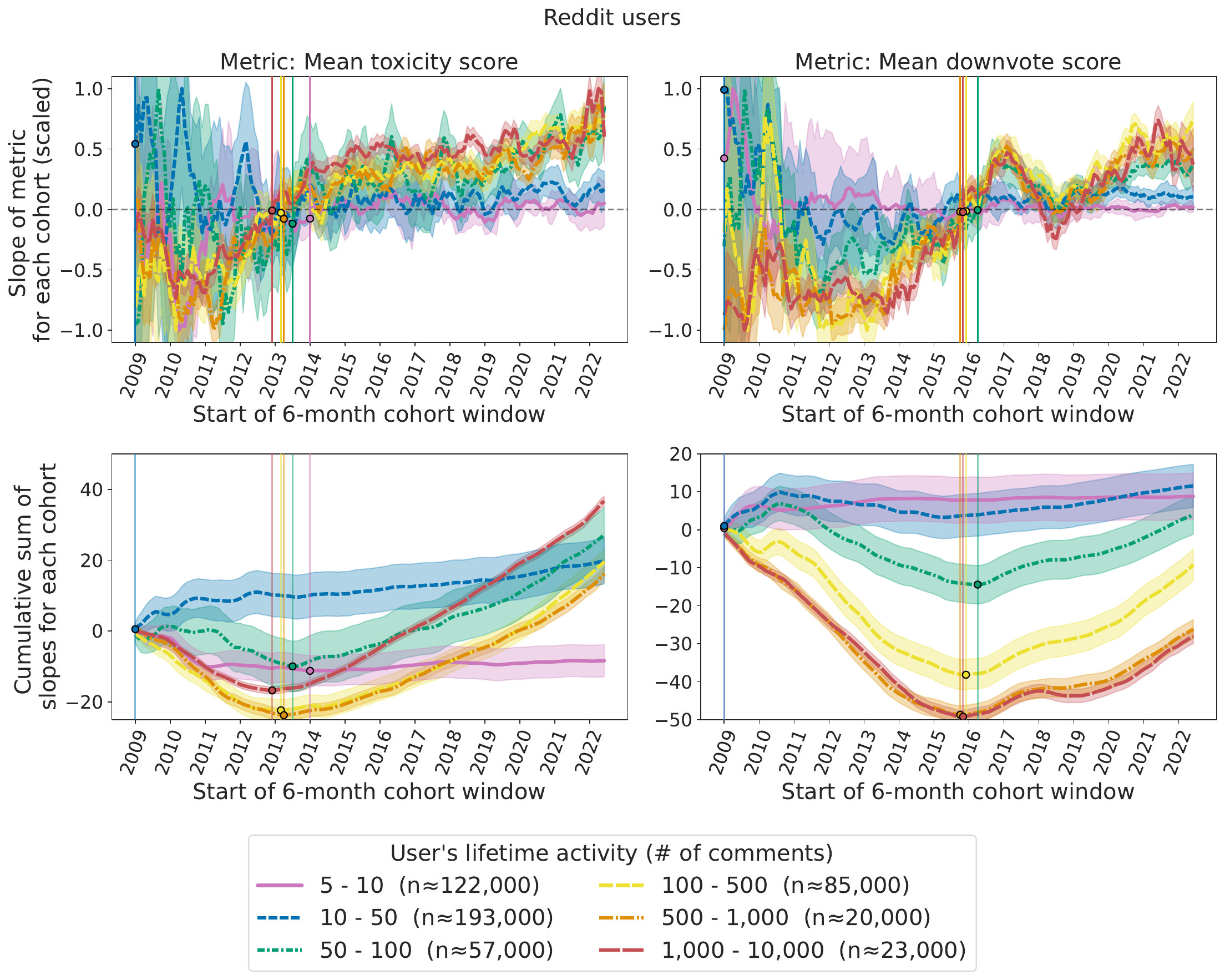} 
    \caption{
    Calculation of zero-crossing point for Figure \ref{fig:D_slope}.
    \cbrk
    The top row is a reproduction of the slope plots from Fig. \ref{fig:D_slope}.
    \cbrk    
    Below each slope plot is a plot of the \enquote{integral} (cumulative sum) of those slopes. The minimum point of each line on this plot is marked by a vertical line. This point is a good proxy for the point at which the corresponding line on the upper plot crosses zero, which is marked by a matching vertical line.
    \cbrk
    Shaded areas on the integral plots represent two versions of 95\% confidence intervals. The lighter, wider intervals are very conservative, while the narrower, darker intervals are very optimistic. See above for explanation.
    }
    \label{fig_supp:SHa_integral_1mo}
\end{figure}

Some lines cross zero only once - for example, the red line on the top left plot in Figure \ref{fig_supp:SHa_integral_1mo}
 (the top two plots in this figure are a reproduction of Fig. \ref{fig:D_slope}). But for more noisy lines, the \enquote{zero-crossing} point might be less clear.

The quantitative definition we use for a line's zero-crossing point is: the minimum point of the line's integral. This definition makes good intuitive sense, since this is the point at which the cumulative sum of values is at its most negative - meaning most values to the left of that point are negative, and most values to the right of that point are positive. In other words, it's the point that best separates the negative and positive values.

Confidence intervals for the cumulative sums were calculated as $\sqrt{\sum{r_i^2}}$ , where $r_i$ are the radii of confidence intervals for each of the summed values. This makes the optimistic assumption that errors are perfectly uncorrelated.\cite{pere2016answer} However, these are just quick and easy calculations since we already have a better sense of confidence from the bootstrapping in Section \ref{section_supp:SB1_bootstrap}. (We could calculate confidence intervals directly from the bootstrapped sums instead, but it would be very computationally intensive to do more bootstrapping runs.)


\begin{figure}[h!]
    \centering
    \includegraphics[width=0.55\columnwidth]{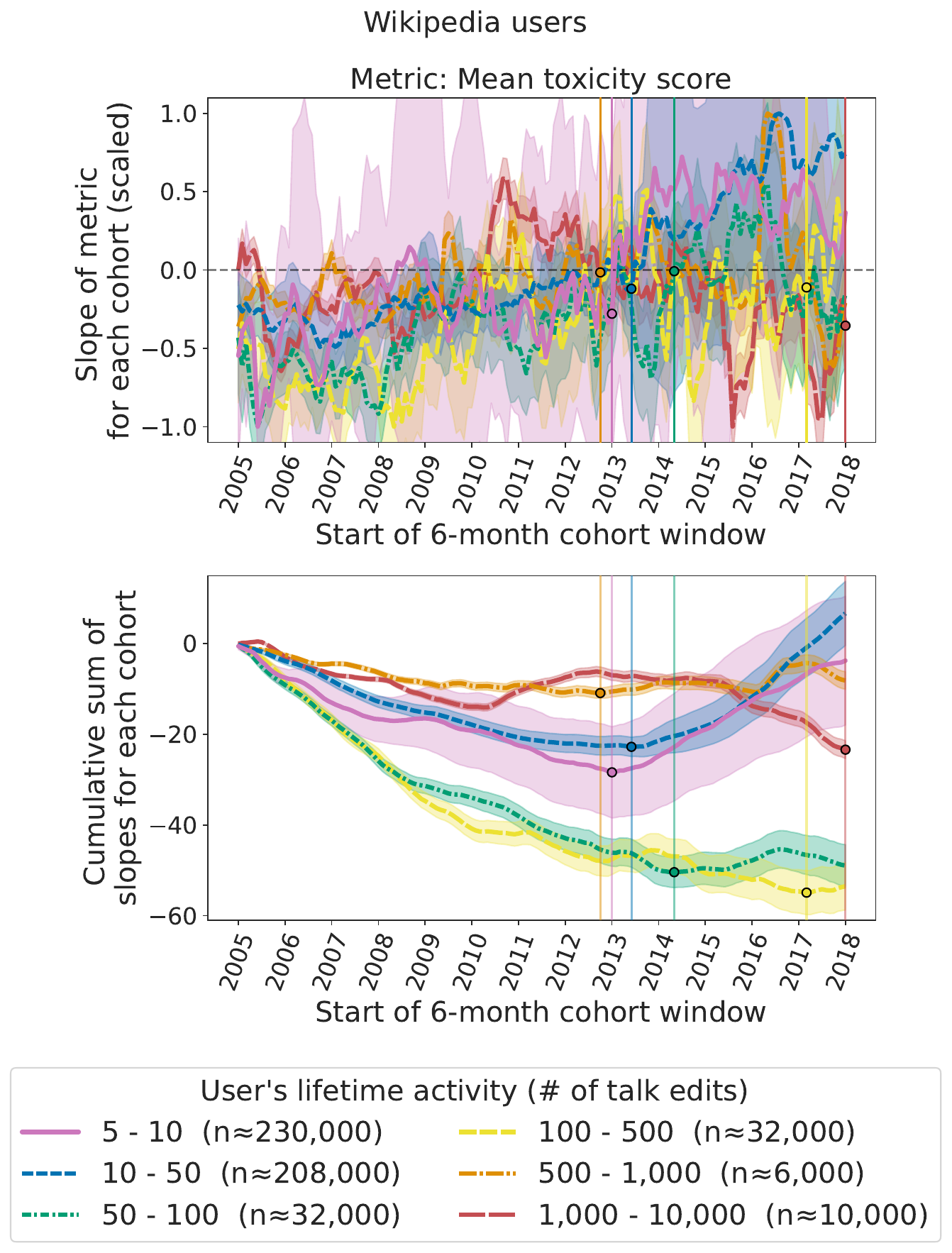} 
    \caption{
    Calculation of zero-crossing point for Figure \ref{fig:WB_slope}.
    \cbrk
    The upper plot is a reproduction of the slope plot from Fig. \ref{fig:WB_slope}.
    \cbrk    
    Below is a plot of the \enquote{integral} (cumulative sum) of those slopes. The minimum point of each line on this plot is marked by a vertical line. This point is a good proxy for the point at which the corresponding line on the upper plot crosses zero, which is marked by a matching vertical line.
    \cbrk
    Shaded areas on the integral plots represent two versions of 95\% confidence intervals. The lighter, wider intervals are very conservative, while the narrower, darker intervals are very optimistic. See above for explanation.
    }
    \label{fig_supp:SHb_integral_figWB_1mo}
\end{figure}

\clearpage
\section{Linear instead of log scale for comment order (x-axis)}
\label{section_supp:C_linear_scale}

\sisetup{round-mode=none, 
}  

\subsection{Mean score figures}

\begin{figure}[h!]
    \centering
    \includegraphics[width=0.95\columnwidth]{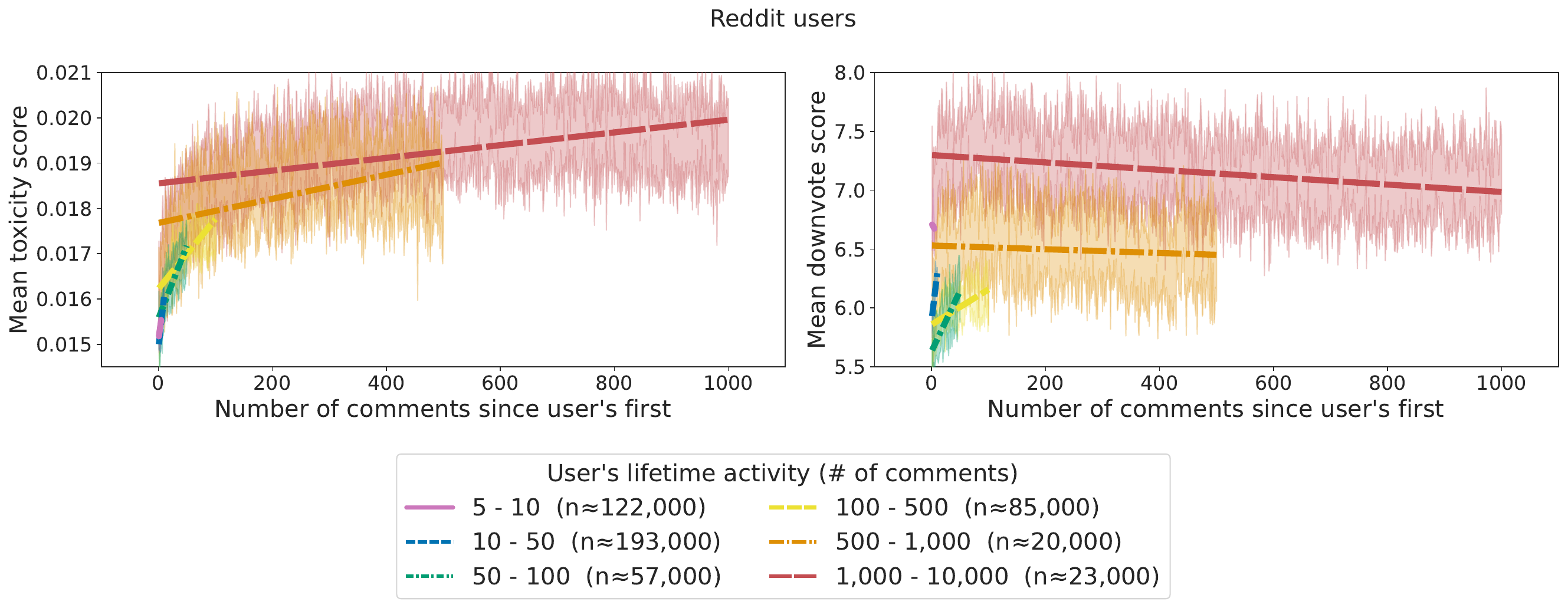} 
    \caption{Version of Figure \ref{fig:B_rainbow} with a linear, instead of logarithmic, x-axis. Trend lines are also fit to linear x-values instead of logarithmic.}
    \label{fig_supp:SCa0_rainbow_auth}
\end{figure}

\begin{figure}[h!]
    \centering
    \includegraphics[width=0.45\columnwidth]{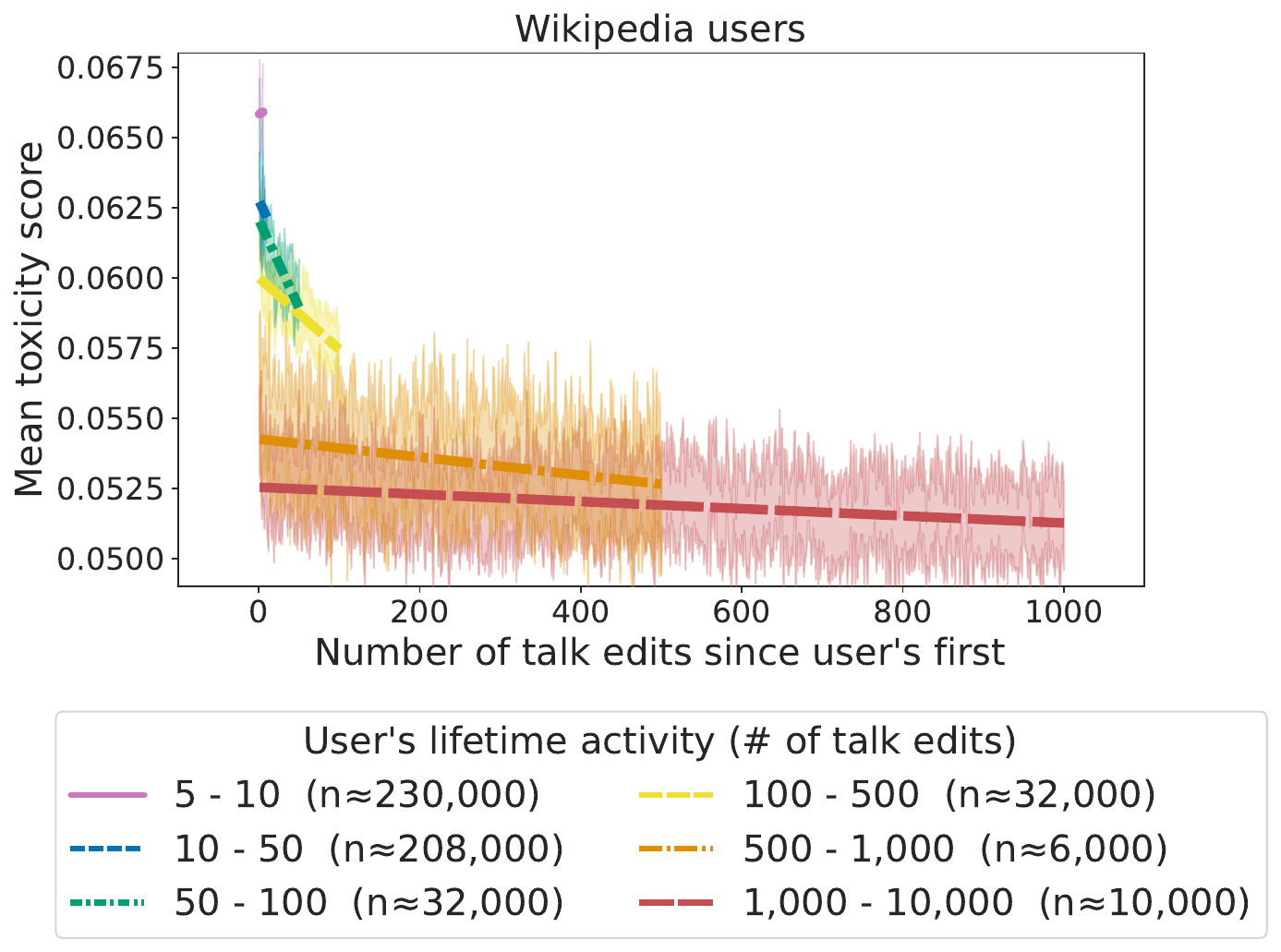} 
    \caption{Version of Figure \ref{fig:WA1_rainbow} with a linear, instead of logarithmic, x-axis. Trend lines are also fit to linear x-values instead of logarithmic.}
    \label{fig_supp:SCb0_rainbow_wiki_auth}
\end{figure}

\begin{figure}[h!]
    \centering
    \includegraphics[width=0.95\columnwidth]{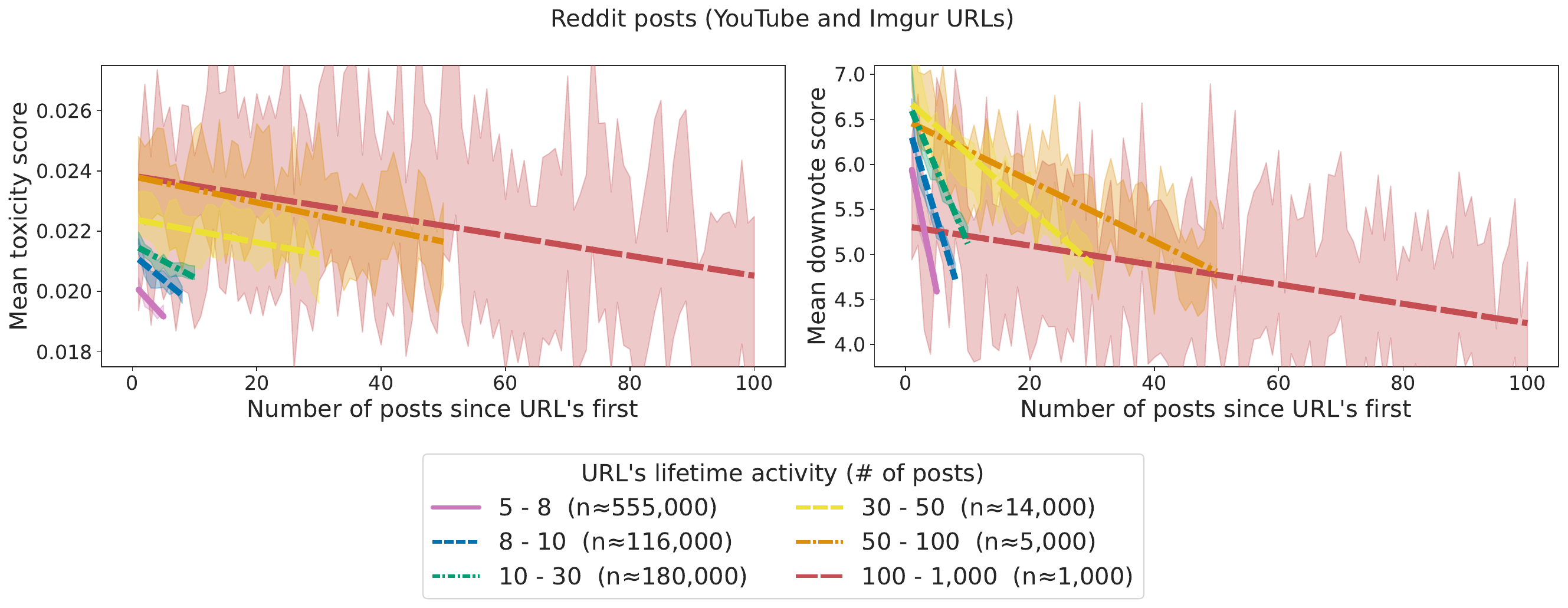} 
    \caption{Version of Figure \ref{fig:LA1_rainbow_reddit_links} with a linear, instead of logarithmic, x-axis. Trend lines are also fit to linear x-values instead of logarithmic.}
    \label{fig_supp:SCa1_rainbow_yt_img}
\end{figure}

\begin{figure}[h!]
    \centering
    \includegraphics[width=0.45\columnwidth]{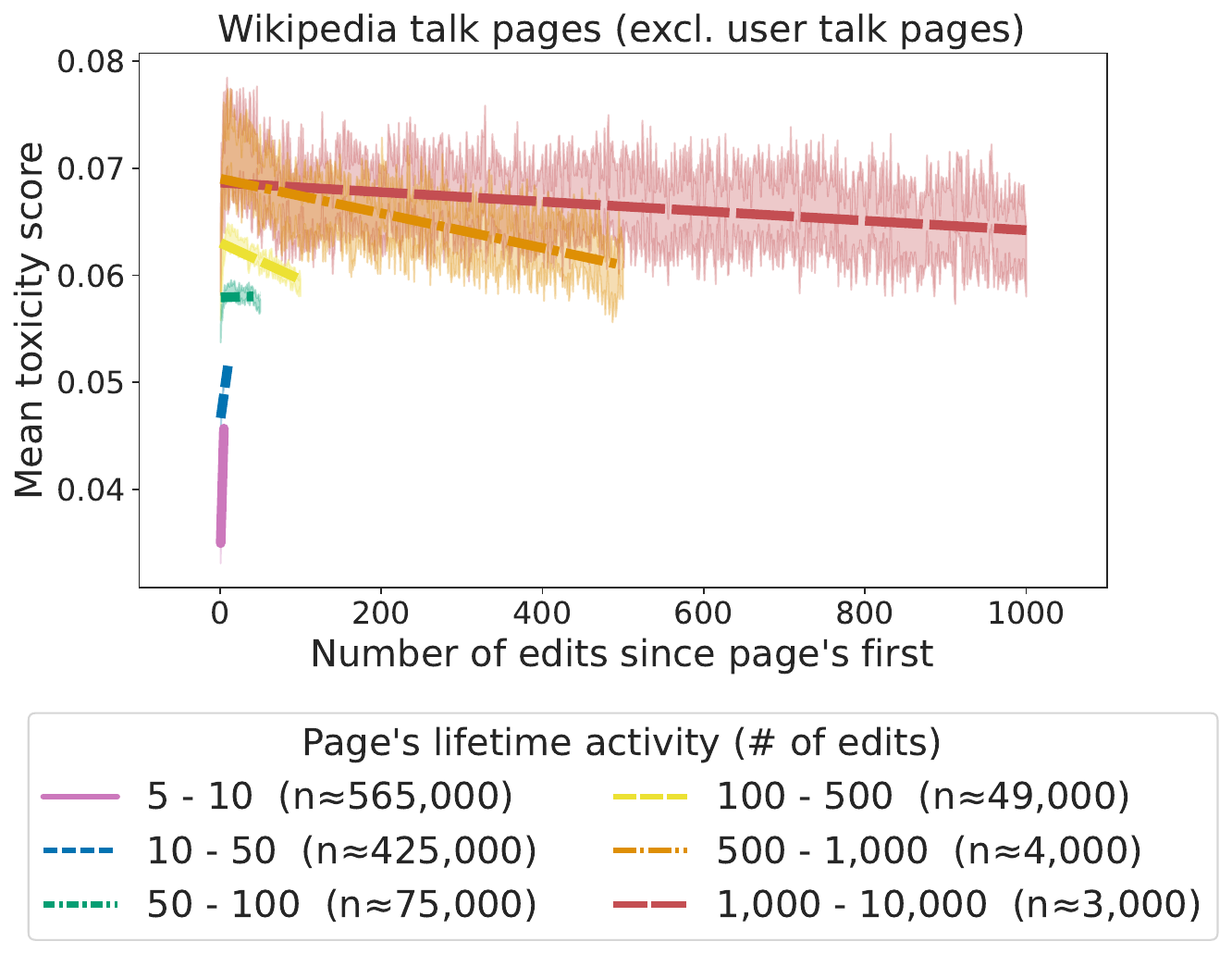} 
    \caption{Version of Figure \ref{fig:LA2_rainbow_wiki_page} with a linear, instead of logarithmic, x-axis. Trend lines are also fit to linear x-values instead of logarithmic.}
    \label{fig_supp:SCb1_rainbow_wiki_page_other}
\end{figure}

\begin{figure}[h!]
    \centering
    \includegraphics[width=0.45\columnwidth]{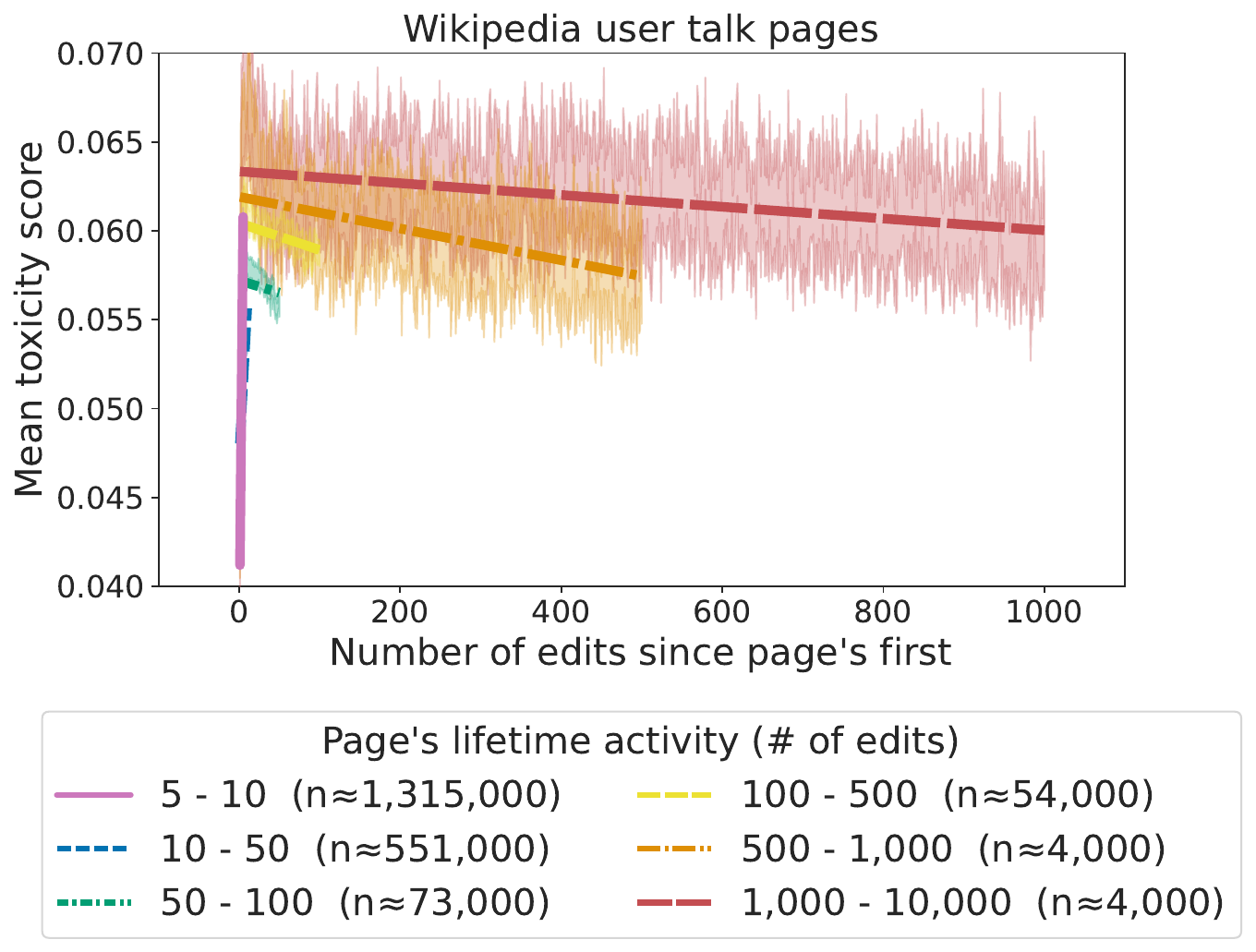} 
    \caption{Version of Figure \ref{fig_supp:SAc1_rainbow_wiki_page_usertalk} with a linear, instead of logarithmic, x-axis. Trend lines are also fit to linear x-values instead of logarithmic.}
    \label{fig_supp:SCb2_rainbow_wiki_page_user}
\end{figure}

\clearpage
\subsection{Slope figures}
\label{section_supp:SC2_linear_slopes}

For slope plots  in the main text (figures similar to Fig. \ref{fig:D_slope}), the \enquote{slope} is of a line fitted to a logarithmic x-axis, for consistency with the mean-score figures such as Figures \ref{fig:B_rainbow} and \ref{fig:C_rainbow_grid}.

In this section, slopes are instead fit to the data without log-transforming either axis, simply fitting to $y = m x + b$.

(To be exact, fitting on the log-transformed x-axis was not done using linear regression on log-transformed data. Instead, we used nonlinear least squares regression to fit the equivalent function $y = m \; log(x) + b$. However, results are very similar for both fitting methods. See Section \ref{section_supp:SB2_pval_tables} for more discussion.) 

\begin{figure}[h!]
    \centering
    \includegraphics[width=0.95\columnwidth]{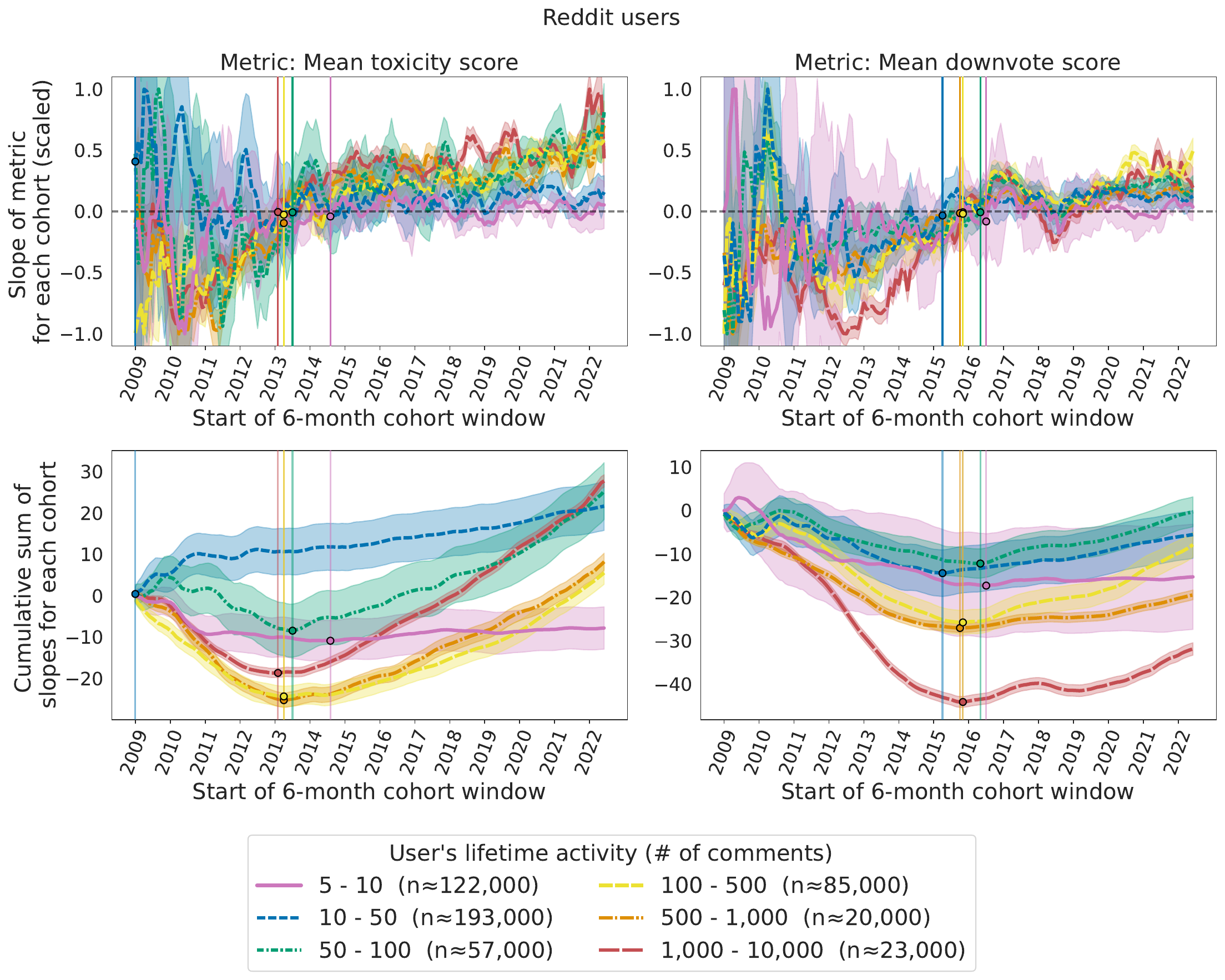} 
    \caption{Version of Figure \ref{fig:D_slope} where slopes are fit to a linear, instead of logarithmic, x-axis.}
    \label{fig_supp:SCc0_slope_auth_newdata_linearscale}
\end{figure}

\begin{figure}[h!]
    \centering
    \includegraphics[width=0.45\columnwidth]{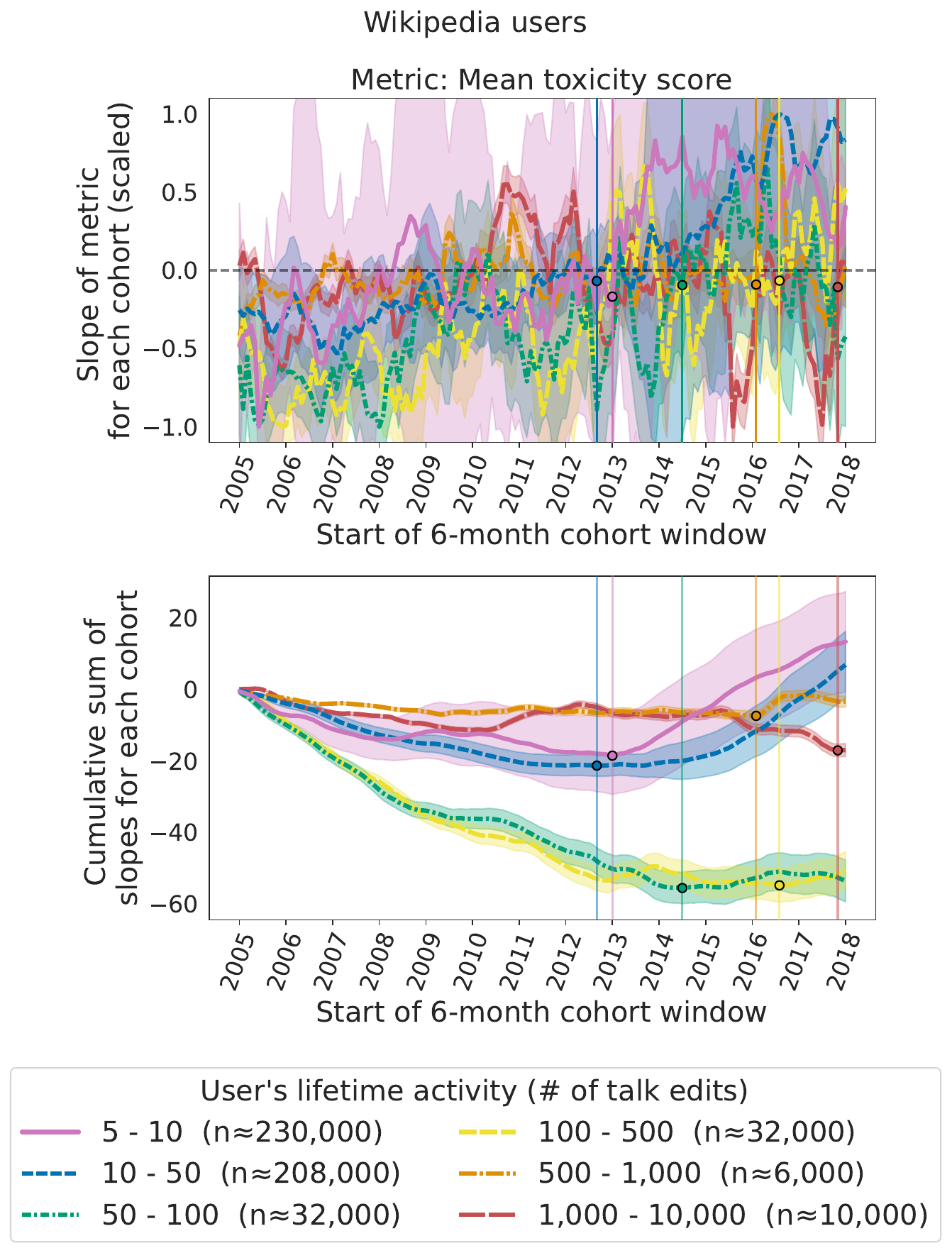} 
    \caption{Version of Figure \ref{fig:WB_slope} where slopes are fit to a linear, instead of logarithmic, x-axis.}
    \label{fig_supp:SCd0_slope_wiki_auth_linearscale}
\end{figure}

\begin{figure}[h!]
    \centering
    \includegraphics[width=0.95\columnwidth]{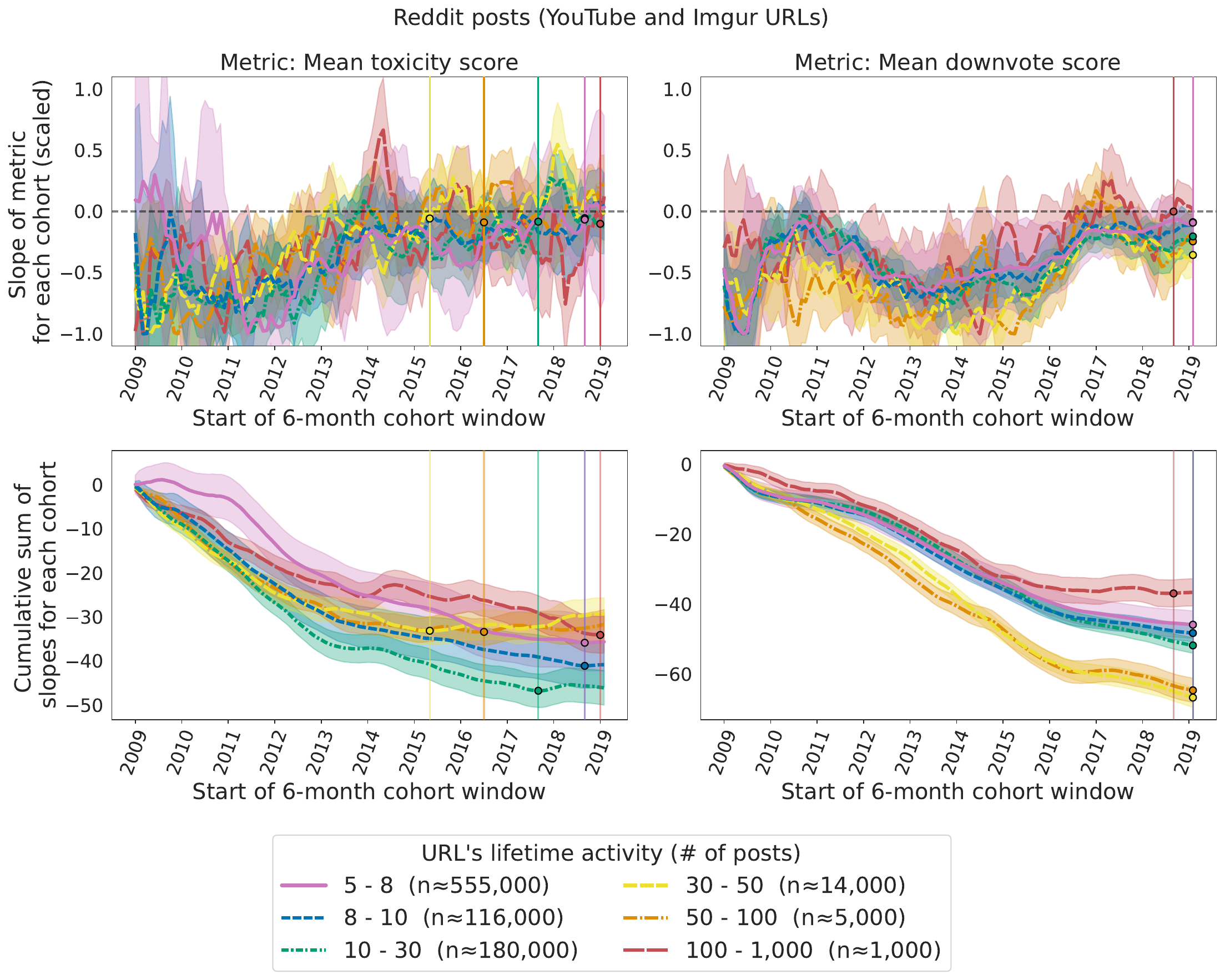} 
    \caption{Version of Figure \ref{fig_supp:SAc2a_slope_yt_img} where slopes are fit to a linear, instead of logarithmic, x-axis.}
    \label{fig_supp:SCc1_slope_yt_img_combined_newscores_linearscale}
\end{figure}

\begin{figure}[h!]
    \centering
    \includegraphics[width=0.45\columnwidth]{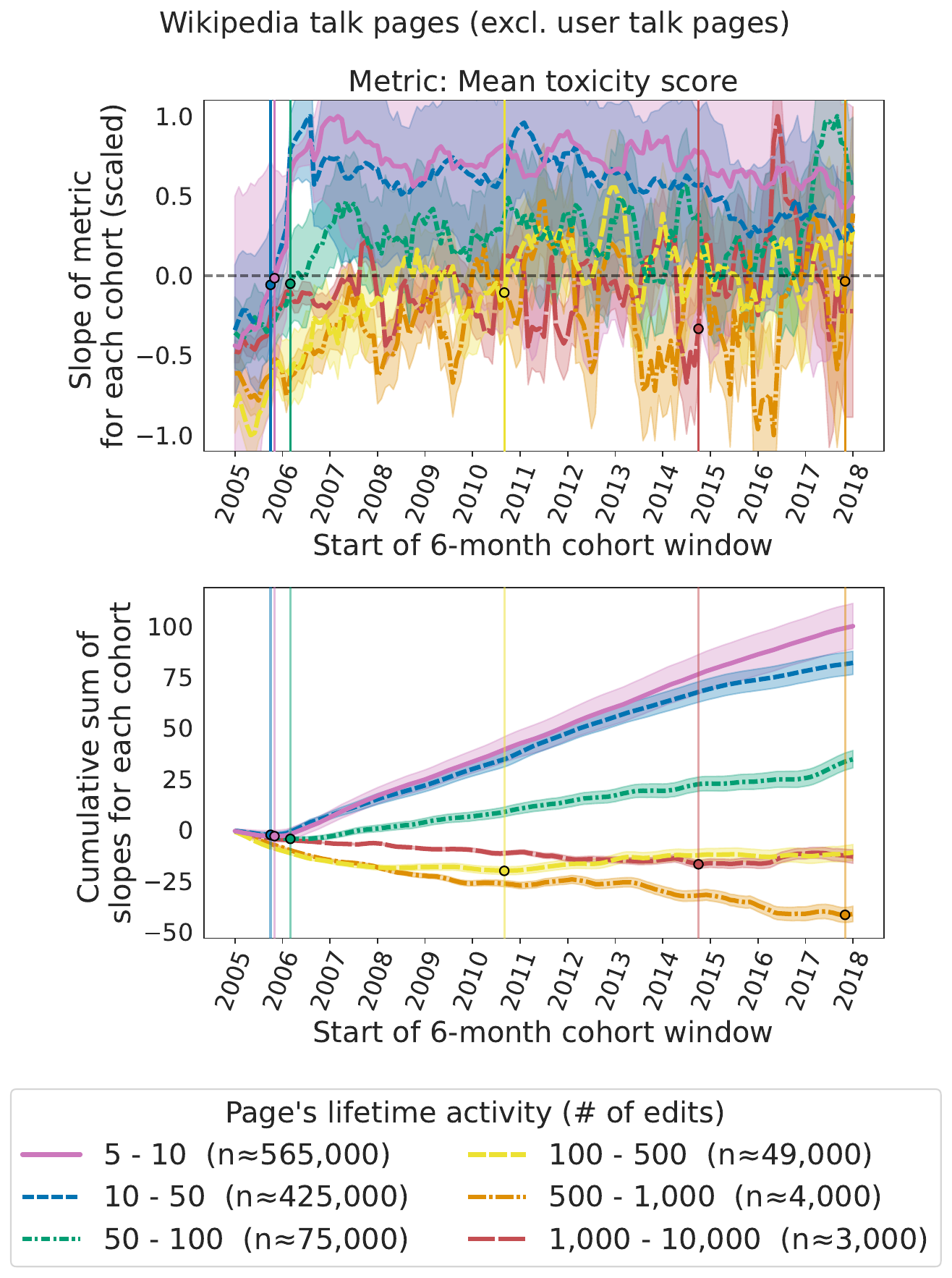} 
    \caption{Version of Figure \ref{fig_supp:SAc2b_wiki_page_slopes_other} where slopes are fit to a linear, instead of logarithmic, x-axis.}
    \label{fig_supp:SCd1_slope_wiki_page_othertalk_linearscale}
\end{figure}

\begin{figure}[h!]
    \centering
    \includegraphics[width=0.45\columnwidth]{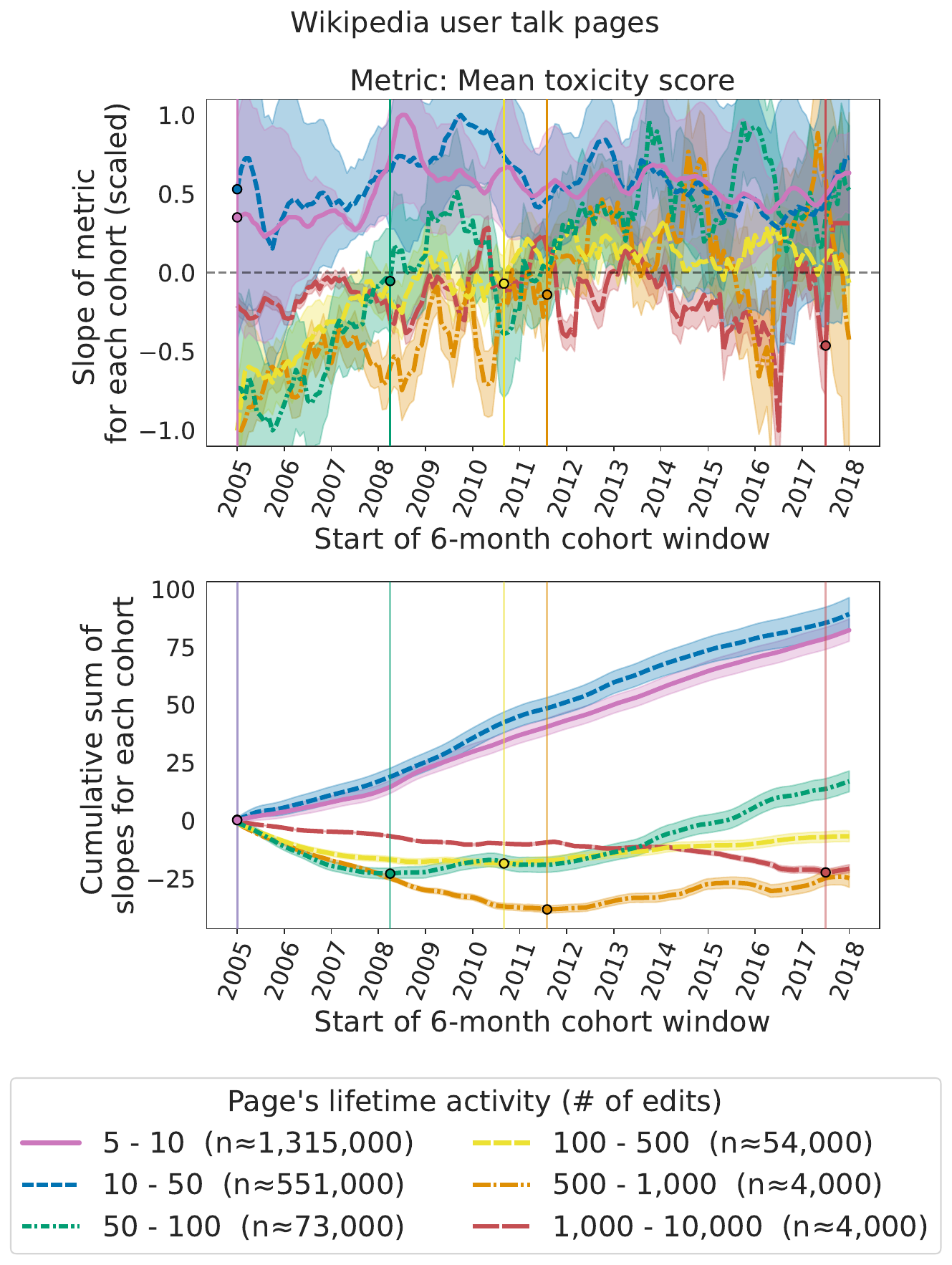} 
    \caption{Version of Figure \ref{fig_supp:SAc2c_wiki_page_slopes_user} where slopes are fit to a linear, instead of logarithmic, x-axis.}
    \label{fig_supp:SCd2_slope_wiki_page_usertalk_linearscale}
\end{figure}

\clearpage
\section{Metrics: Distributions and correlations}

\subsection{Metric distributions}
\label{section_supp:F0_metric_dists}

\begin{figure}[h!]
    \centering
    \includegraphics[width=0.45\columnwidth]{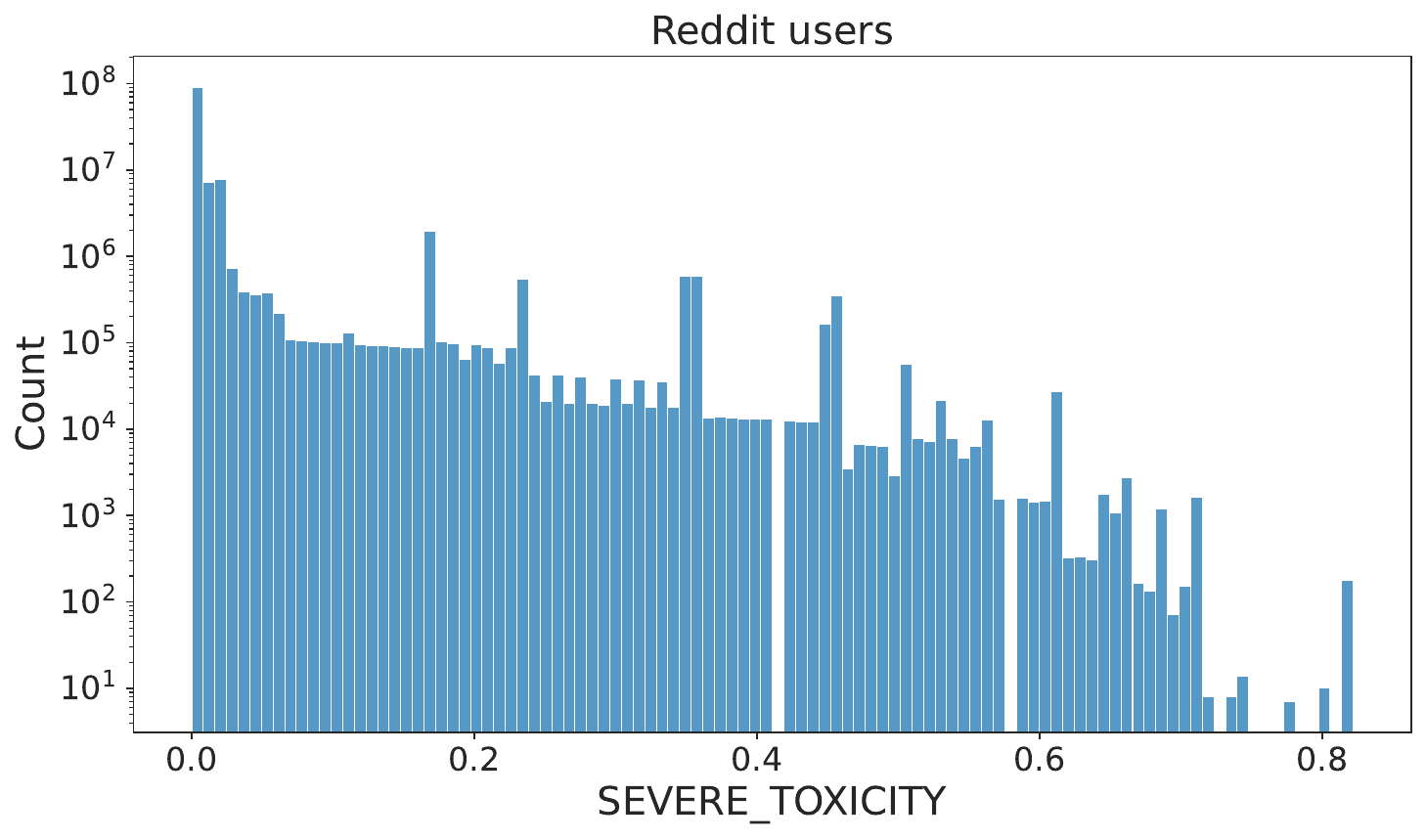} 
    \caption{Distribution of toxicity scores for Reddit user dataset.}
    \label{fig_supp:SF0a2_tox_hist_linear}
\end{figure}

\begin{figure}[h!]
    \centering
    \includegraphics[width=0.45\columnwidth]{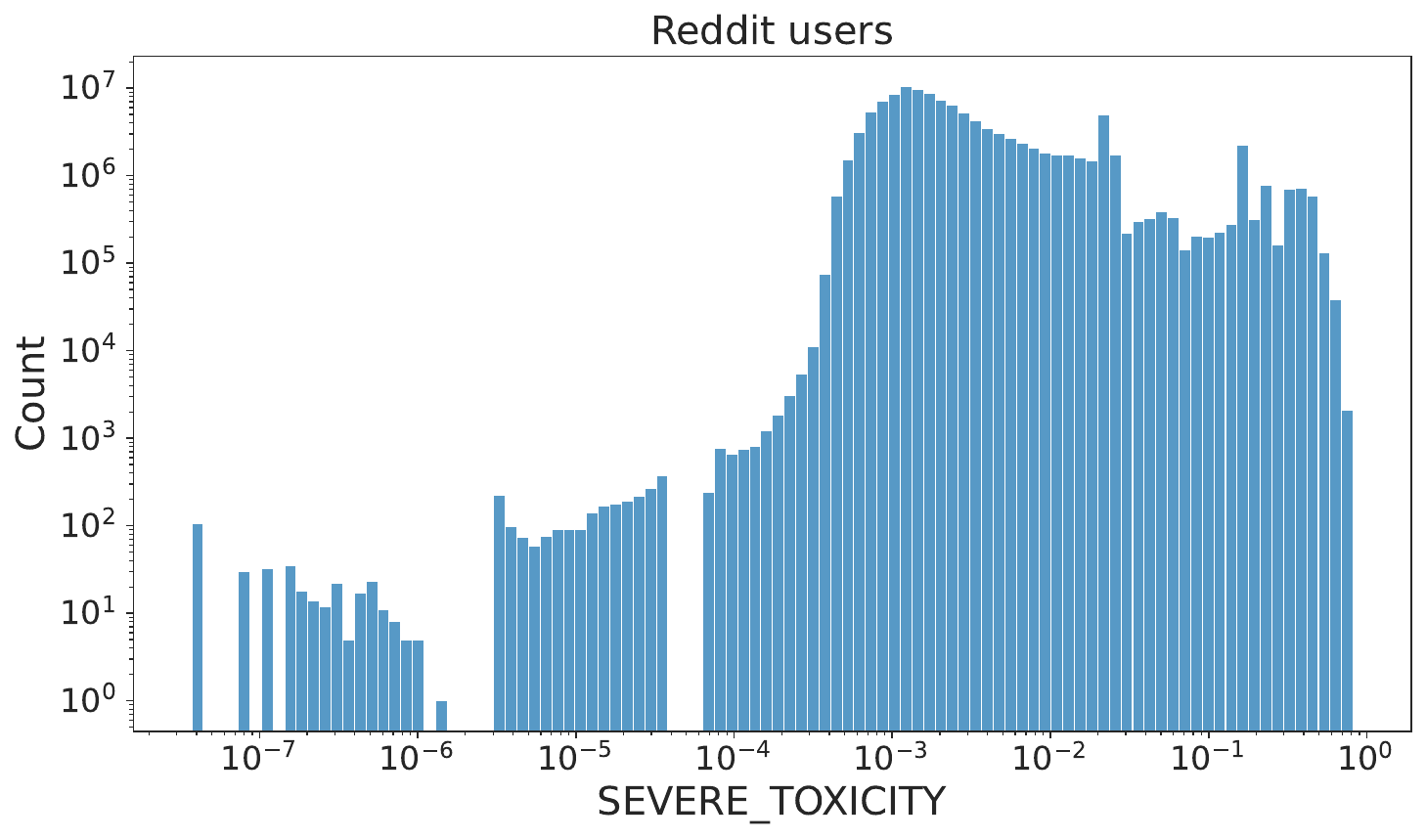} 
    \caption{Distribution of toxicity scores for Reddit user dataset, using logarithmic bins.}
    \label{fig_supp:SF0a_tox_hist}
\end{figure}

\begin{figure}[h!]
    \centering
    \includegraphics[width=0.45\columnwidth]{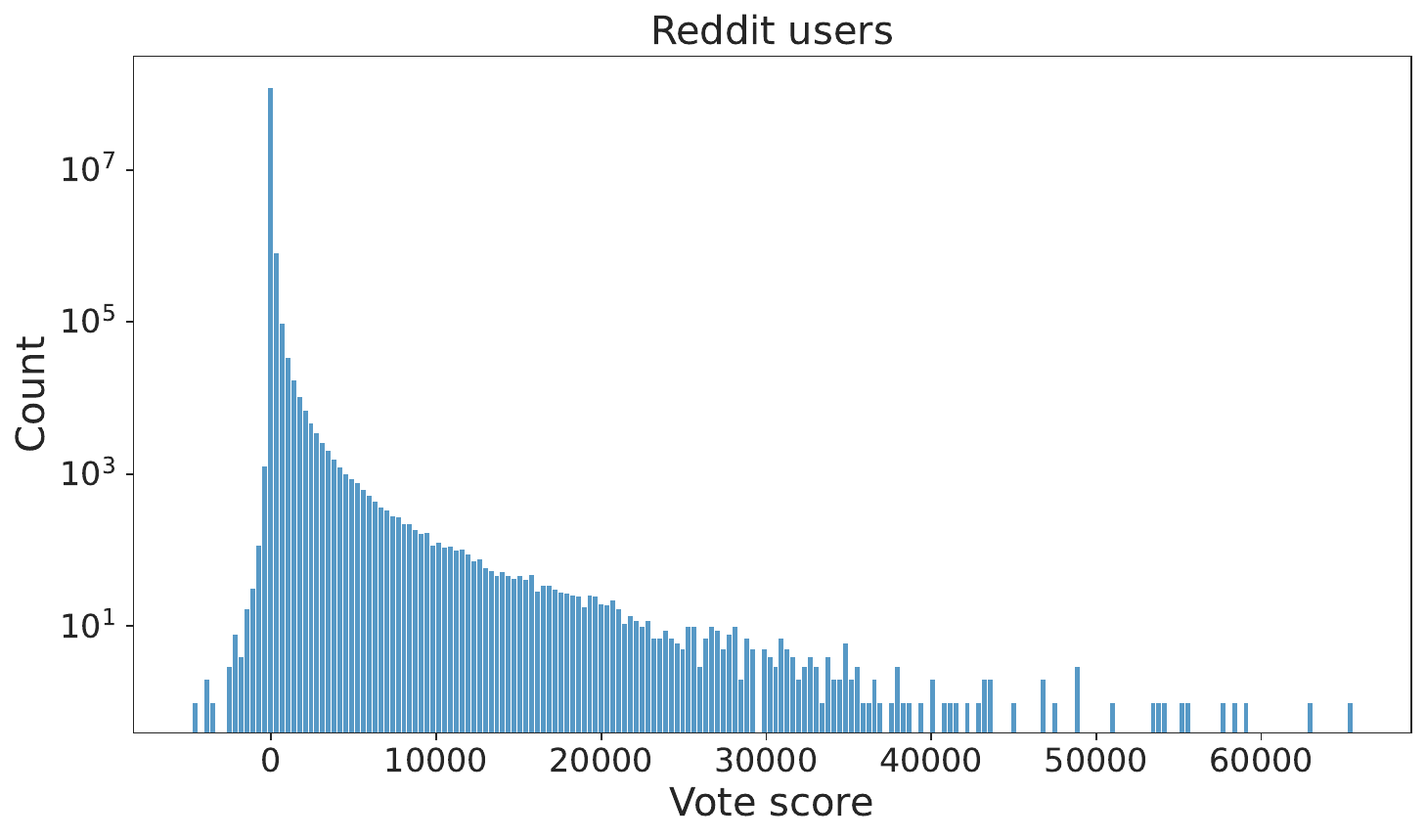} 
    \caption{Distribution of vote scores for Reddit user dataset.}
    \label{fig_supp:SF0b2_vote_hist_linear}
\end{figure}

\begin{figure}[h!]
    \centering
    \includegraphics[width=0.45\columnwidth]{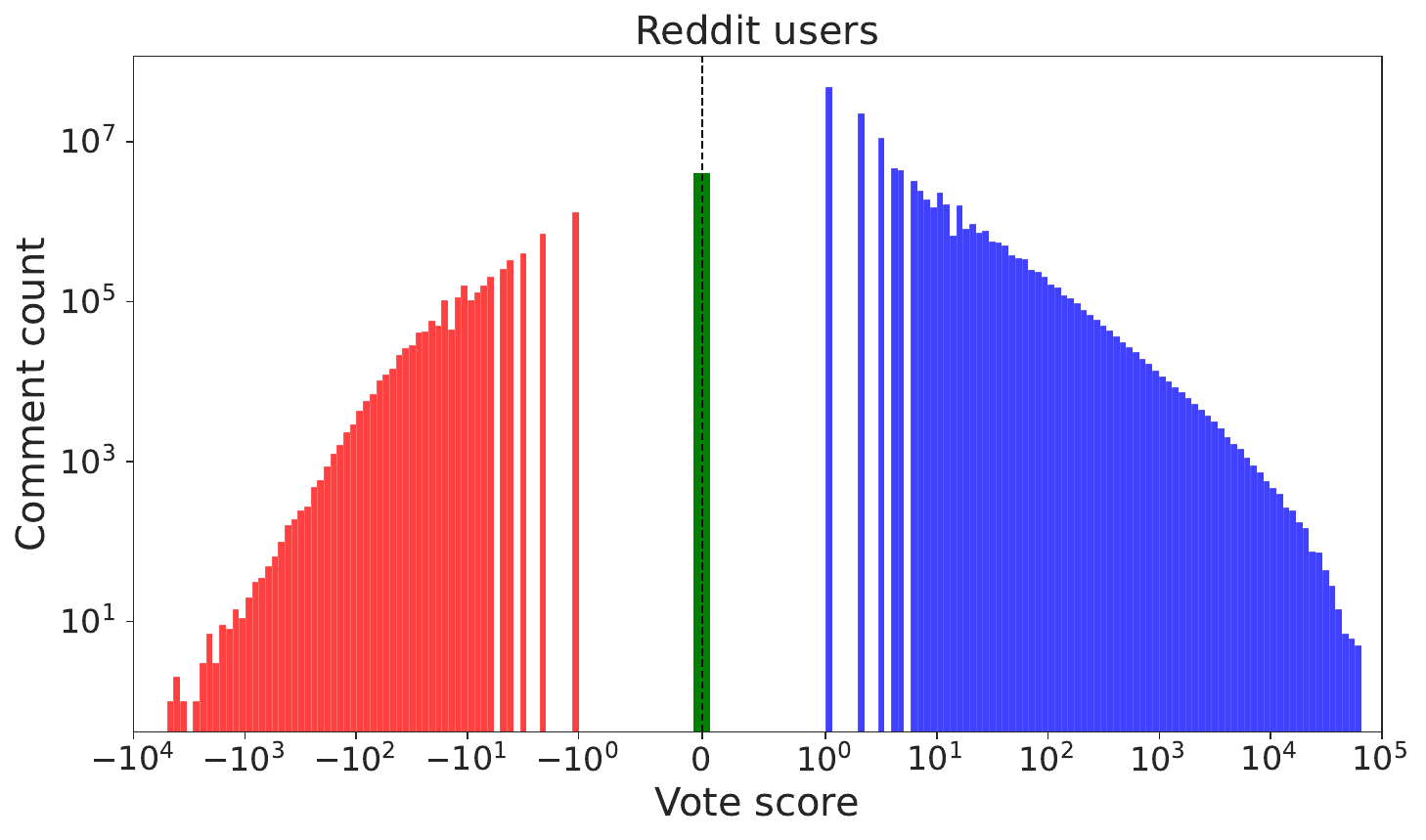} 
    \caption{Distribution of vote scores for Reddit user dataset, using symmetrical logarithmic bins.}
    \label{fig_supp:SF0b_tox_vote}
\end{figure}

\clearpage
\subsection{
    Skewed toxicity scores: quantized scores
}

\begin{figure}[h]
    \centering

    \begin{subfigure}[b]{0.24\columnwidth}
        \centering
        \includegraphics[width=1\columnwidth]{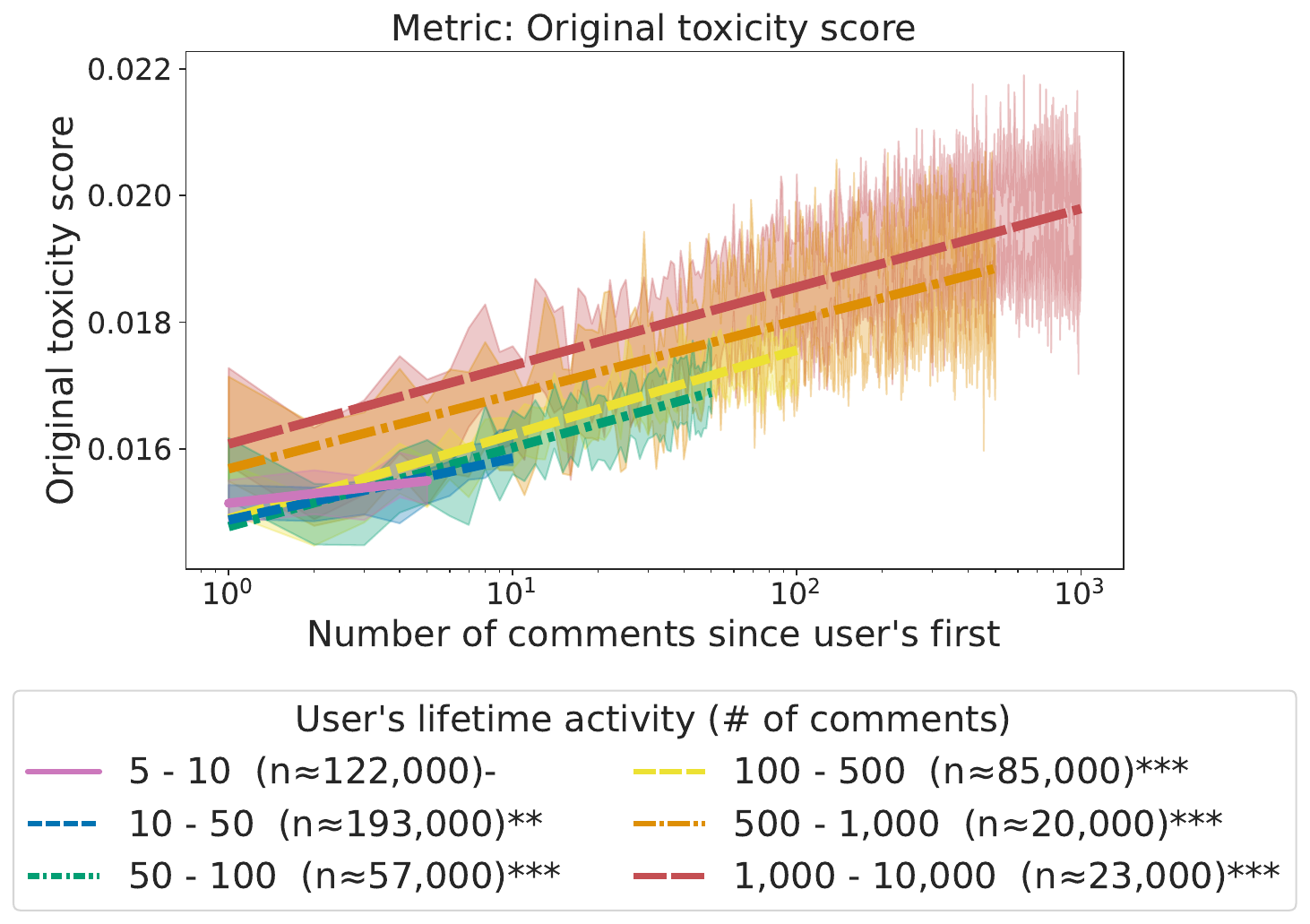} 
        \caption{}
    \end{subfigure}
    \begin{subfigure}[b]{0.24\columnwidth}
        \centering
        \includegraphics[width=1\columnwidth]{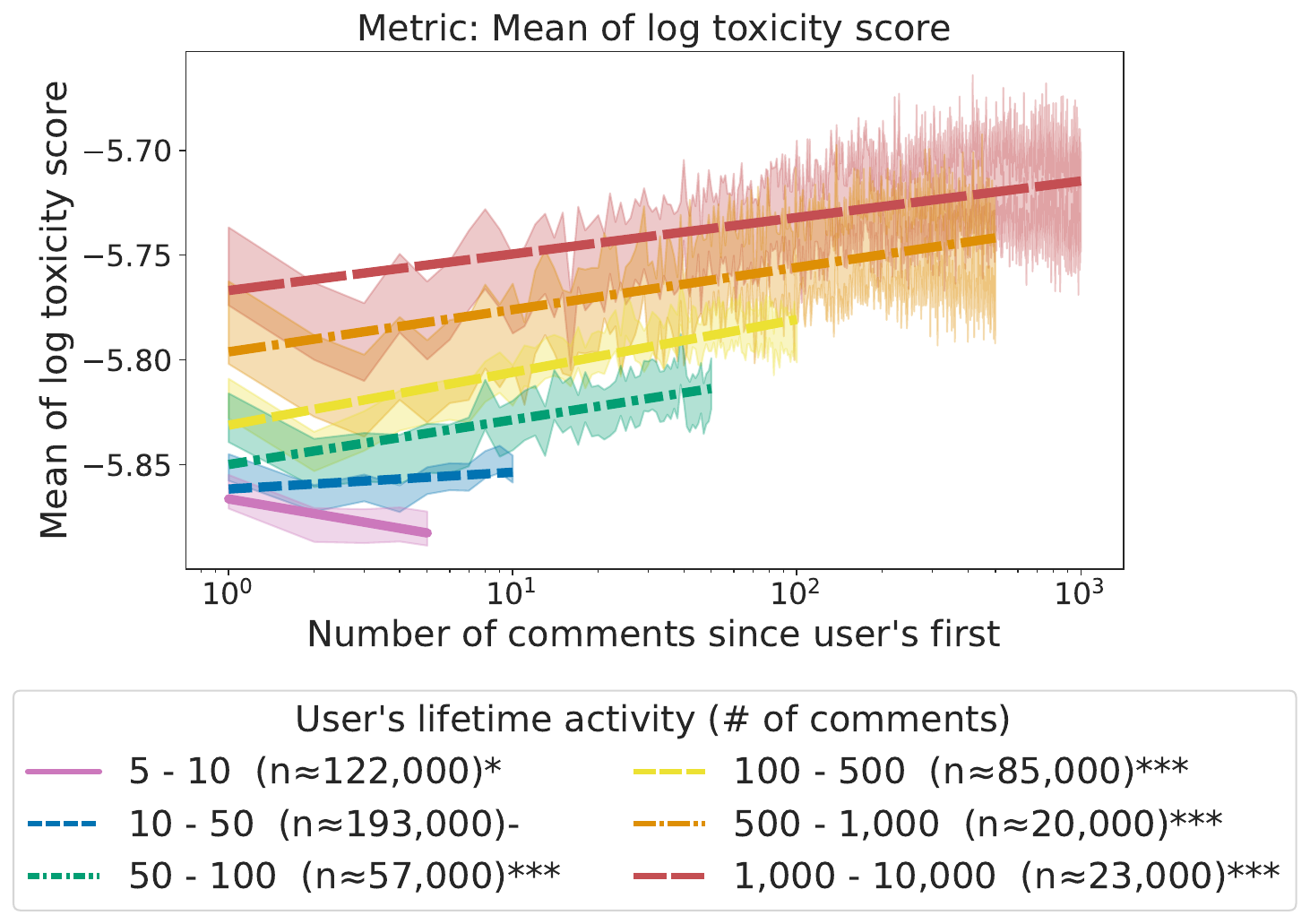} 
        \caption{}
    \end{subfigure}
    
    \begin{subfigure}[b]{0.24\columnwidth}
        \centering
        \includegraphics[width=1\columnwidth]{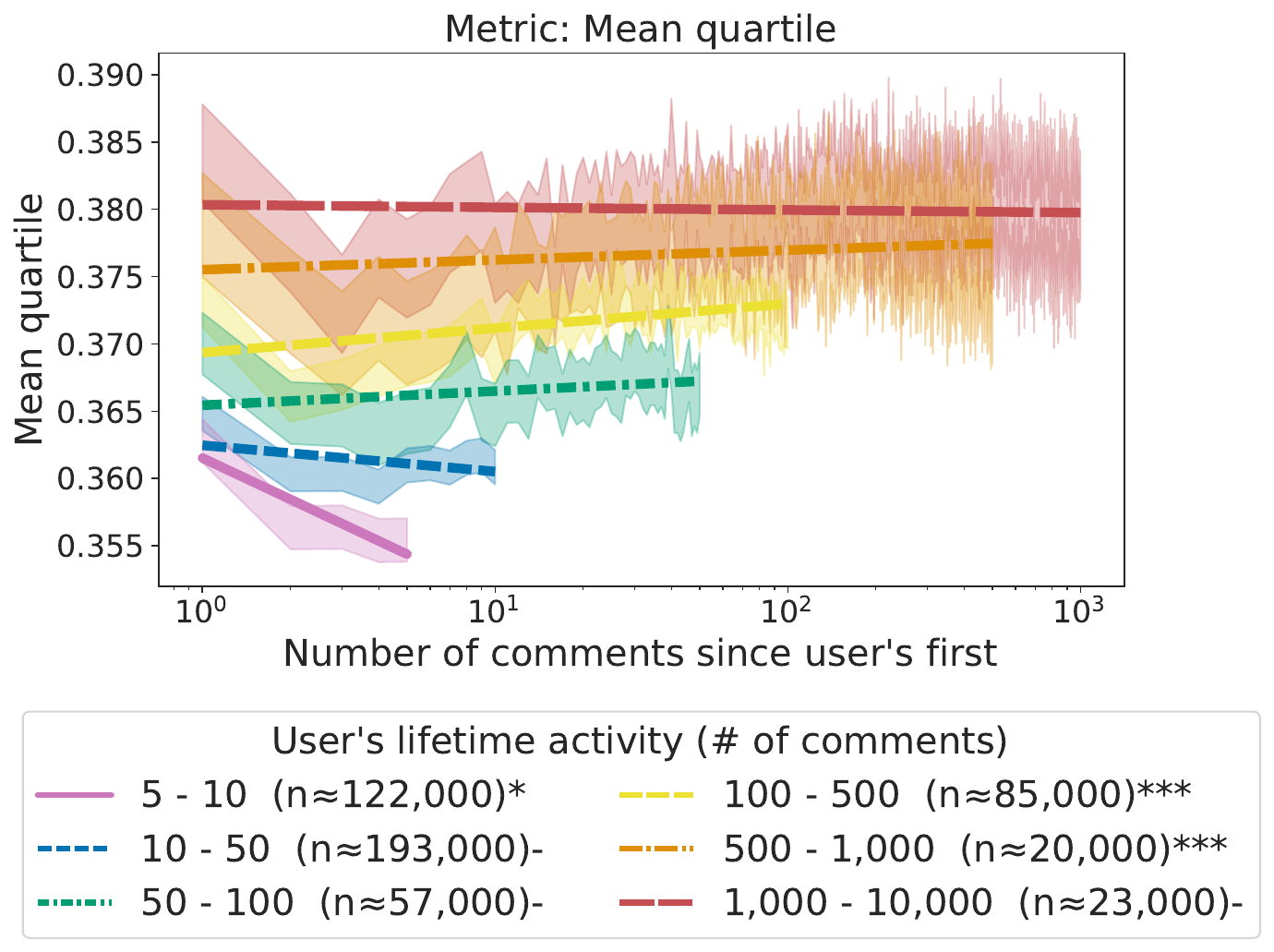} 
        \caption{}
        \label{subfig_supp:SF1_rainbow_c}
    \end{subfigure}
    \begin{subfigure}[b]{0.24\columnwidth}
        \centering
        \includegraphics[width=1\columnwidth]{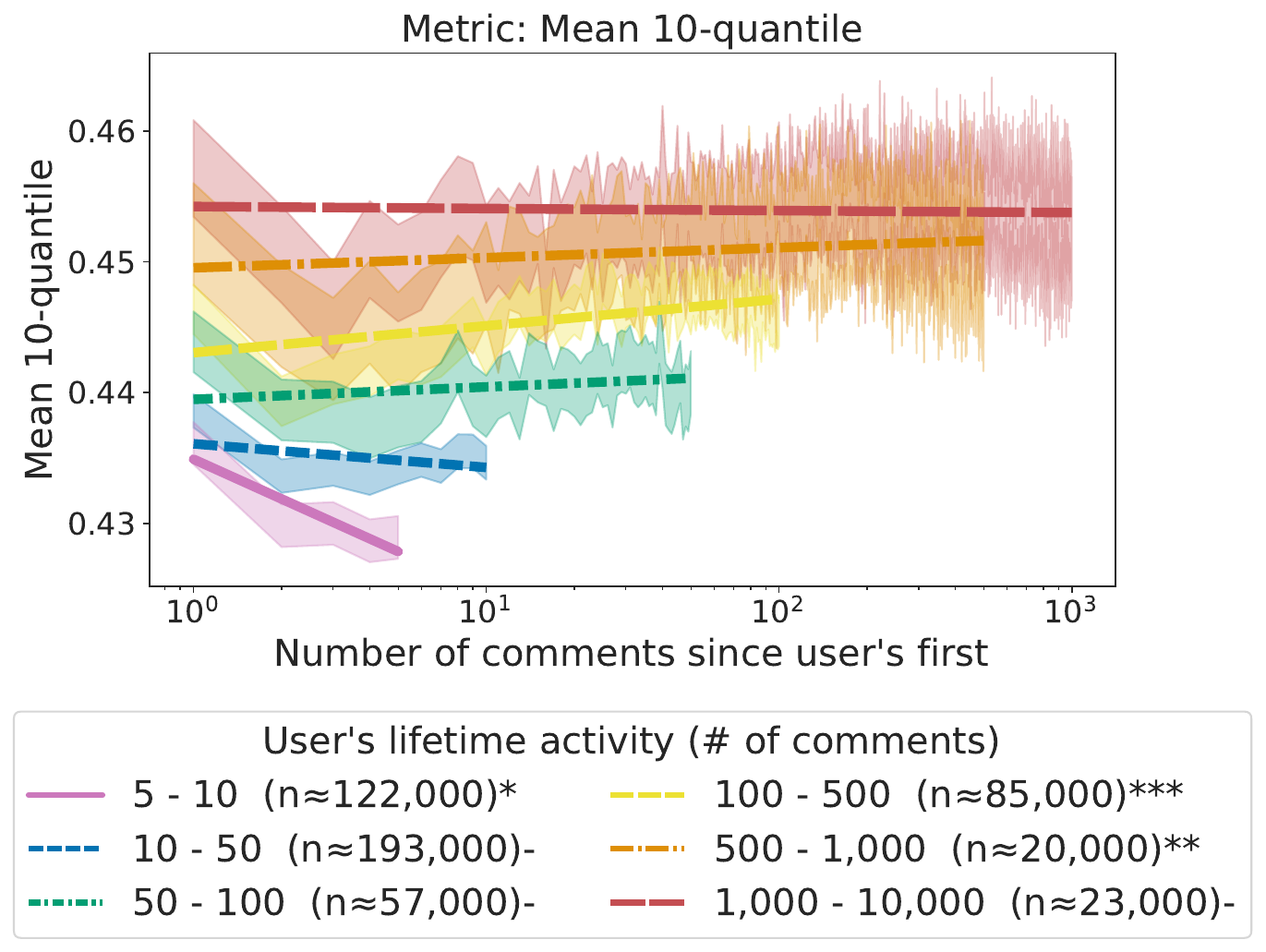}
        \caption{}
    \end{subfigure}
    \begin{subfigure}[b]{0.24\columnwidth}
        \centering
        \includegraphics[width=1\columnwidth]{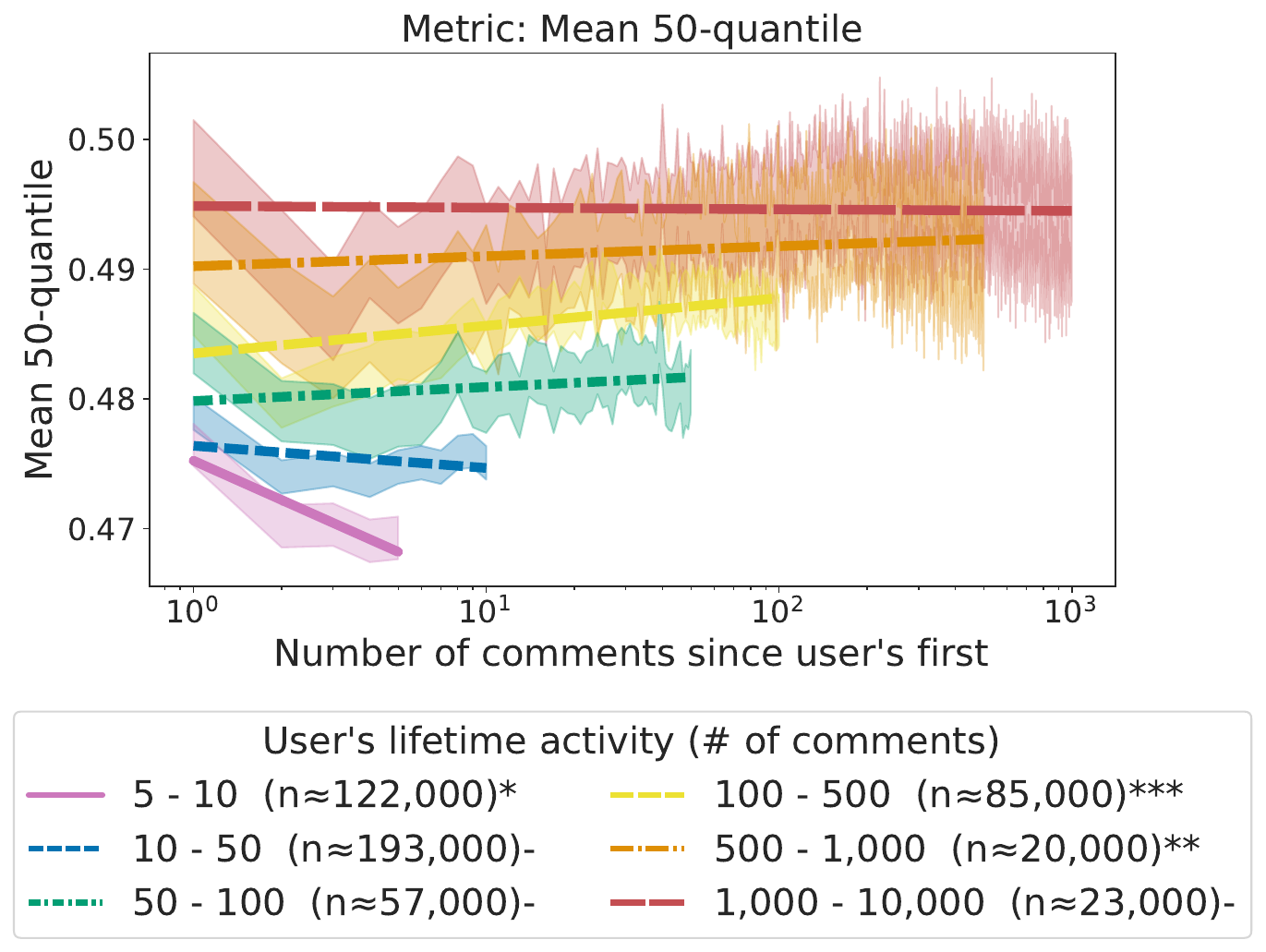}
        \caption{}
        \label{subfig_supp:SF1_rainbow_e}
    \end{subfigure}

    \begin{subfigure}[b]{0.19\columnwidth}
        \centering
        \includegraphics[width=1\columnwidth]{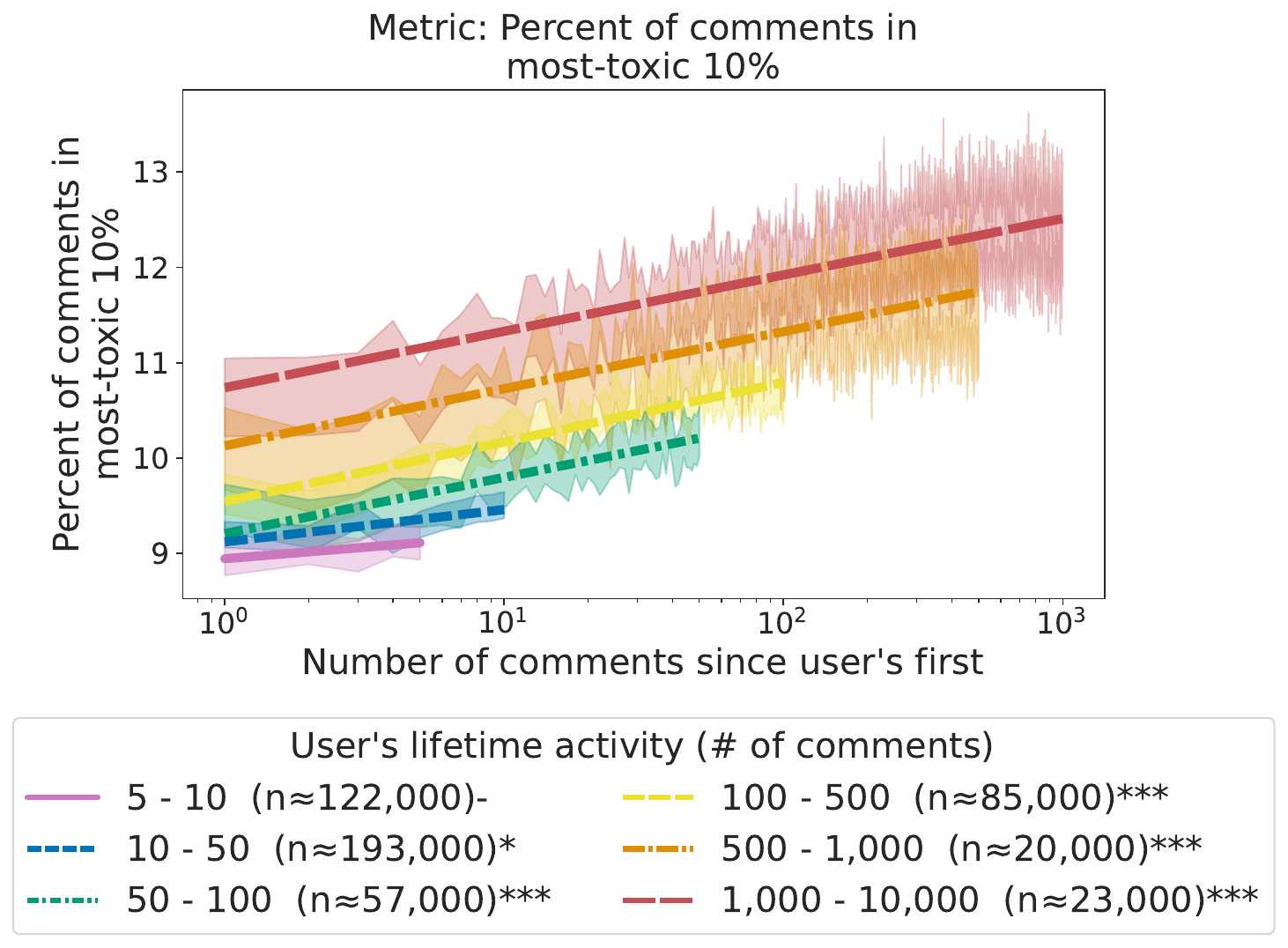} 
        \caption{}
        \label{subfig_supp:SF1_rainbow_f}
    \end{subfigure}
    \begin{subfigure}[b]{0.19\columnwidth}
        \centering
        \includegraphics[width=1\columnwidth]{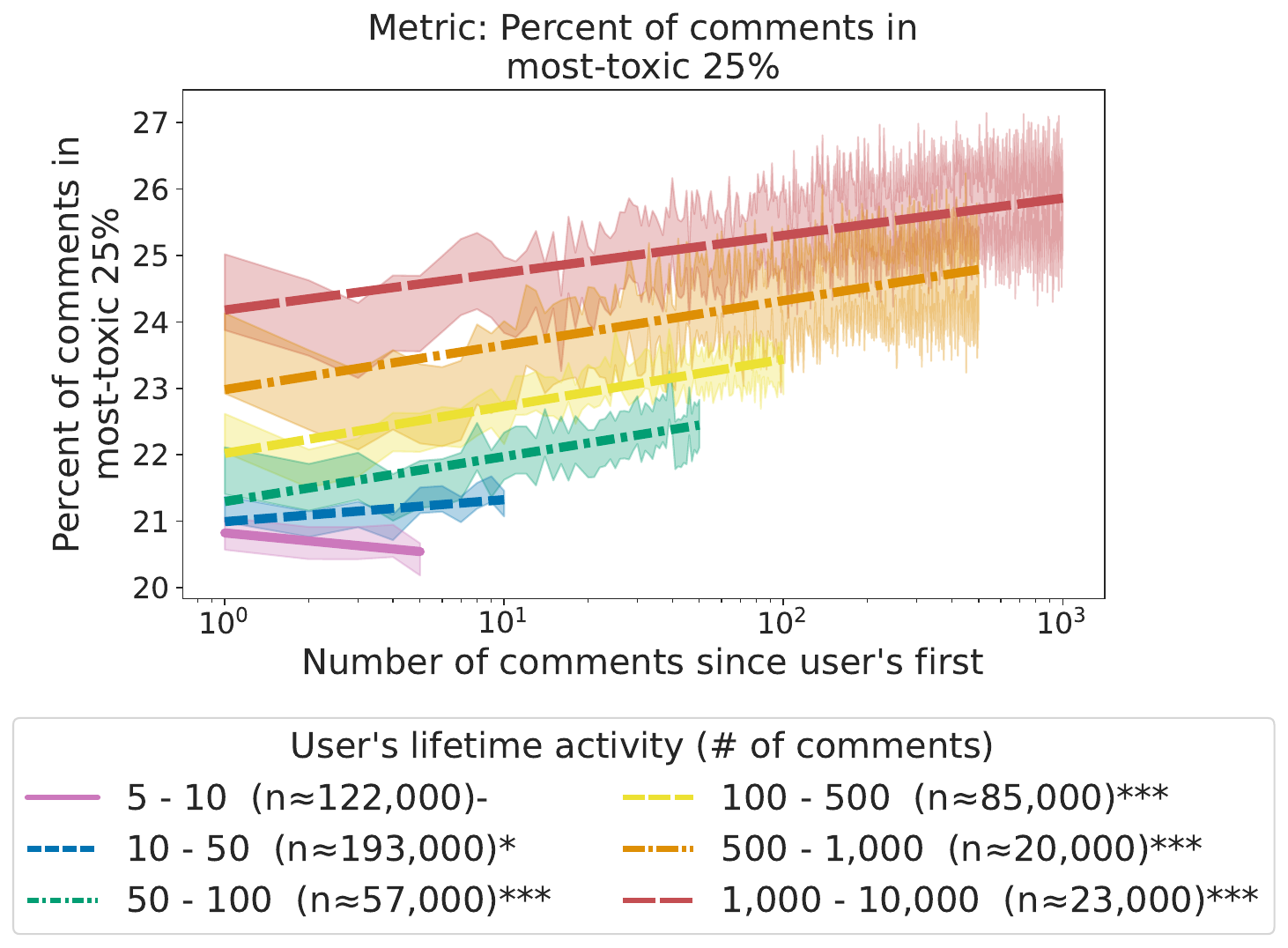} 
        \caption{}
    \end{subfigure}
    \begin{subfigure}[b]{0.19\columnwidth}
        \centering
        \includegraphics[width=1\columnwidth]{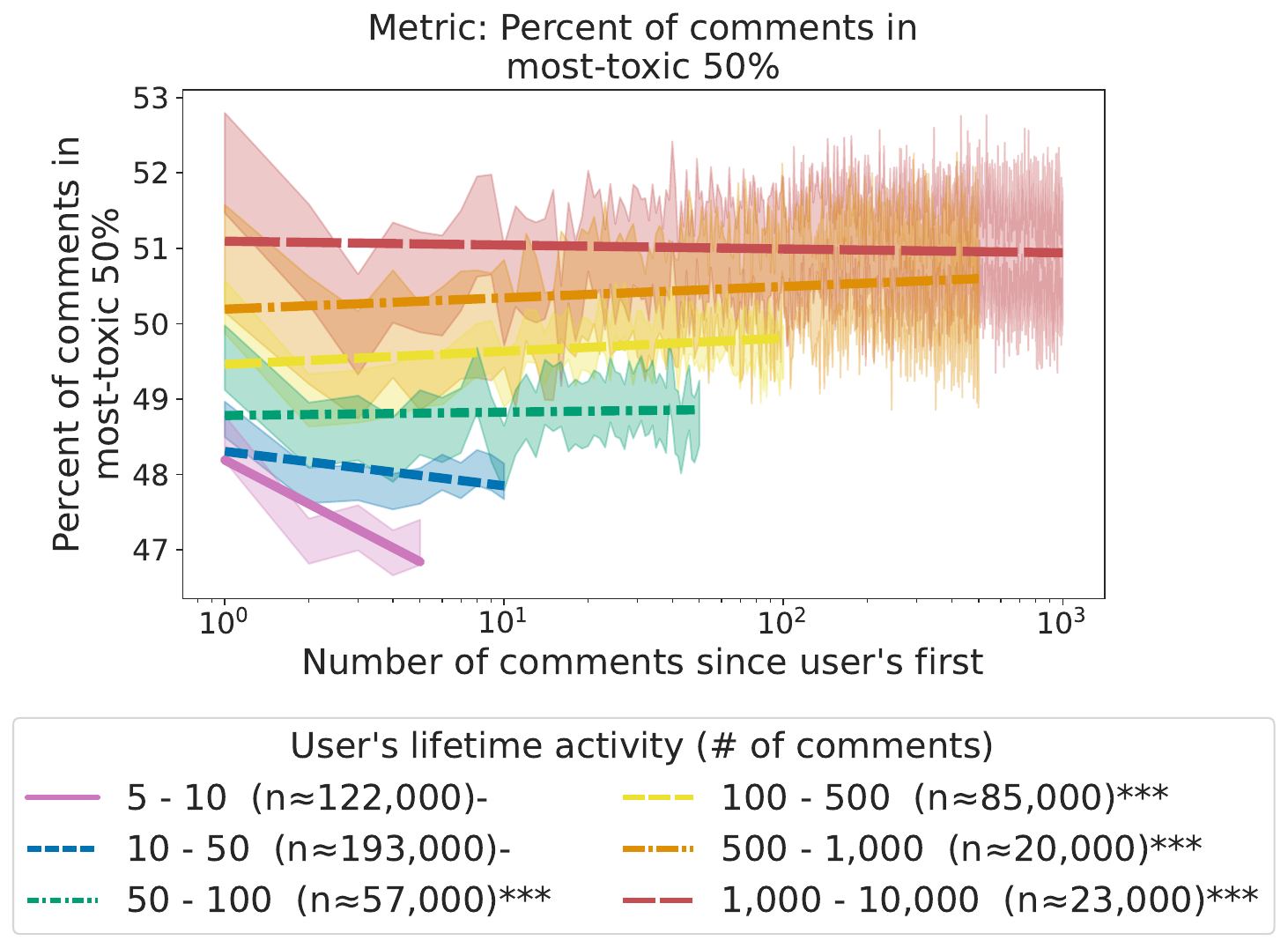} 
        \caption{}
    \end{subfigure}
    \begin{subfigure}[b]{0.19\columnwidth}
        \centering
        \includegraphics[width=1\columnwidth]{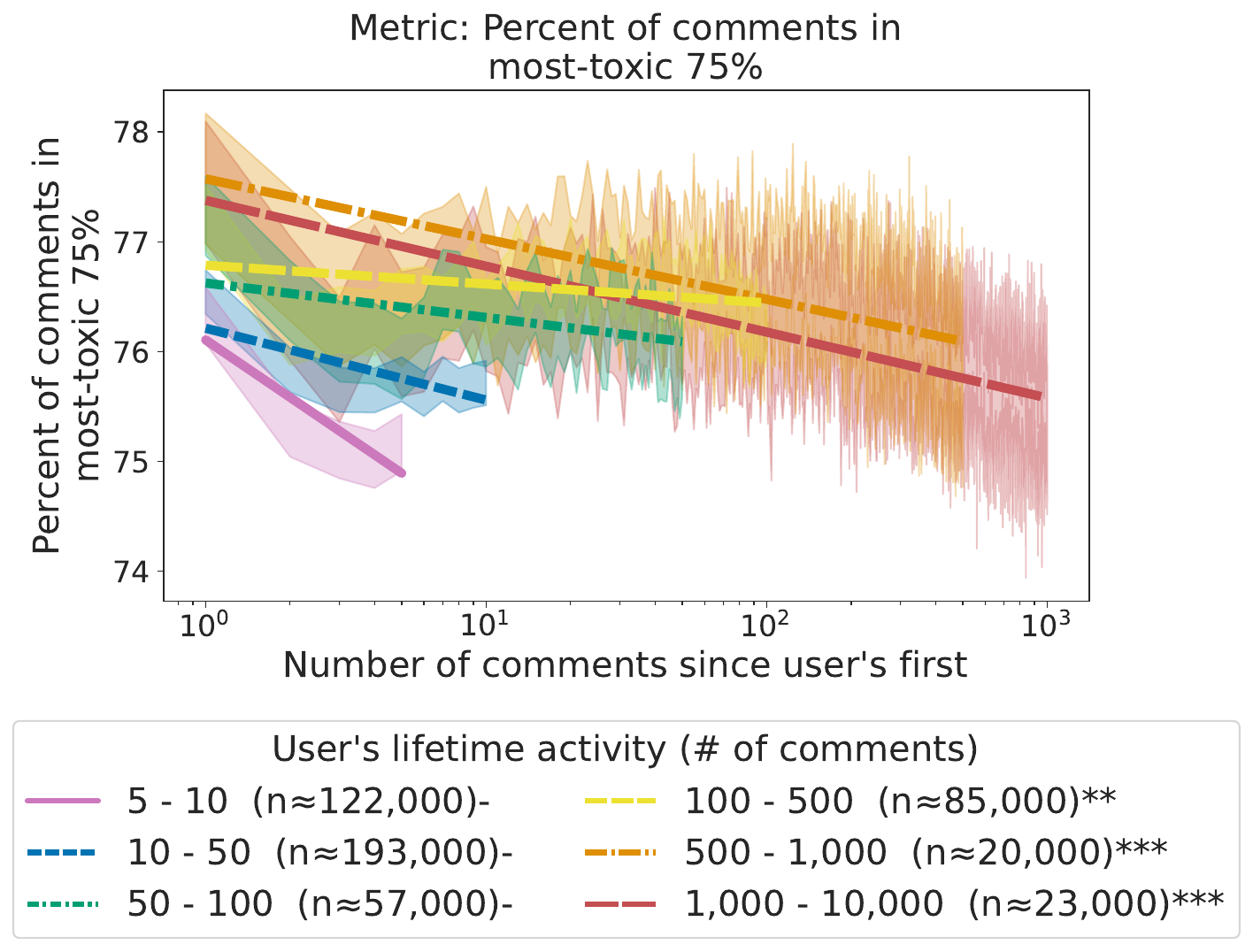} 
        \caption{}
    \end{subfigure}
    \begin{subfigure}[b]{0.19\columnwidth}
        \centering
        \includegraphics[width=1\columnwidth]{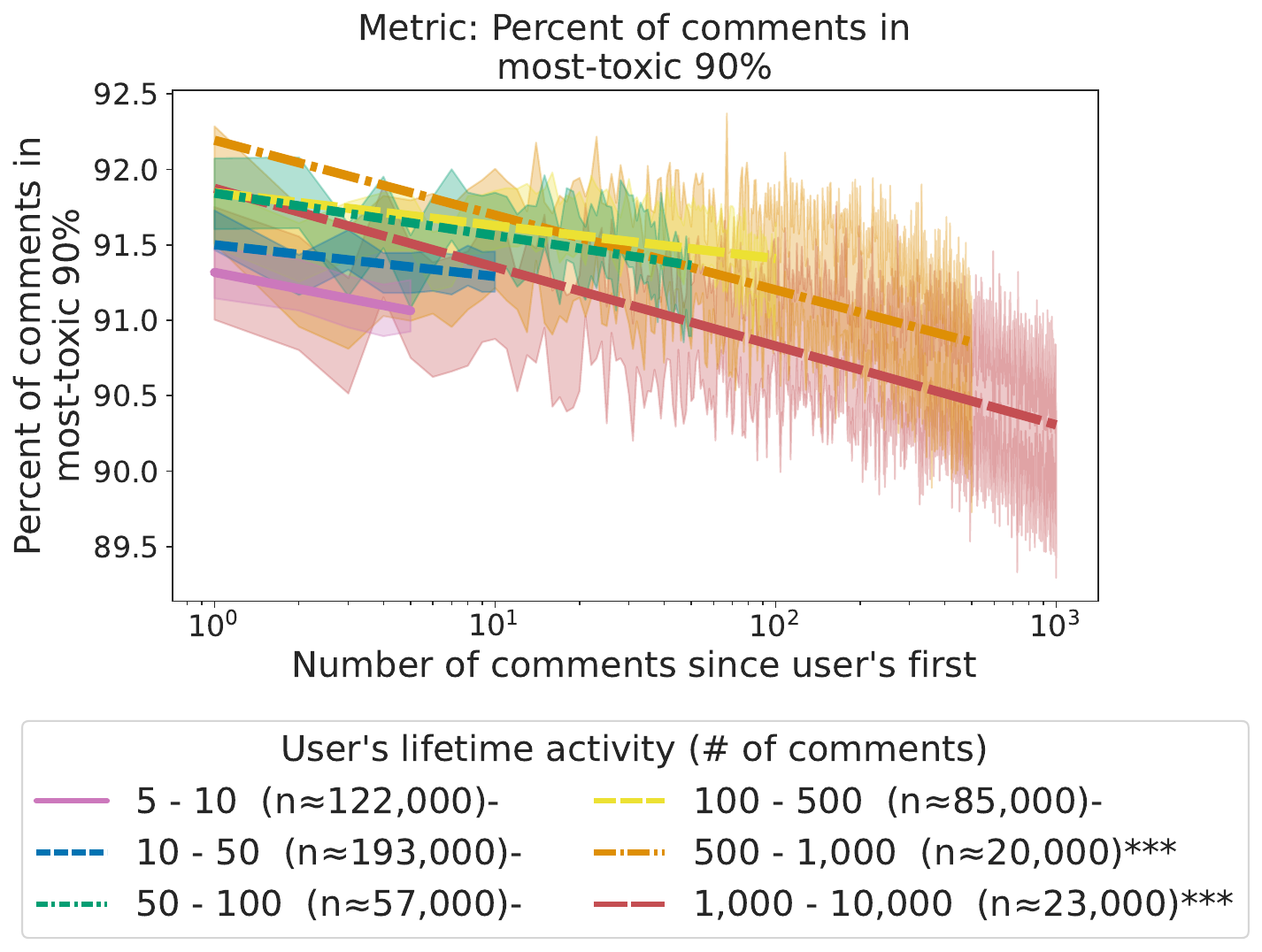} 
        \caption{}
        \label{subfig_supp:SF1_rainbow_j}
    \end{subfigure}

    \caption{Trends over user lifetimes, for various versions of the SEVERE\_TOXICITY score. 
    \cbrk
    The first row shows the original SEVERE\_TOXICITY score and the natural log of that score, which are both skewed metrics (see Figures \ref{fig_supp:SF0a2_tox_hist_linear} and \ref{fig_supp:SF0a_tox_hist}). 
    The second row shows trends for the SEVERE\_TOXICITY score transformed into quantiles at various resolution levels, creating uniform distributions of scores.
    The third row shows trends for binarized versions of the SEVERE\_TOXICITY score. This metric represents the percentage of comments in a given group that had toxicity scores in the top N\% of all comments in the dataset.}
    \label{fig_supp:SF1_rainbow_skewedmetrics}
\end{figure}

\begin{figure}[h]
    \centering

    \begin{subfigure}[b]{0.24\columnwidth}
        \centering
        \includegraphics[width=1\columnwidth]{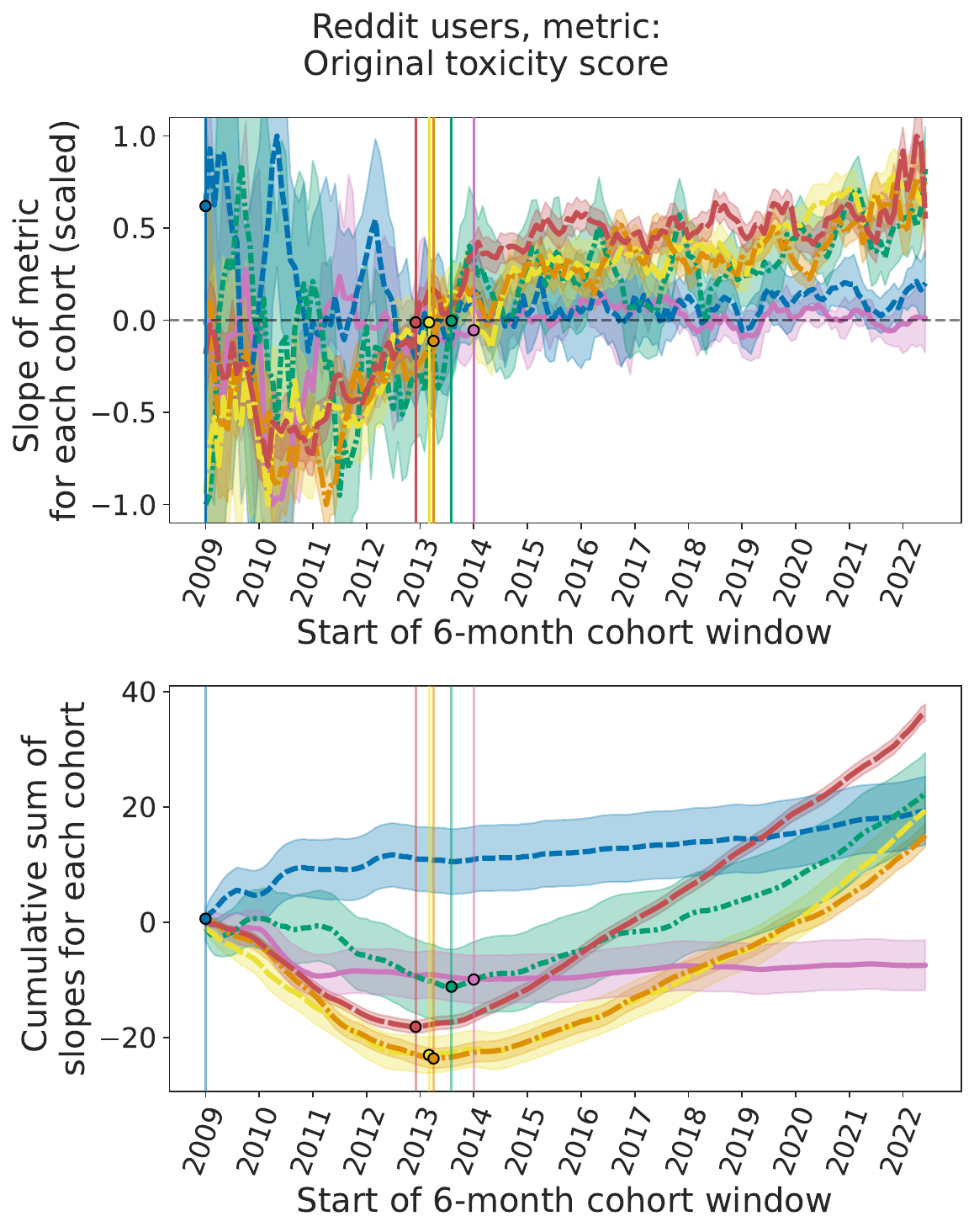} 
        \caption{}
    \end{subfigure}
    \begin{subfigure}[b]{0.24\columnwidth}
        \centering
        \includegraphics[width=1\columnwidth]{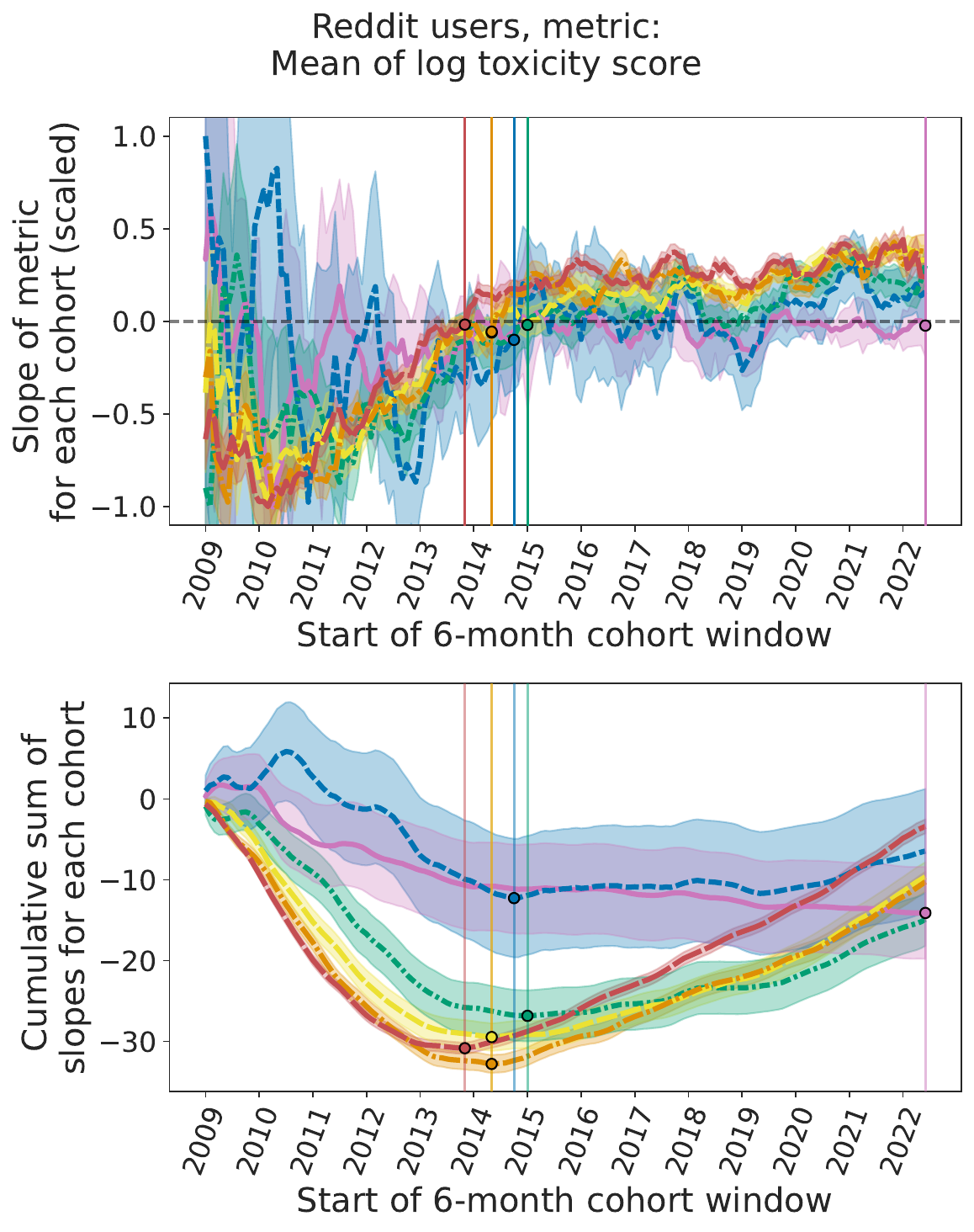} 
        \caption{}
    \end{subfigure}
    
    \begin{subfigure}[b]{0.24\columnwidth}
        \centering
        \includegraphics[width=1\columnwidth]{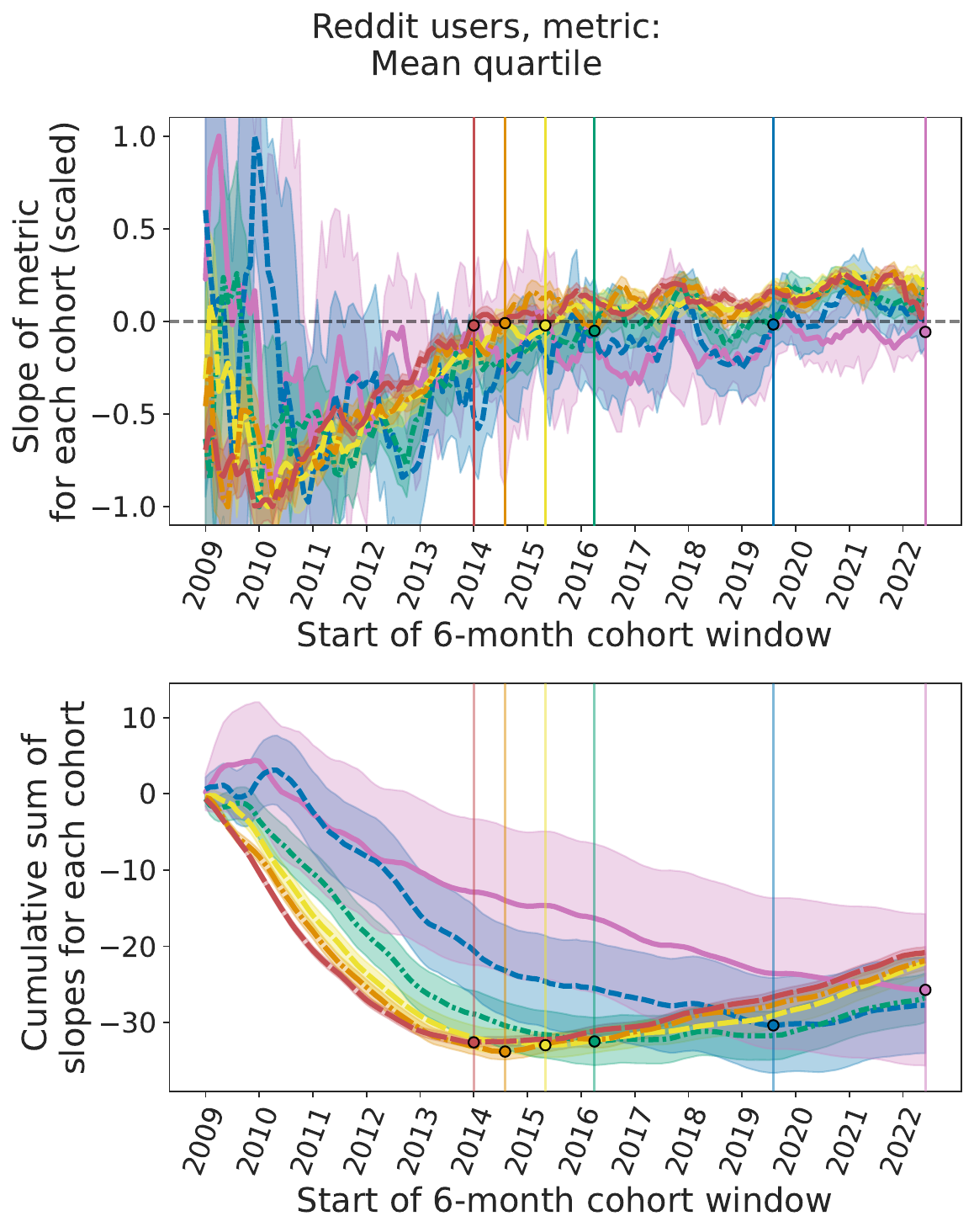} 
        \caption{}
        \label{subfig_supp:SF1_slope_c}
    \end{subfigure}
    \begin{subfigure}[b]{0.24\columnwidth}
        \centering
        \includegraphics[width=1\columnwidth]{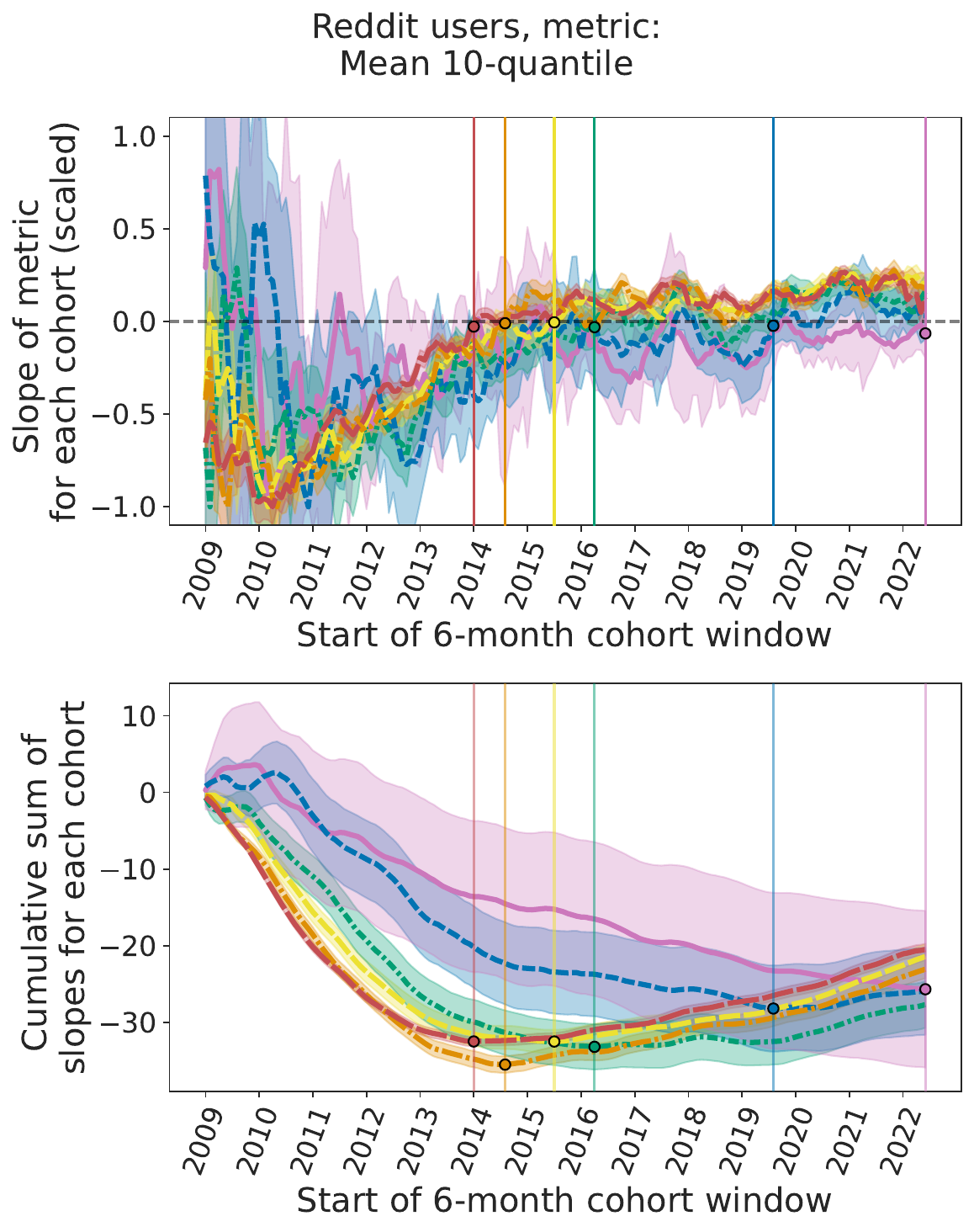}
        \caption{}
    \end{subfigure}
    \begin{subfigure}[b]{0.24\columnwidth}
        \centering
        \includegraphics[width=1\columnwidth]{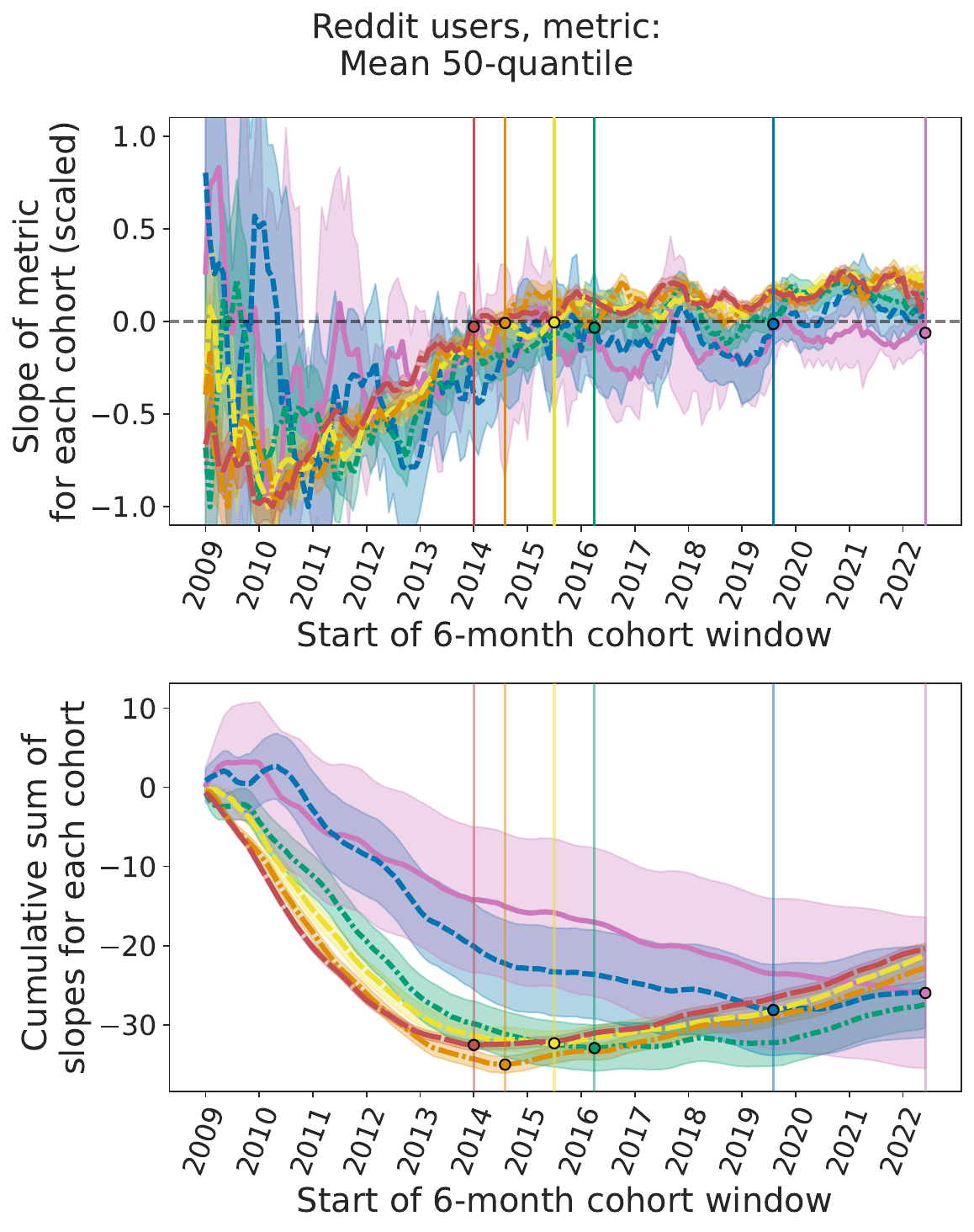}
        \caption{}
        \label{subfig_supp:SF1_slope_e}
    \end{subfigure}

    \begin{subfigure}[b]{0.19\columnwidth}
        \centering
        \includegraphics[width=1\columnwidth]{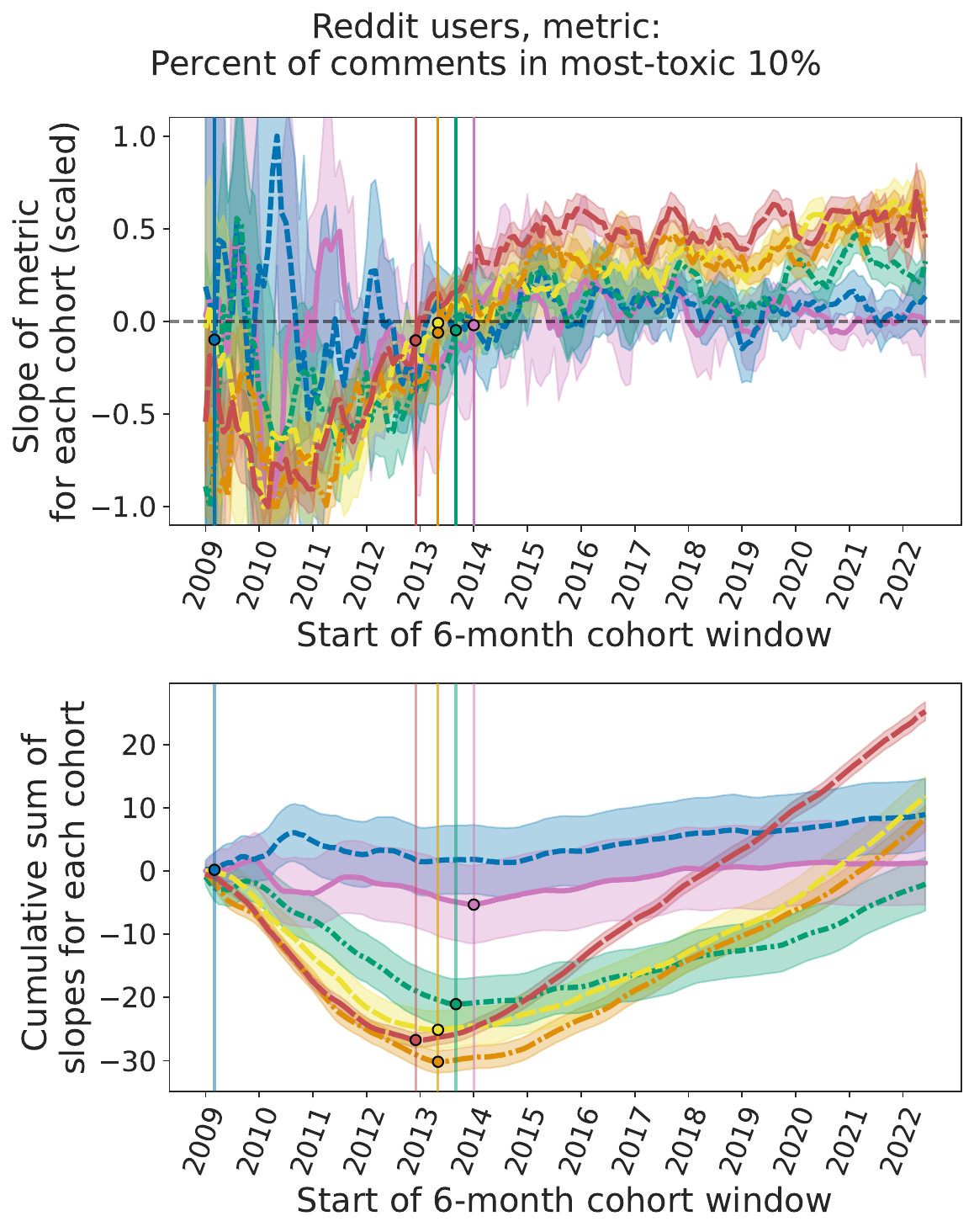} 
        \caption{}
        \label{subfig_supp:SF1_slope_f}
    \end{subfigure}
    \begin{subfigure}[b]{0.19\columnwidth}
        \centering
        \includegraphics[width=1\columnwidth]{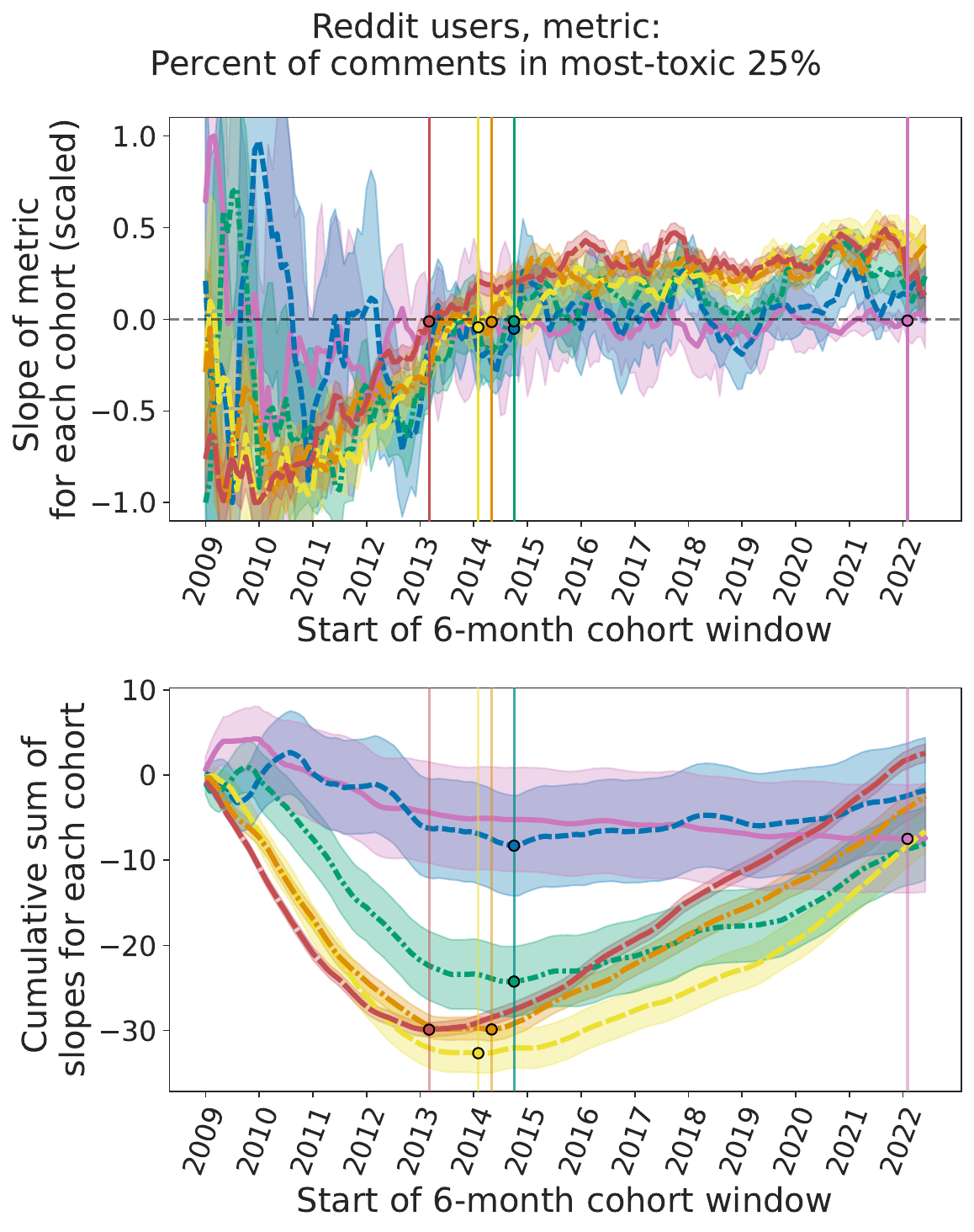} 
        \caption{}
    \end{subfigure}
    \begin{subfigure}[b]{0.19\columnwidth}
        \centering
        \includegraphics[width=1\columnwidth]{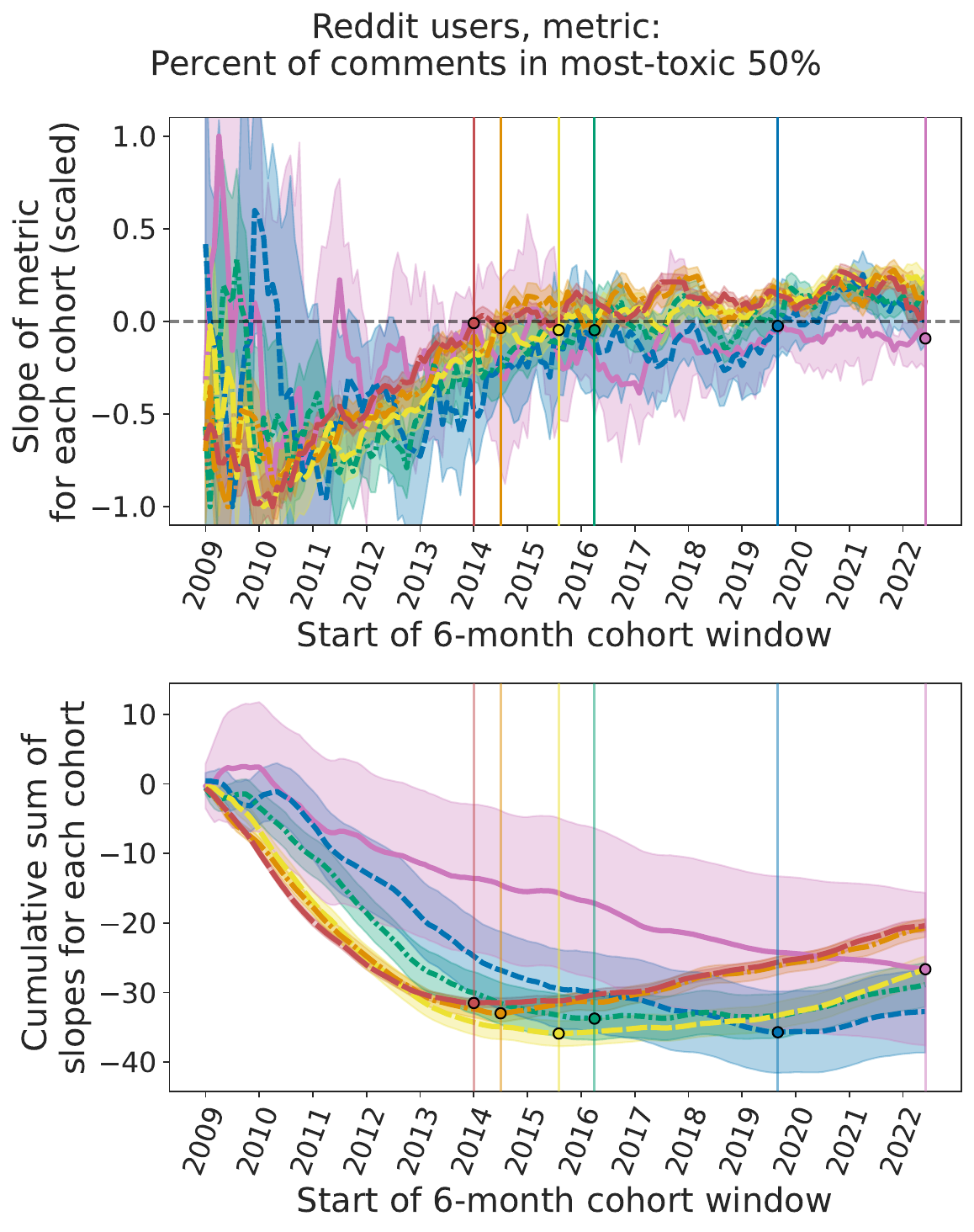} 
        \caption{}
    \end{subfigure}
    \begin{subfigure}[b]{0.19\columnwidth}
        \centering
        \includegraphics[width=1\columnwidth]{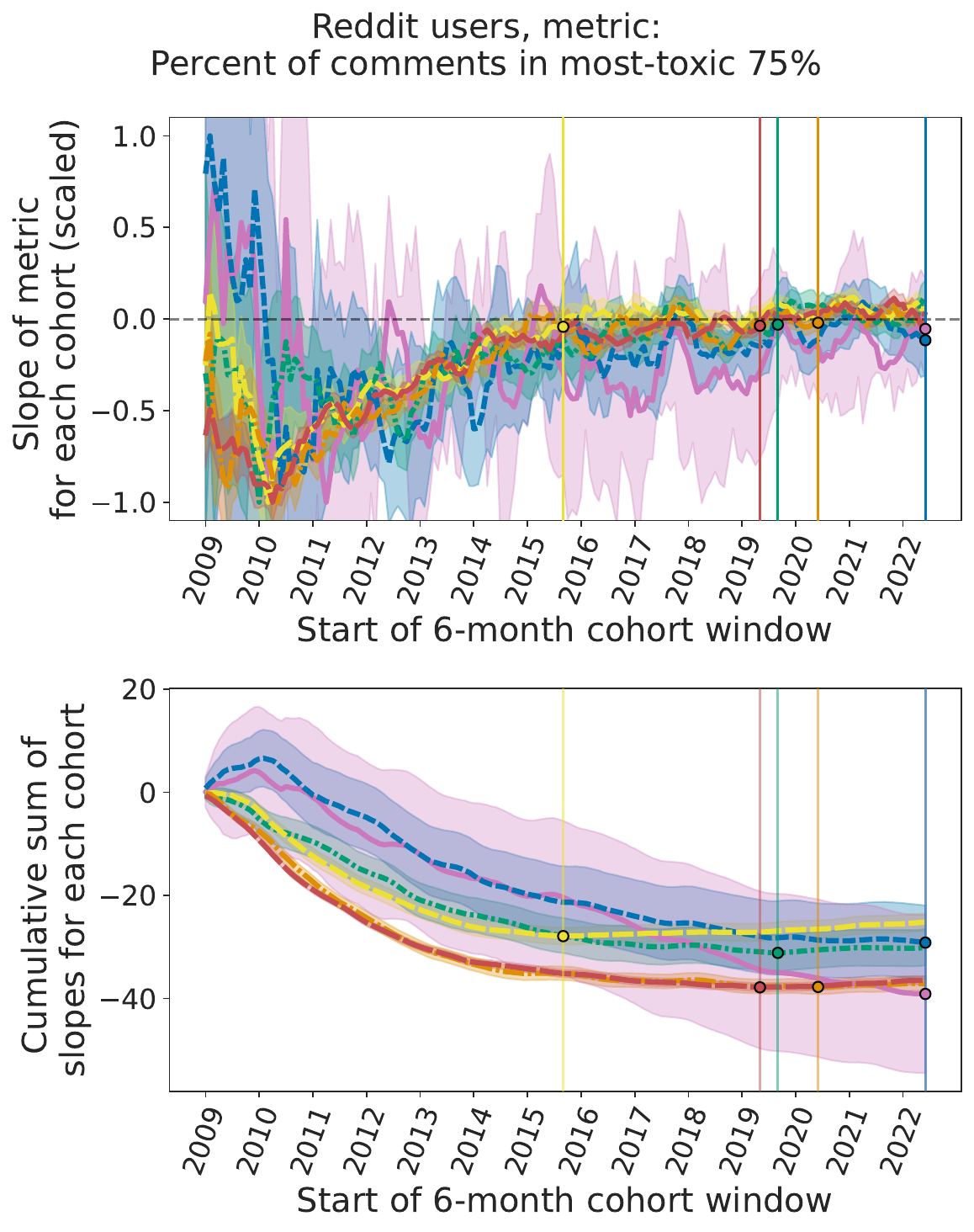} 
        \caption{}
    \end{subfigure}
    \begin{subfigure}[b]{0.19\columnwidth}
        \centering
        \includegraphics[width=1\columnwidth]{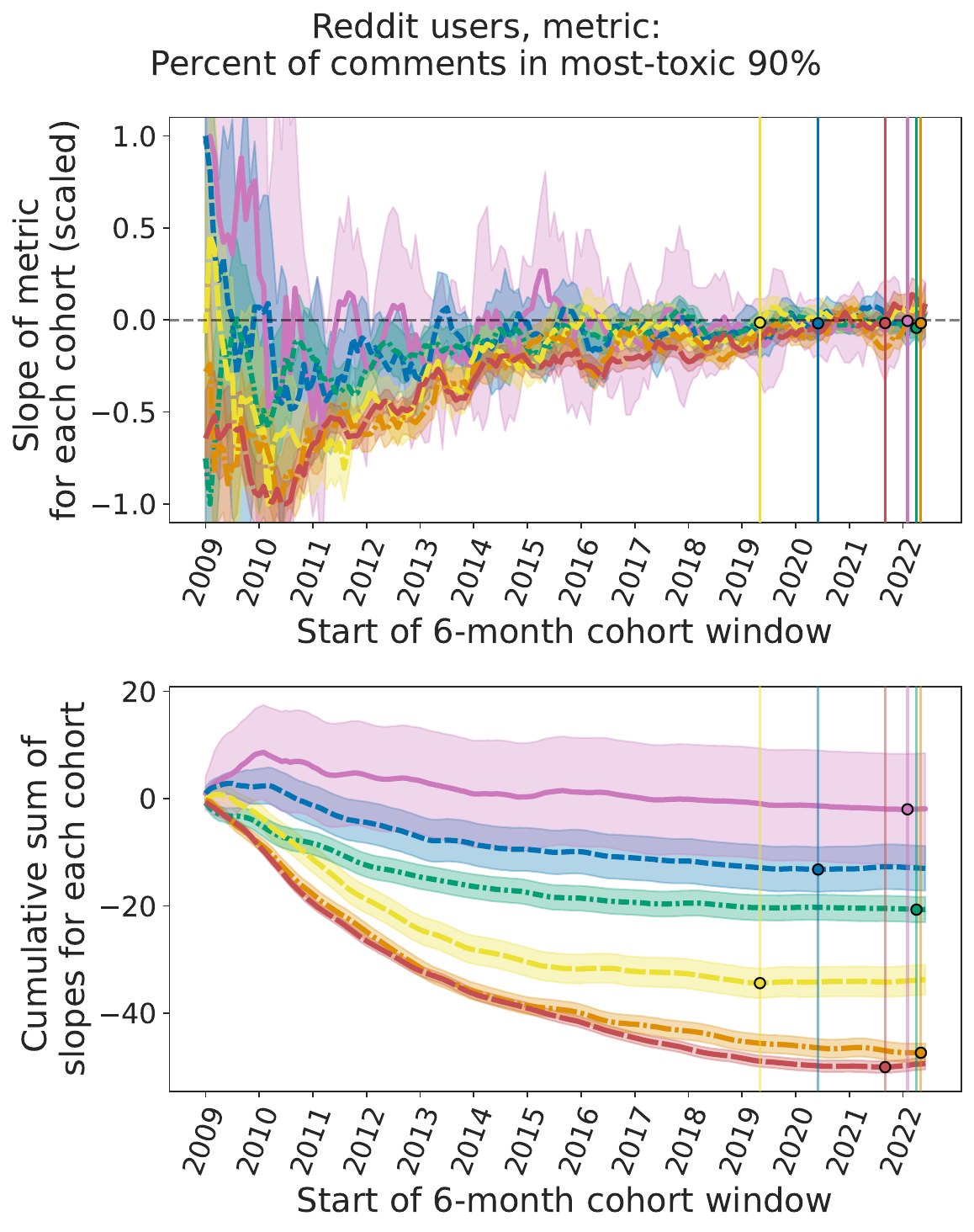} 
        \caption{}
        \label{subfig_supp:SF1_slope_j}
    \end{subfigure}

    \caption{Changes over time in user lifetime trends, for various versions of the SEVERE\_TOXICITY score. 
    \cbrk
    The first row shows the original SEVERE\_TOXICITY score and the natural log of that score, which are both skewed metrics (see Figures \ref{fig_supp:SF0a2_tox_hist_linear} and \ref{fig_supp:SF0a_tox_hist}). 
    The second row shows trends for the SEVERE\_TOXICITY score transformed into quantiles at various resolution levels, creating uniform distributions of scores.
    The third row shows trends for binarized versions of the SEVERE\_TOXICITY score. This metric represents the percentage of comments in a given group that had toxicity scores in the top N\% of all comments in the dataset.
    \cbrk
    See Section \ref{section_supp:SH_integral_plots} for explanation of the cumulative sum plots.
    Legends are omitted for space, as they're identical to that in Figure \ref{fig:D_slope}.
    }
    \label{fig_supp:SF1_slope_skewedmetrics}
\end{figure}

As seen in Section \ref{section_supp:F0_metric_dists}, Perspective API's SEVERE\_TOXICITY score doesn't have a uniform distribution. To assess the effect of metric skew, we transformed the SEVERE\_TOXICITY score into two different kinds of non-skewed versions of the metric, and found similar results. First, we used various versions of quantiles (Figures \ref{subfig_supp:SF1_rainbow_c}-\subref{subfig_supp:SF1_rainbow_e} and \ref{subfig_supp:SF1_slope_c}-\subref{subfig_supp:SF1_slope_e}), mapping toxicity scores into a 0-1 range based on where that score ranked among all toxicity scores in the dataset. Second, we used a binarized version of the score (Fig. \ref{subfig_supp:SF1_rainbow_f}-\subref{subfig_supp:SF1_rainbow_j} and \ref{subfig_supp:SF1_slope_f}-\subref{subfig_supp:SF1_slope_j}), analogous to the downvote score, where a toxicity score in the top N\% of all scores in the dataset gets a score of 1, and the other 100-N\% of comments get a score of 0. 

Interestingly, the binarized scores show similar results to the original when N is lower; in other words, only when the metric is isolating the most toxic comments, rather than the least toxic comments. This tracks with what we see in Section \ref{section_supp:D_alt_metrics_controls}, that the patterns we observe are present only for metrics that are more specific to toxic behavior.

\clearpage
\subsection{Are toxicity and downvote scores correlated?} 

Note that \enquote{comment vote score} here is different from \enquote{downvote score}. \enquote{Comment vote score} is the vote score for a comment as displayed on Reddit, calculated as total number of upvotes minus downvotes. \enquote{Downvote score} measures the percentage of comments that were \enquote{downvoted} (received more downvotes than upvotes), calculated as whether the comment's vote score was 0 or lower. The downvote score threshold is 0, rather than -1, because all comments start with a score of 1 (representing an upvote from the author's own account).

\begin{figure}[h!]
    \centering
    \includegraphics[width=0.45\columnwidth]{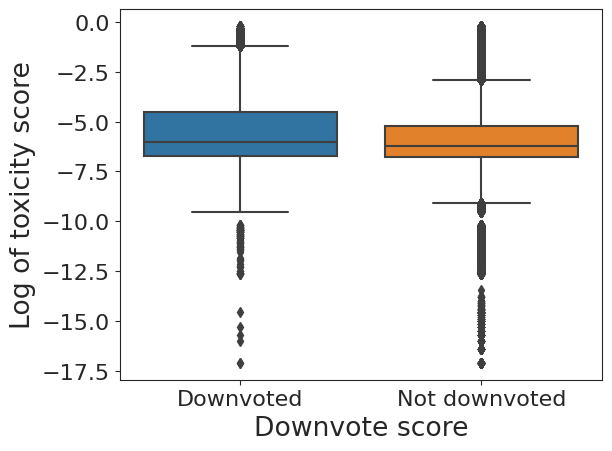} 
    \caption{Distribution of toxicity scores for both values of the binary downvote metric.}
    \label{fig_supp:SFa_boxplot_downvote_score}
\end{figure}

\begin{figure}[h!]
    \centering
    \includegraphics[width=0.45\columnwidth]{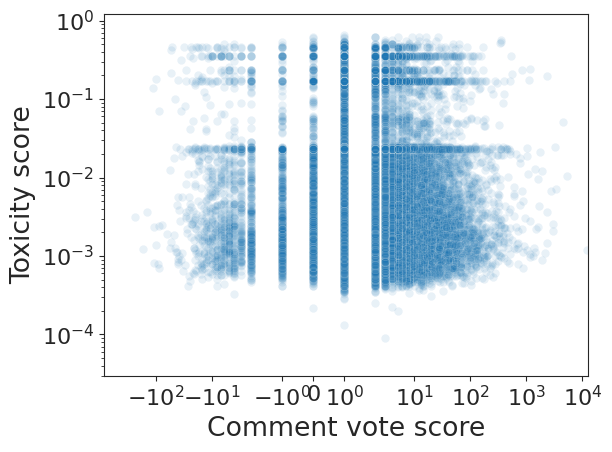} 
    \caption{Comment vote score vs. toxicity score.}
    \label{fig_supp:SFb_scatter_vote_score}
\end{figure}

\clearpage
\subsection{Are toxicity scores correlated with comment length?}

\begin{figure}[h!]
    \centering
    \includegraphics[width=0.45\columnwidth]{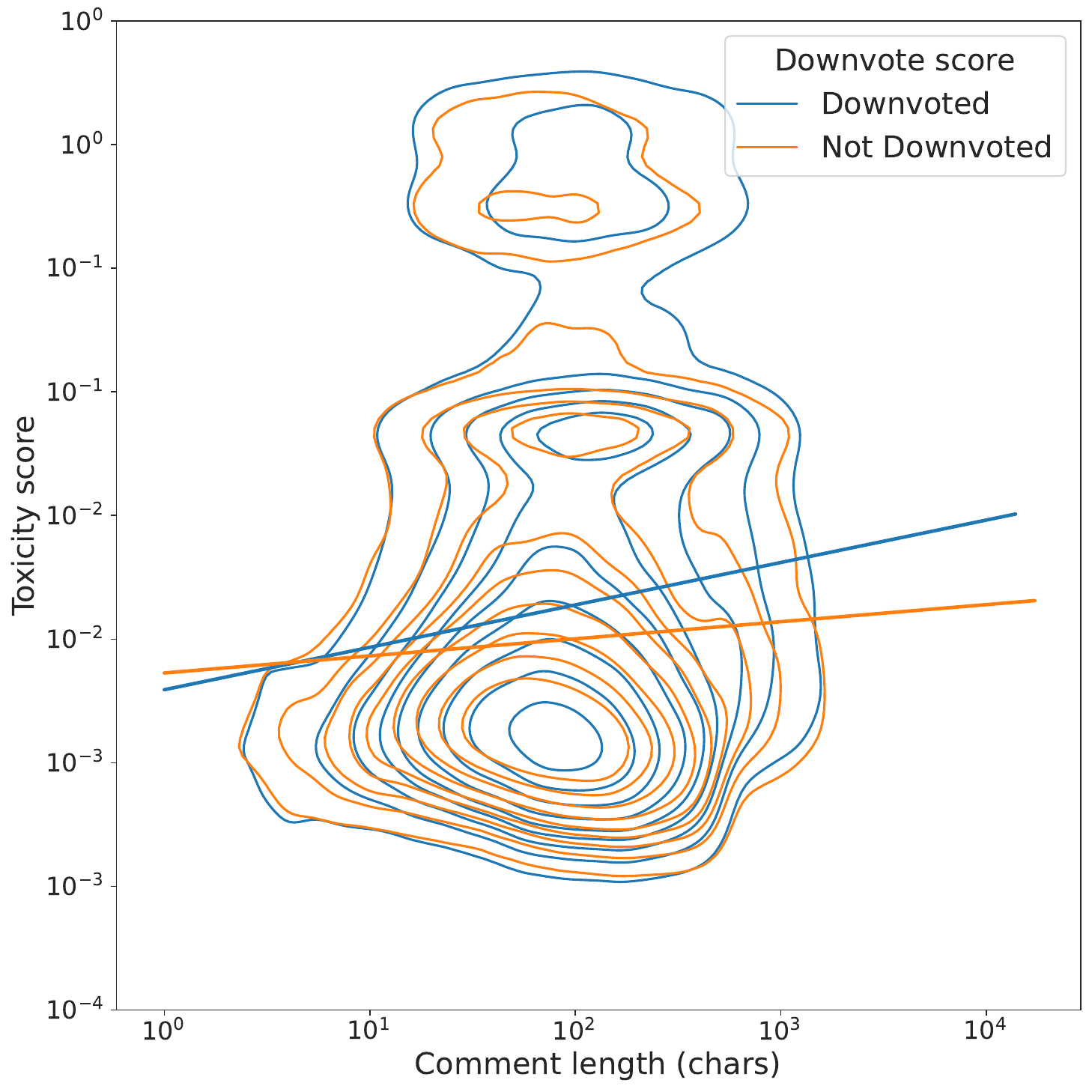} 
    \caption{Distribution of comment length vs. toxicity score. Each density plot generated from a sample of 100,000 comments. Regression lines generated from entire dataset, with 100 bootstraps used for the confidence intervals shown.}
    \label{fig_supp:SK_comment_tox_jointplot}
\end{figure}

\begin{figure}[h!]
    \centering
    \includegraphics[width=0.45\columnwidth]{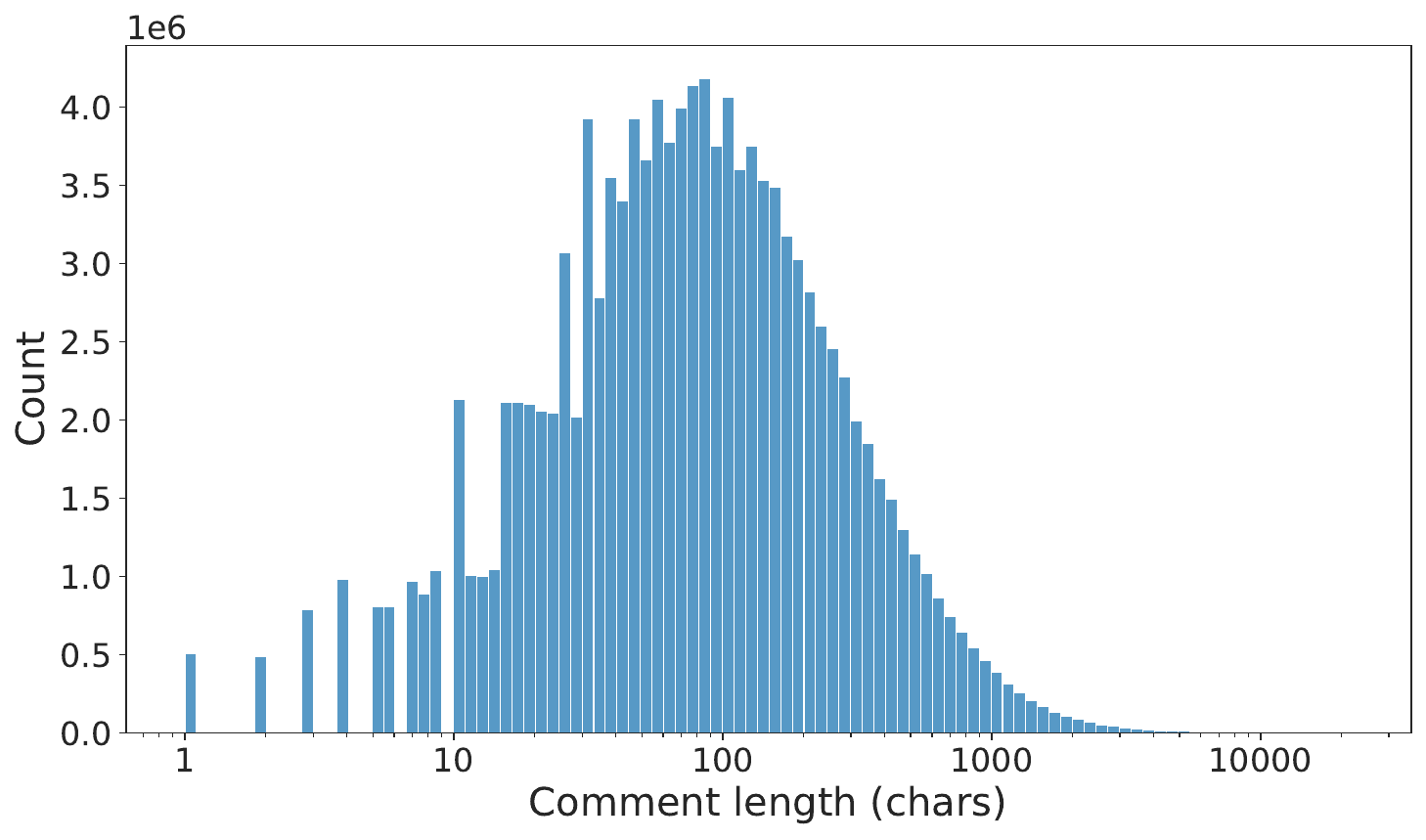} 
    \caption{Distribution of comment lengths.}
    \label{fig_supp:SK_comment_hist}
\end{figure}

\clearpage
\subsection{Do mean scores change over time?}
\label{section_supp:F3_total_tox_over_time}

\begin{figure}[h]
    \centering
    
    \begin{subfigure}[b]{0.45\columnwidth}
        \centering
        \includegraphics[width=1\columnwidth]{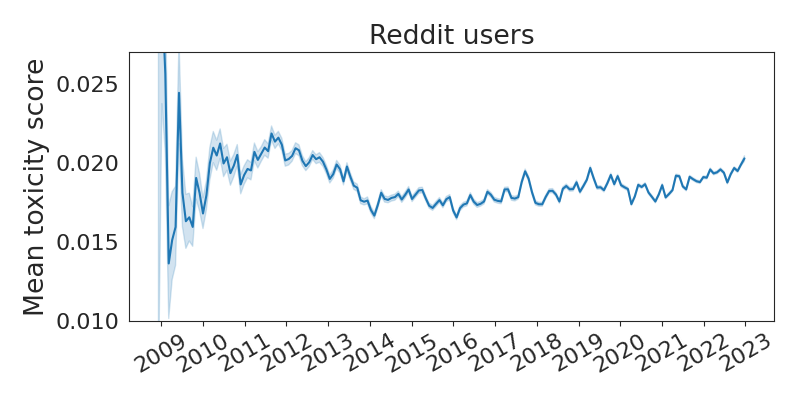} 
        \caption{Mean toxicity score over time}
    \end{subfigure}
    \begin{subfigure}[b]{0.45\columnwidth}
        \centering
        \includegraphics[width=1\columnwidth]{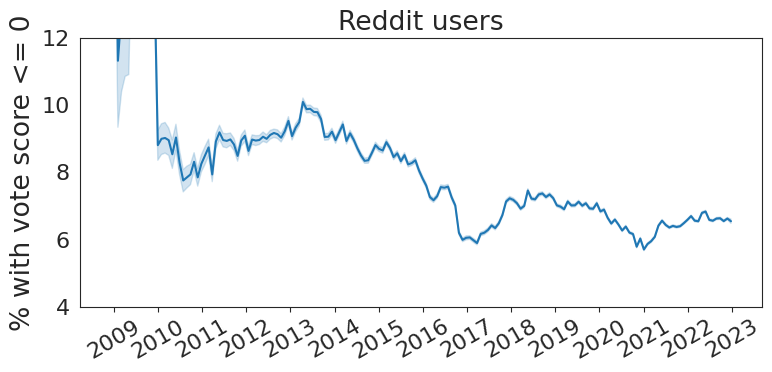} 
        \caption{Mean downvote score over time}
    \end{subfigure}

    \caption{Mean toxicity and downvote scores per month, over all 120 million comments made by our sample of 1 million Reddit users.}
    \label{fig_supp:SF3a_mean_scores_all_reddit}
\end{figure}

\clearpage
\section{Metrics: Alternate metrics / negative \enquote{control}}
\label{section_supp:D_alt_metrics_controls}

In this section are versions of Figures \ref{fig:B_rainbow} and \ref{fig:D_slope}, for various metrics besides the toxicity and downvote scores.

First, we look at alternate vote score metrics: similar to the downvote score, but using a different binary threshold for comments' vote scores. Positive thresholds look very different from the downvote score. Negative thresholds look similar to the downvote score when close to 0, but get less similar as thresholds become more negative. This may be because the farther a score is from 1, the more people must have voted on that comment; few comments are seen by large numbers of people, so most comments don't have the opportunity to get many votes. Thus, thresholds with higher absolute value are really measuring two things at once: the perceived value of the comment, and the number of people who saw it. 

\enquote{Controversiality} is a binary measure, which is 1 if the ratio of upvotes to downvotes is close to 1, and 0 otherwise. To prevent manipulation, Reddit provides only this measure, instead of raw numbers of upvotes and downvotes. Reddit also doesn't publish the threshold it uses to determine whether the upvote/downvote ratio is \enquote{close} to 1. Before 2017 at least, when Reddit was open-source, it appears that the criteria were that upvotes were 40-60\% of total votes, and that there were at least 7 total votes.\cite{justachetan2019whatWITHLINK}

Note that all of these metrics are binary, so aggregations over groups of comments are expressed as fractions between 0 and 1. The downvote score from the main text is also a binary vote score threshold, but is expressed as a percentage out of 100, i.e., a fraction similar to these metrics but multiplied by 100.

\subsection{Mean score figures}

\begin{figure}[h!]
    \centering
    \includegraphics[width=0.67\columnwidth]{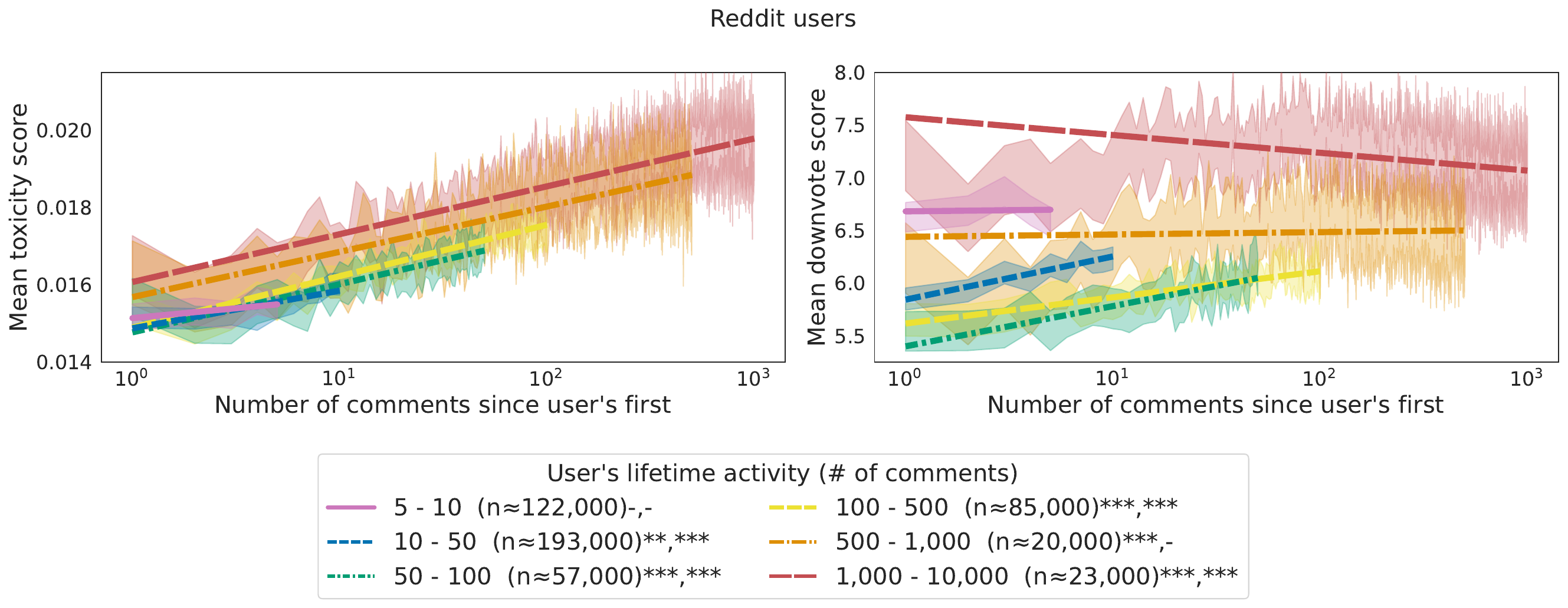} 
    \caption{Figure \ref{fig:B_rainbow} from main text, reproduced for comparison.}
    \label{fig_supp:B_repeat_SDa}
\end{figure}

\begin{figure}[h]
    \centering
    
    \begin{subfigure}[b]{0.33\columnwidth}
        \centering
        \includegraphics[width=1\columnwidth]{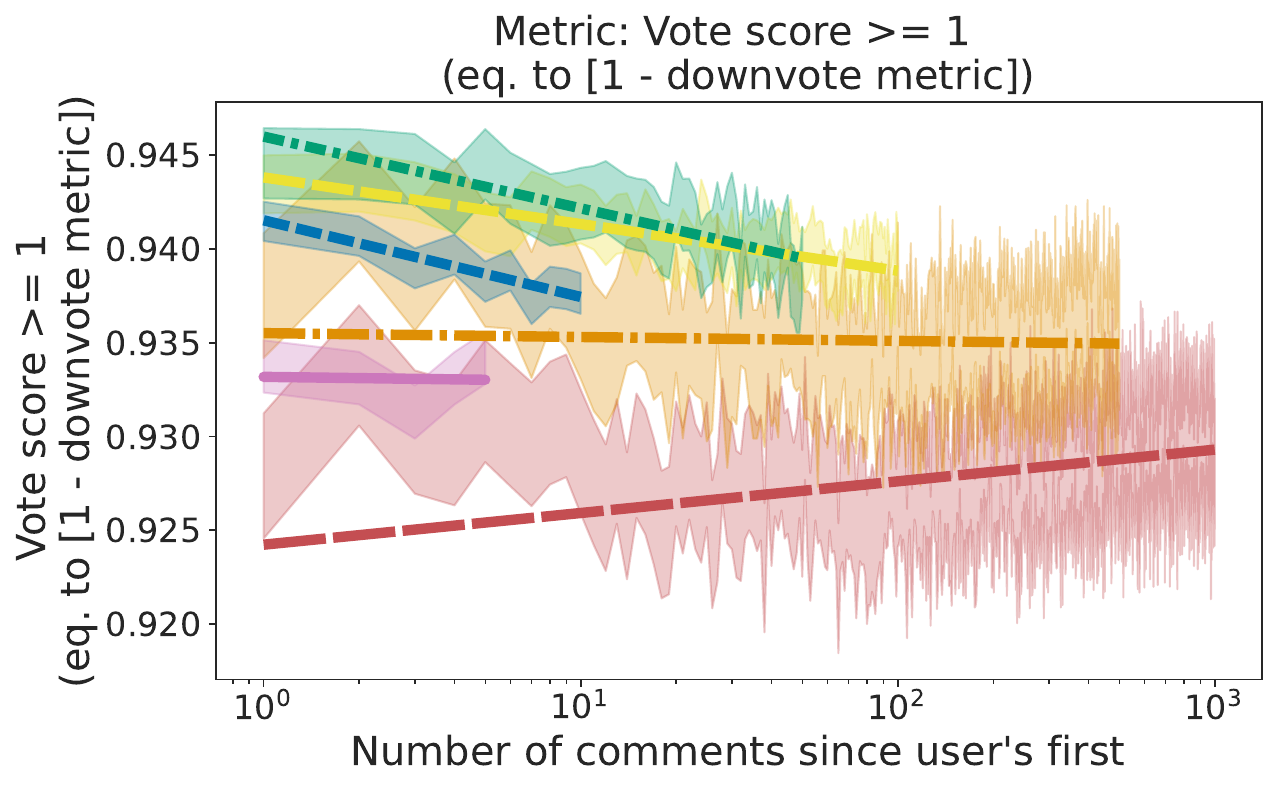} 
        \caption{}
        \label{fig:SDa1_rainbow_metric_ge_1}
    \end{subfigure}
    \begin{subfigure}[b]{0.33\columnwidth}
        \centering
        \includegraphics[width=1\columnwidth]{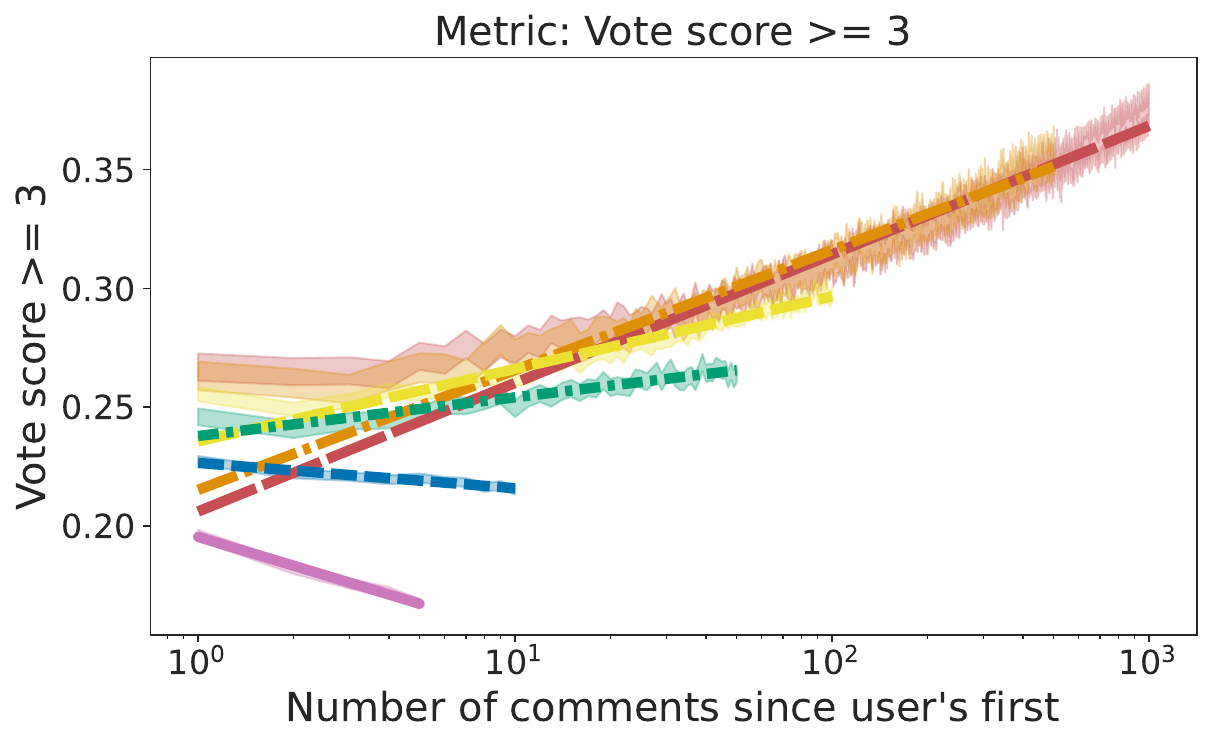} 
        \caption{}
    \end{subfigure}
    \begin{subfigure}[b]{0.33\columnwidth}
        \centering
        \includegraphics[width=1\columnwidth]{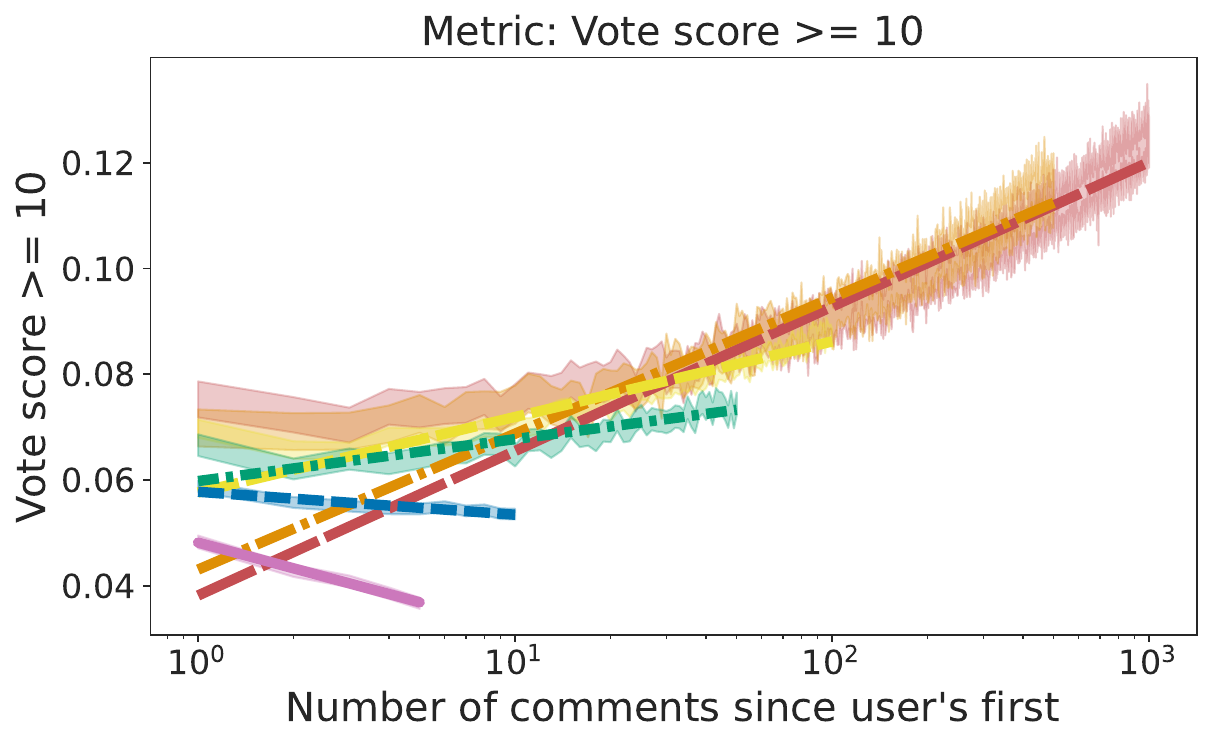} 
        \caption{}
    \end{subfigure}
    
    \caption{Trends over Reddit users' lifetimes for positive vote score thresholds. 
    \cbrk
    Legends omitted for space, as they're identical to that in Figure \ref{fig_supp:B_repeat_SDa}.
    }
    \label{fig_supp:SDa_rainbow_metric_pos_thresholds}
\end{figure}

\begin{figure}[h]
    \centering
    
    \begin{subfigure}[b]{0.33\columnwidth}
        \centering
        \includegraphics[width=1\columnwidth]{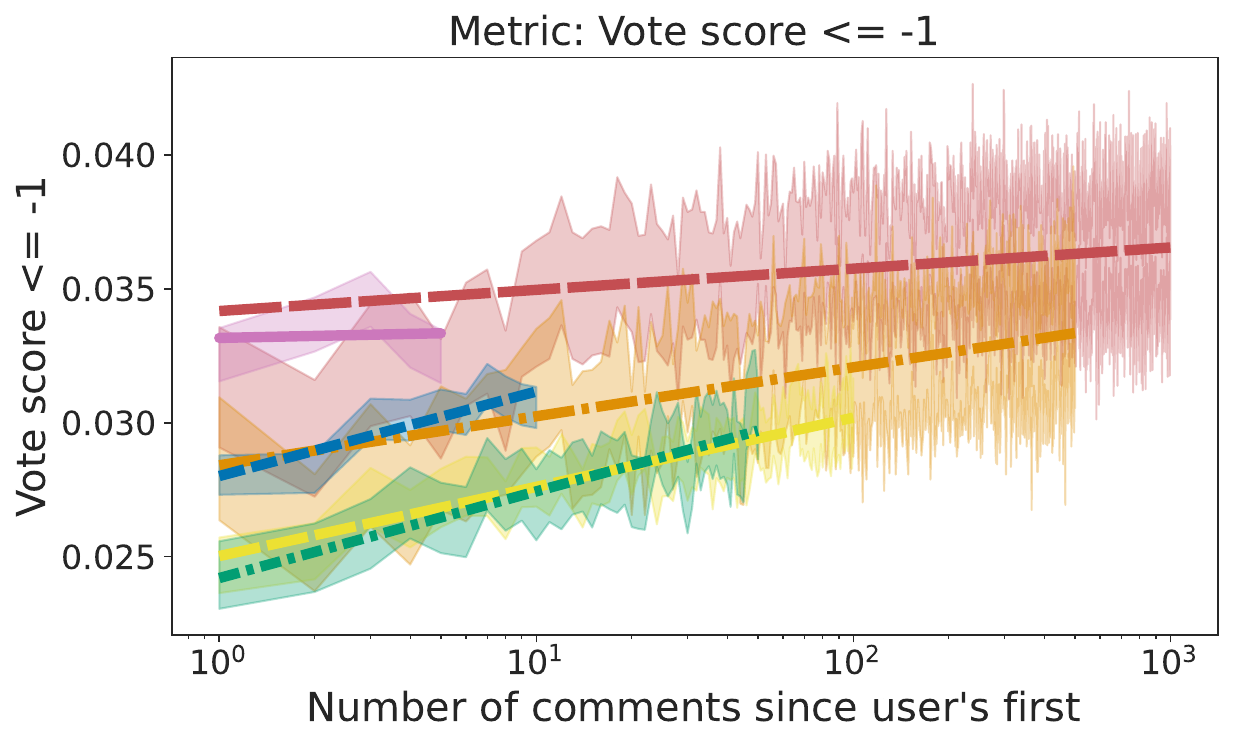} 
        \caption{}
    \end{subfigure}
    \begin{subfigure}[b]{0.33\columnwidth}
        \centering
        \includegraphics[width=1\columnwidth]{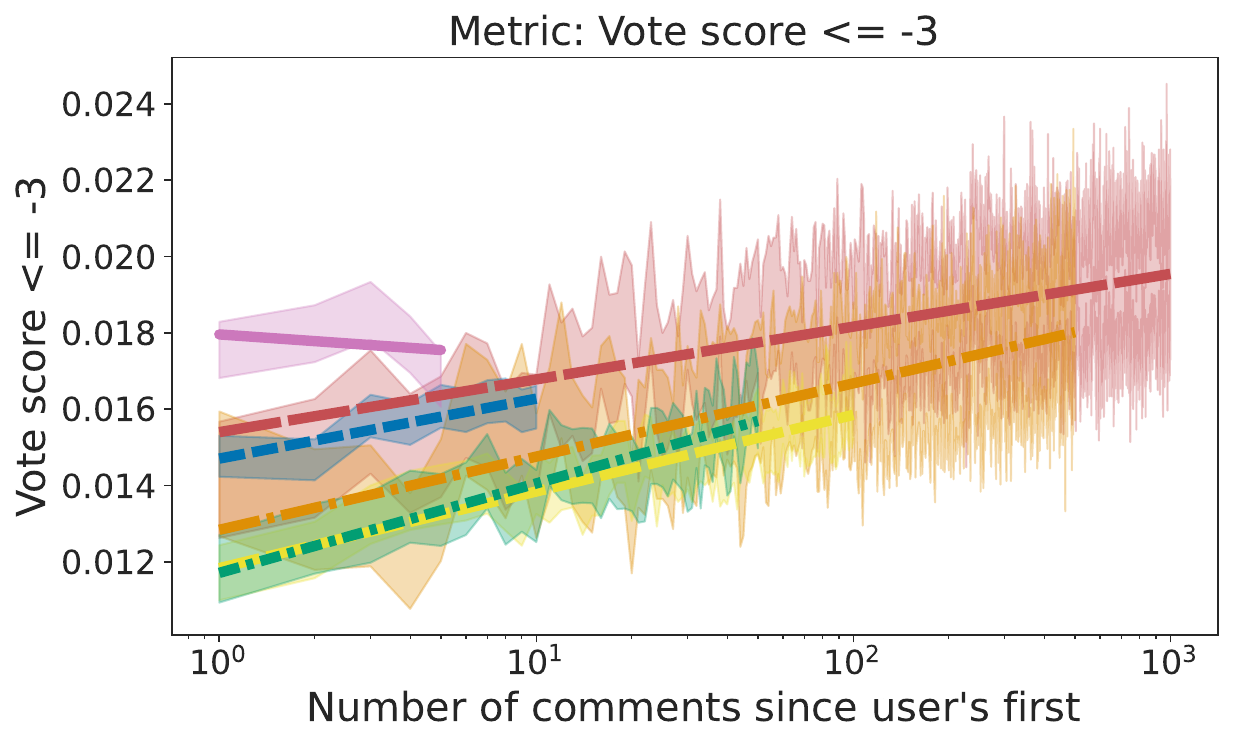} 
        \caption{}
    \end{subfigure}
    \begin{subfigure}[b]{0.33\columnwidth}
        \centering
        \includegraphics[width=1\columnwidth]{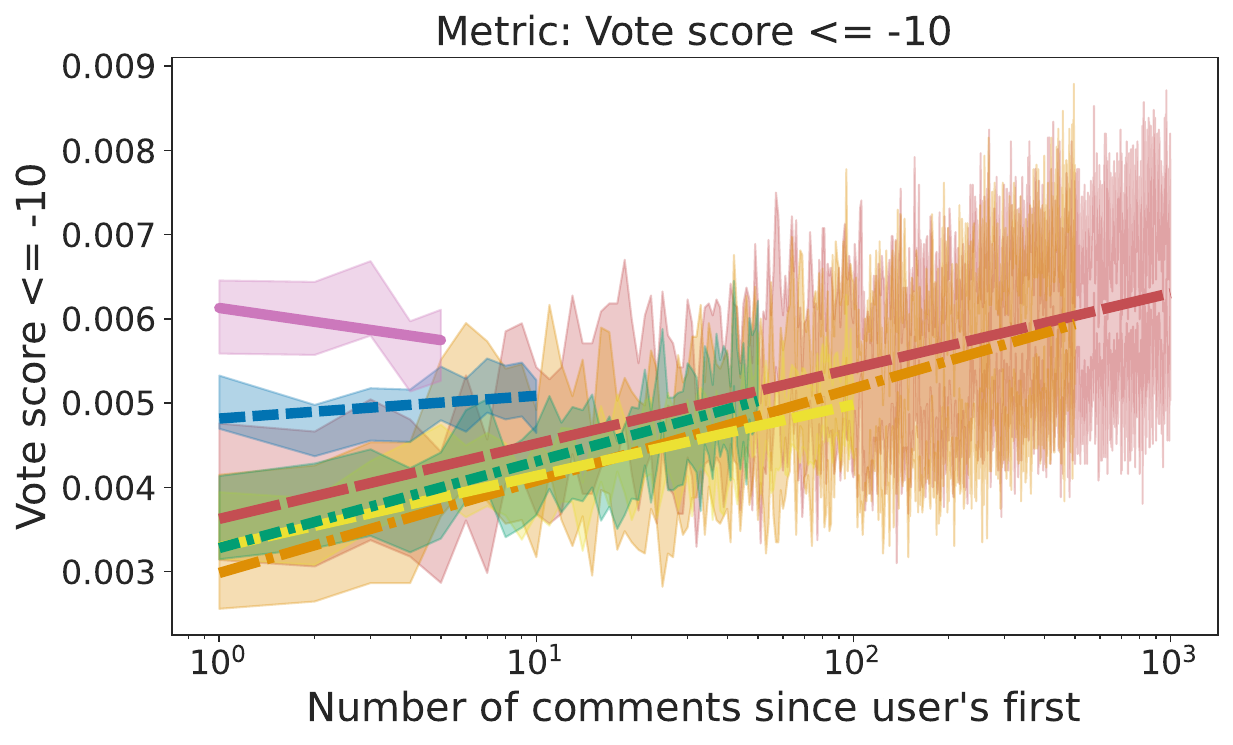} 
        \caption{}
    \end{subfigure}
    
    \caption{Trends over Reddit users' lifetimes for negative vote score thresholds.
    \cbrk
    Legends omitted for space, as they're identical to that in Figure \ref{fig_supp:B_repeat_SDa}.}
    \label{fig_supp:SDa_rainbow_metric_neg_thresholds}
\end{figure}

\begin{figure}[h!]
    \centering
    \includegraphics[width=0.33\columnwidth]{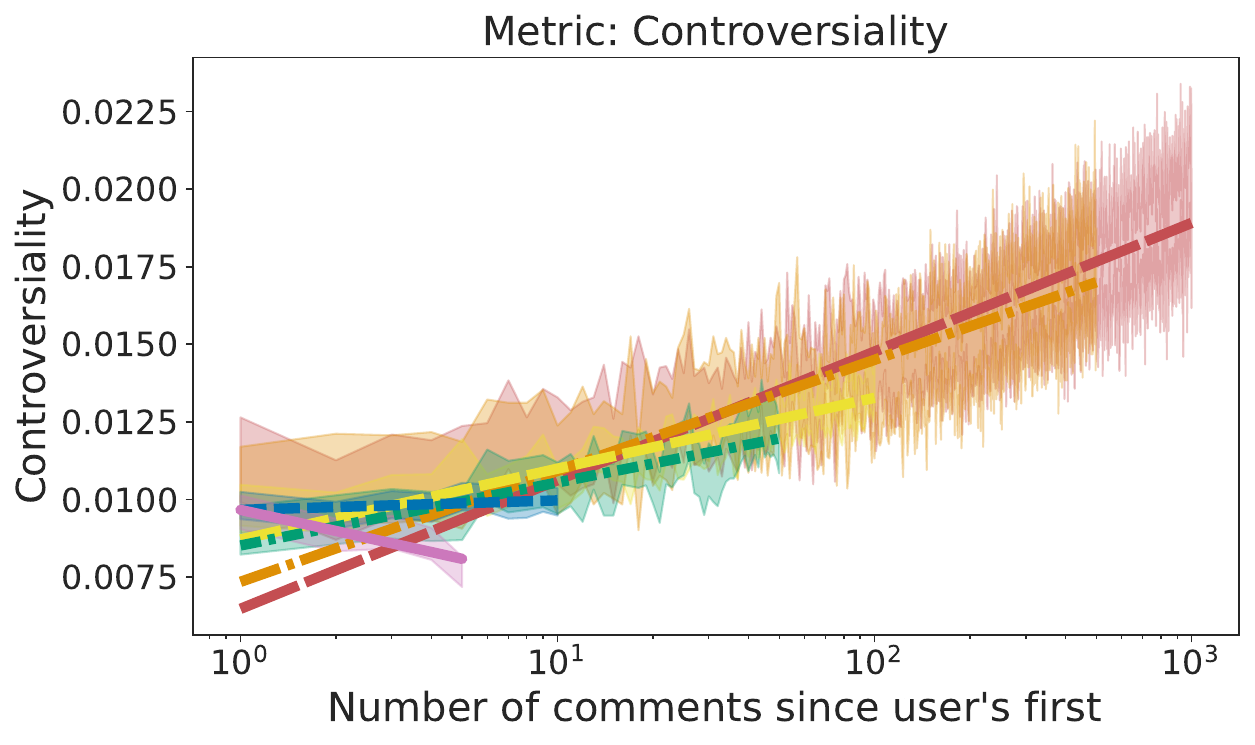} 
    \caption{Trends over Reddit users' lifetimes for \enquote{controversiality}, a binary metric for comments with upvote/downvote ratios close to 1. \cite{justachetan2019whatWITHLINK} 
    \cbrk
    Legends omitted for space, as they're identical to that in Figure \ref{fig_supp:B_repeat_SDa}.}
    \label{fig_supp:SDa_rainbow_metric_controversiality}
\end{figure}

\clearpage
\subsection{Slope figures}

\begin{figure}[h!]
    \centering
    \includegraphics[width=0.6\columnwidth]{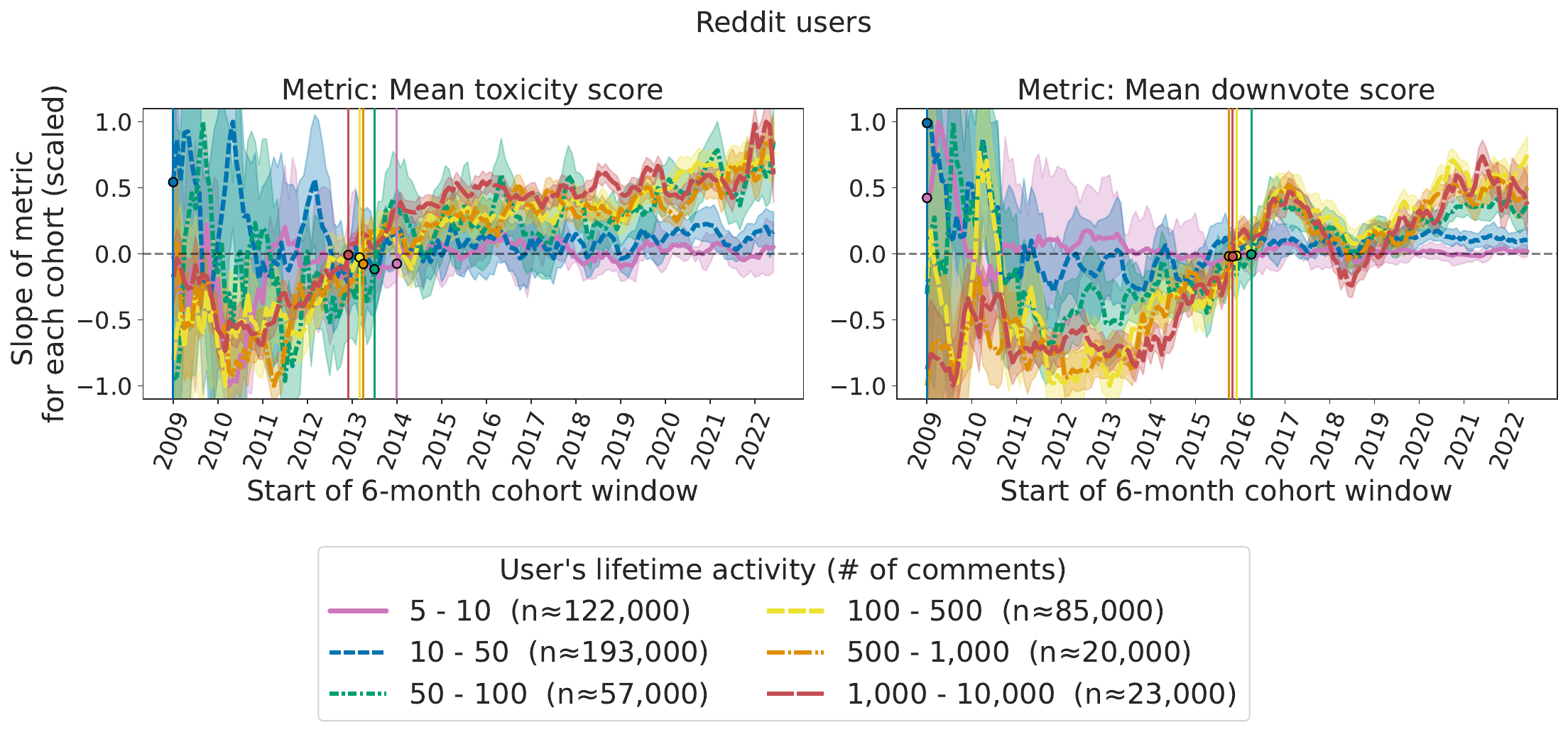} 
    \caption{Figure \ref{fig:D_slope} from main text, reproduced for comparison.}
    \label{fig_supp:D_repeat_SDb}
\end{figure}

\begin{figure}[h]
    \centering
    
    \begin{subfigure}[b]{0.3\columnwidth}
        \centering
        \includegraphics[width=1\columnwidth]{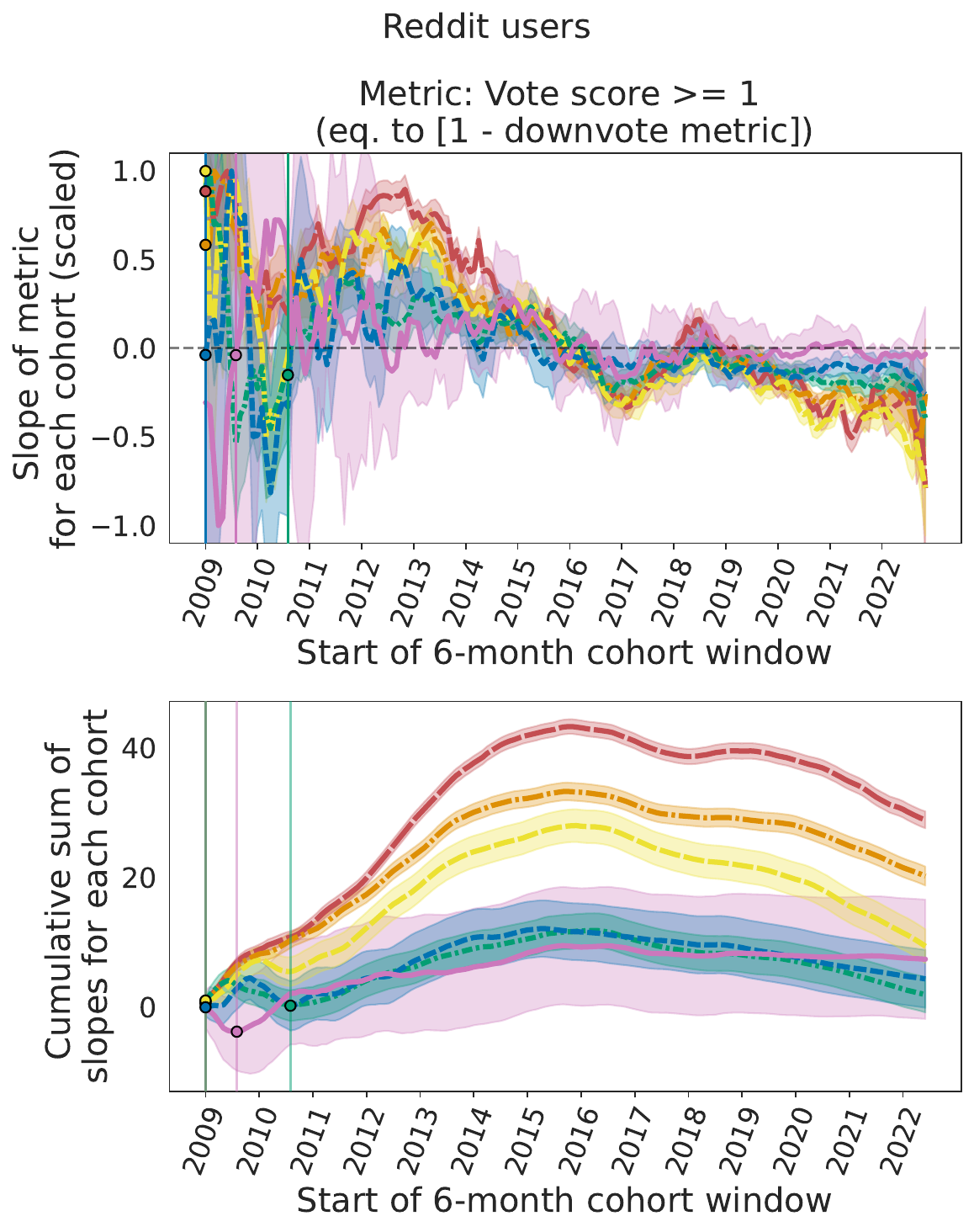} 
        \caption{}
    \end{subfigure}
    \begin{subfigure}[b]{0.3\columnwidth}
        \centering
        \includegraphics[width=1\columnwidth]{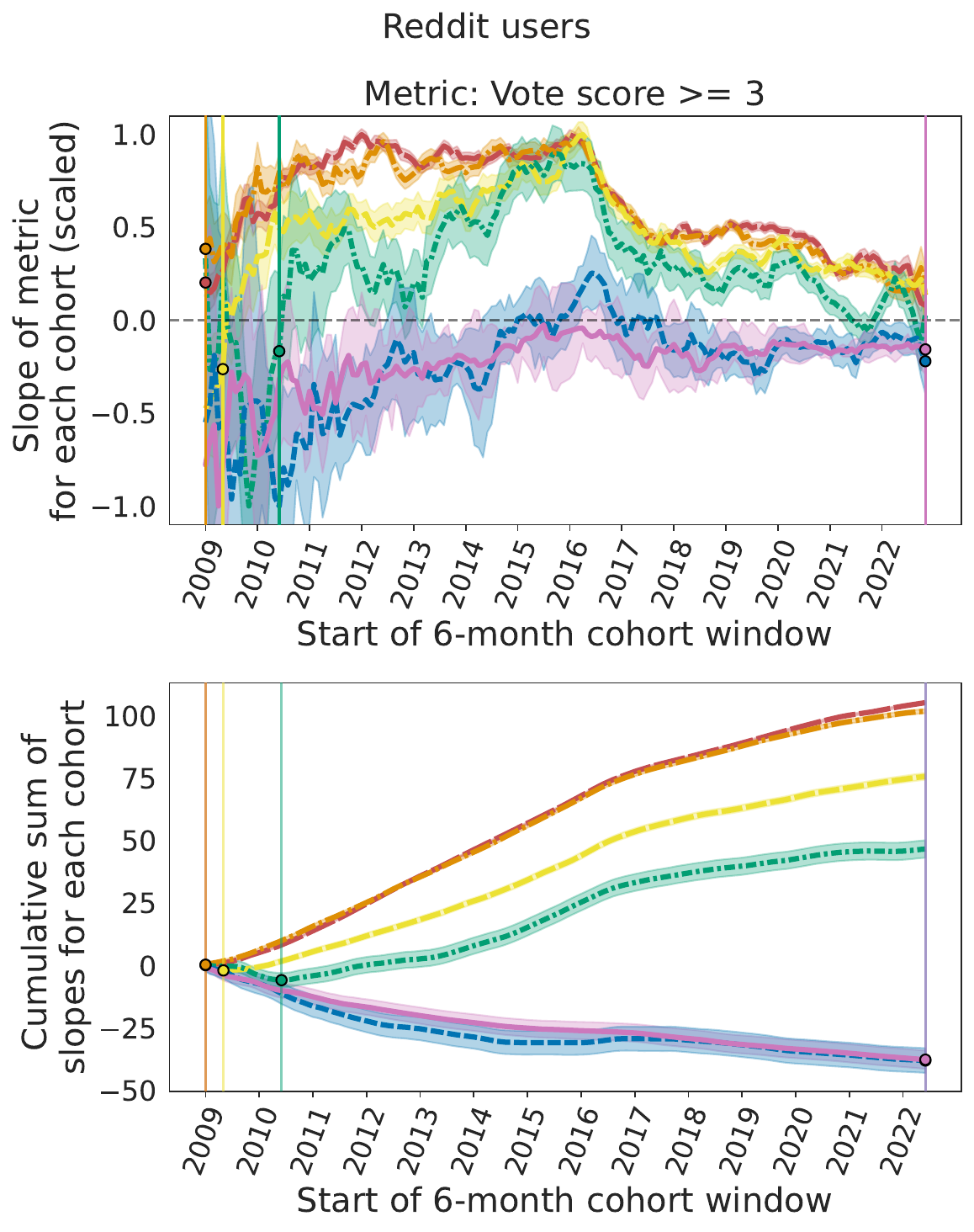} 
        \caption{}
    \end{subfigure}
    \begin{subfigure}[b]{0.3\columnwidth}
        \centering
        \includegraphics[width=1\columnwidth]{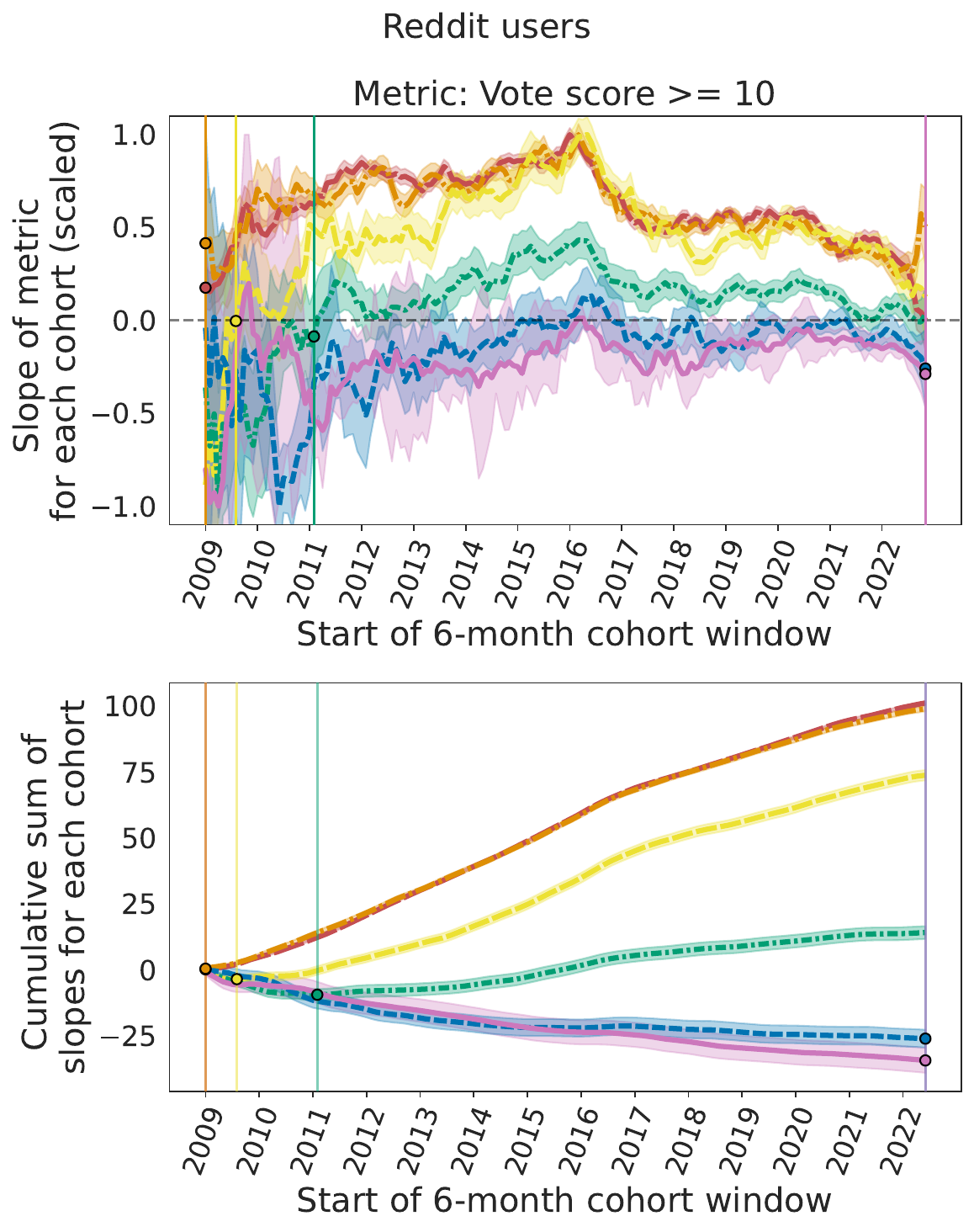} 
        \caption{}
    \end{subfigure}
    
    \caption{Changes over time in user lifetime trends, for negative vote score thresholds.
    \cbrk
    Note that absolute distance from 0 shouldn't be compared between activity level buckets, as each bucket is scaled so that the maximum absolute value is 1.
    \cbrk
    Legends omitted for space, as they're identical to that in Figure \ref{fig_supp:D_repeat_SDb}.
    }
    \label{fig_supp:SDb_slope_metric_pos_thresholds}
\end{figure}

\begin{figure}[h]
    \centering
    
    \begin{subfigure}[b]{0.3\columnwidth}
        \centering
        \includegraphics[width=1\columnwidth]{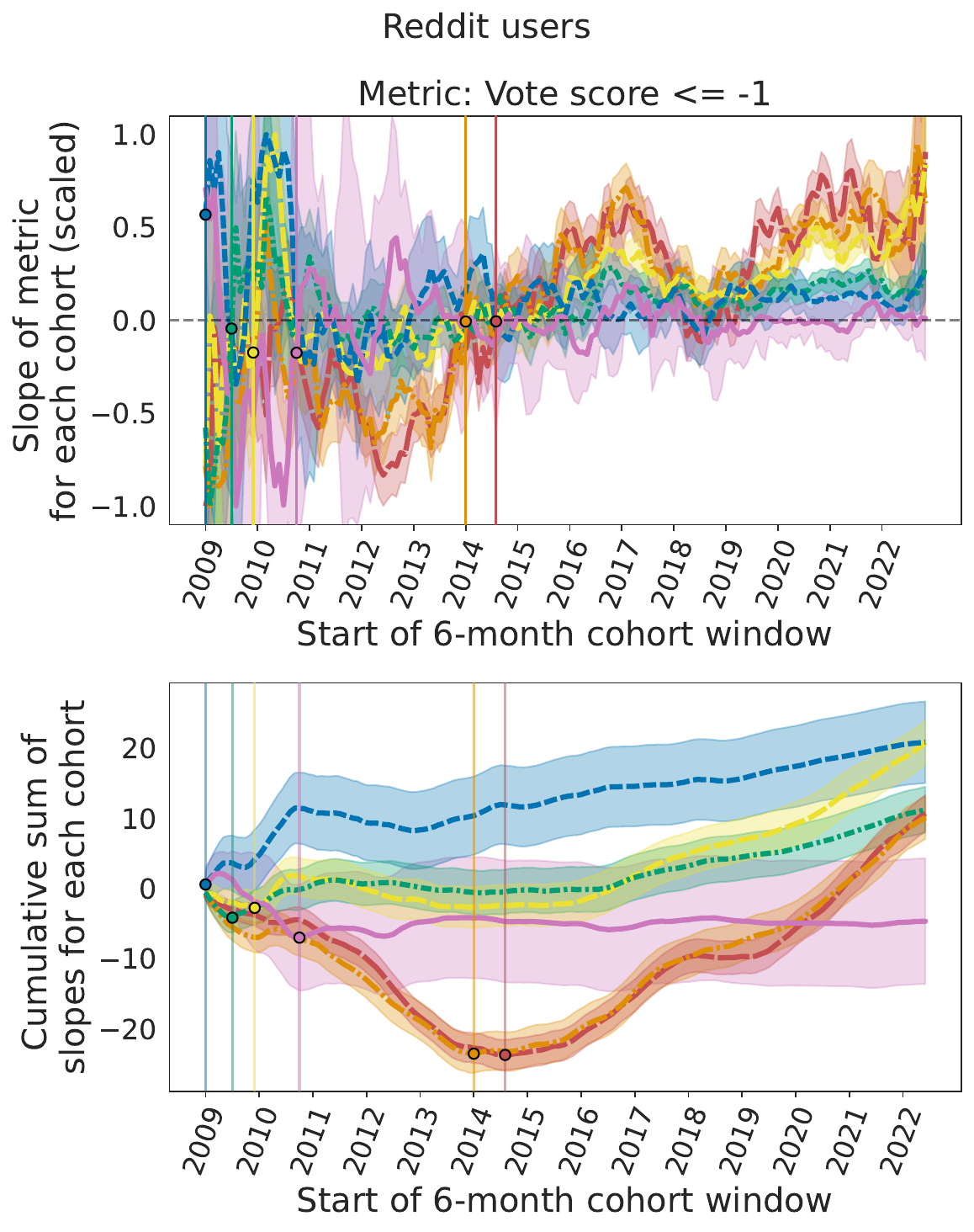} 
        \caption{}
    \end{subfigure}
    \begin{subfigure}[b]{0.3\columnwidth}
        \centering
        \includegraphics[width=1\columnwidth]{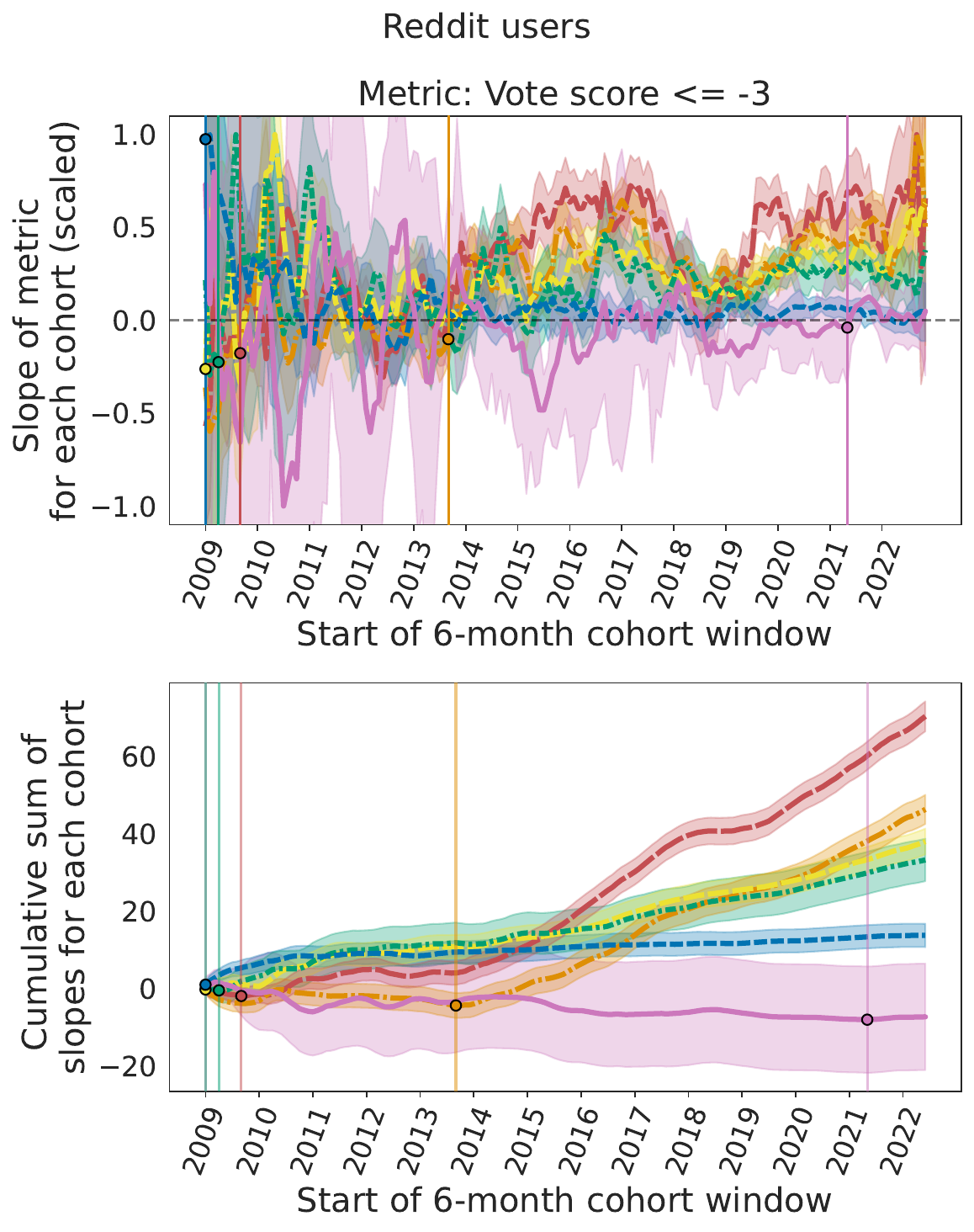} 
        \caption{}
    \end{subfigure}
    \begin{subfigure}[b]{0.3\columnwidth}
        \centering
        \includegraphics[width=1\columnwidth]{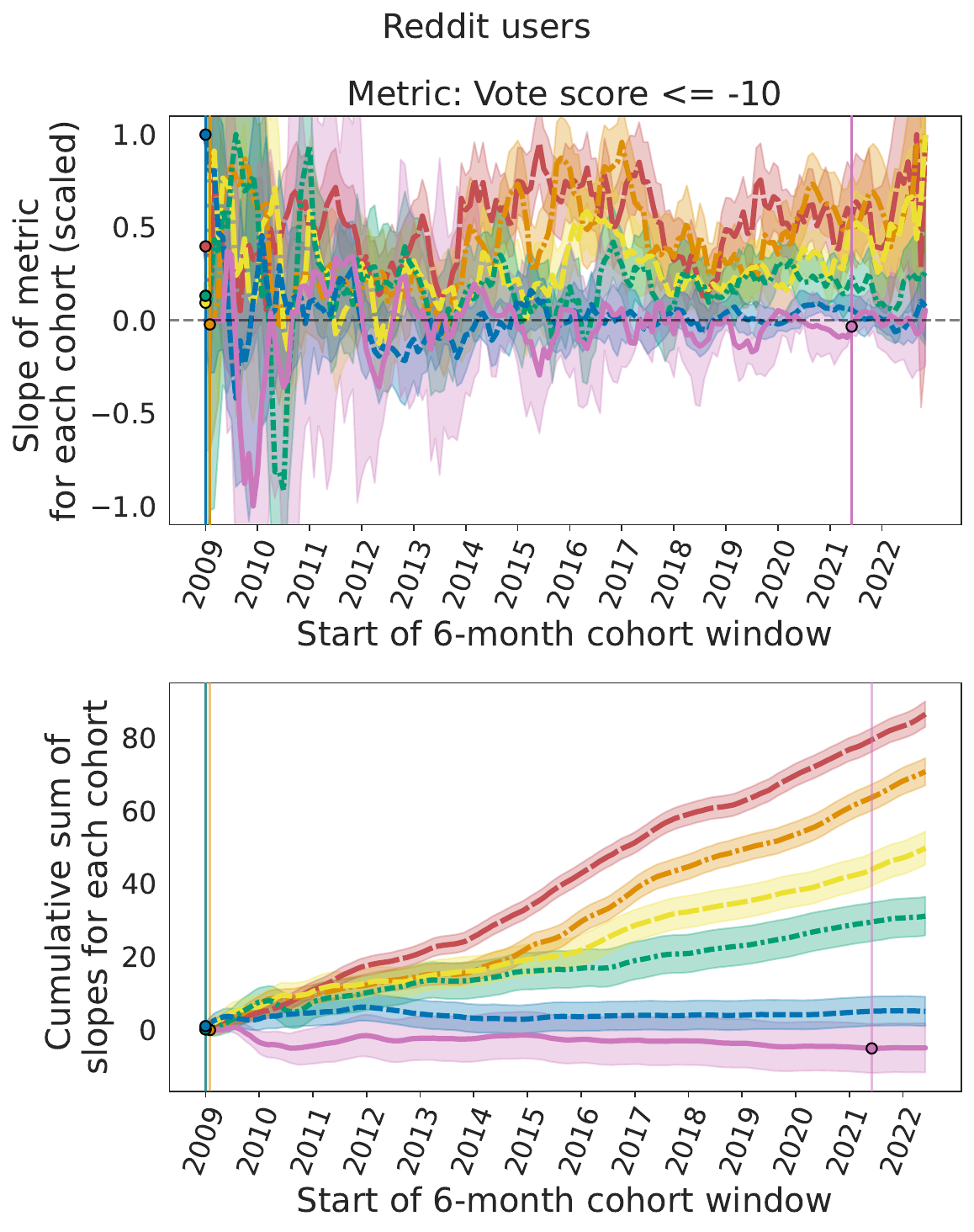} 
        \caption{}
    \end{subfigure}

    \caption{Changes over time in user lifetime trends, for negative vote score thresholds.
    \cbrk
    Legends omitted for space, as they're identical to that in Figure \ref{fig_supp:D_repeat_SDb}.}
    \label{fig_supp:SDb_slope_metric_neg_thresholds}
\end{figure}

\begin{figure}[h!]
    \centering
    \includegraphics[width=0.3\columnwidth]{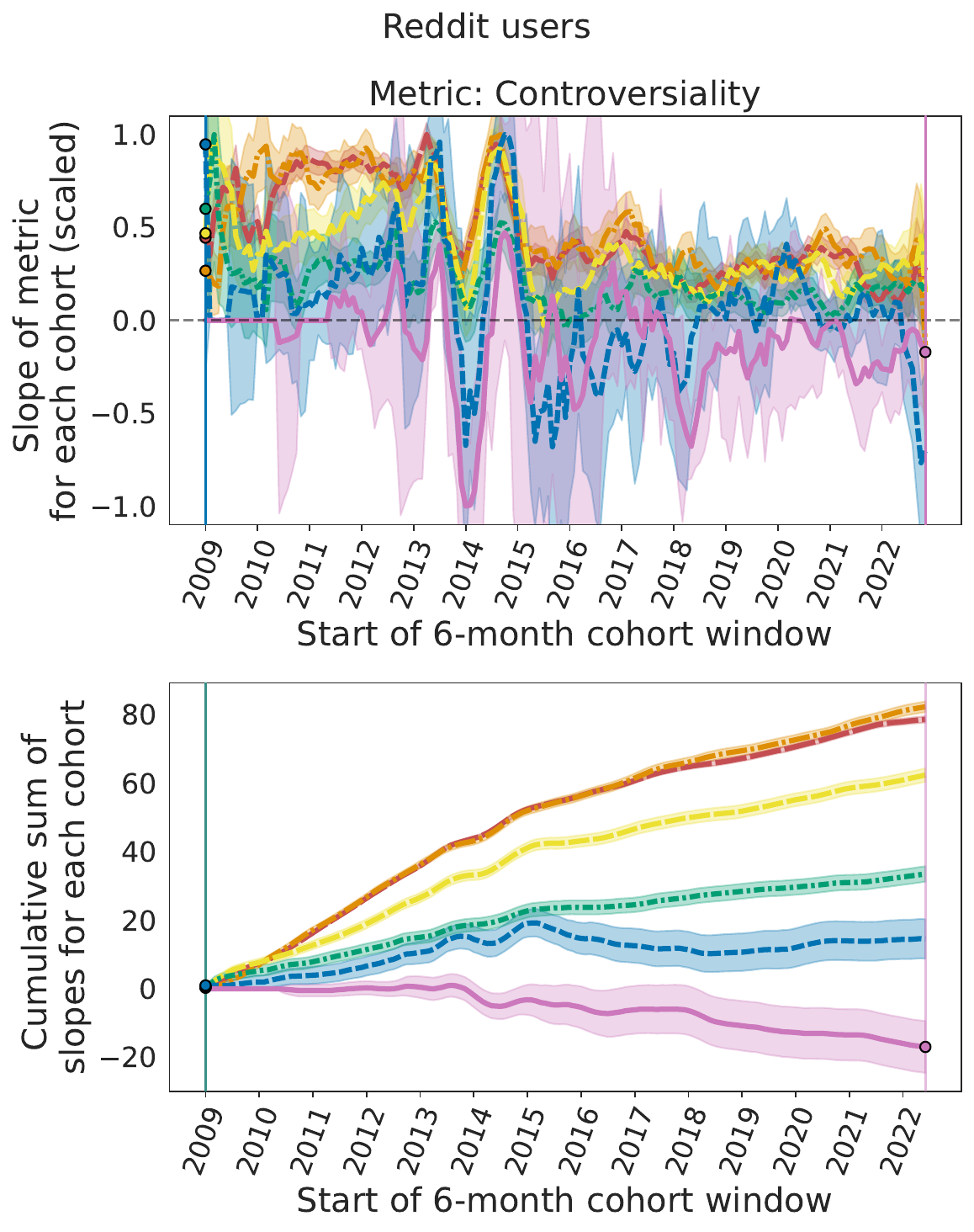} 
    \caption{Changes over time in user lifetime trends, for \enquote{controversiality}, a binary metric for comments with upvote/downvote ratios close to 1. \cite{justachetan2019whatWITHLINK}
    \cbrk
    Controversiality is defined by Reddit, and the exact definition is proprietary and subject to change. The metric was only reported sporadically before 2015, which likely explains the sharp dip just before then.
    \cbrk
    Legends omitted for space, as they're identical to that in Figure \ref{fig_supp:D_repeat_SDb}.
    }
    \label{fig_supp:SDb_slope_metric_controversiality}
\end{figure}

\clearpage
\section{Metrics: Alternate activity-bucketing metrics}
\label{section_supp:E_alt_buckets}

\begin{figure}[h!]
    \centering
    \includegraphics[width=0.95\columnwidth]{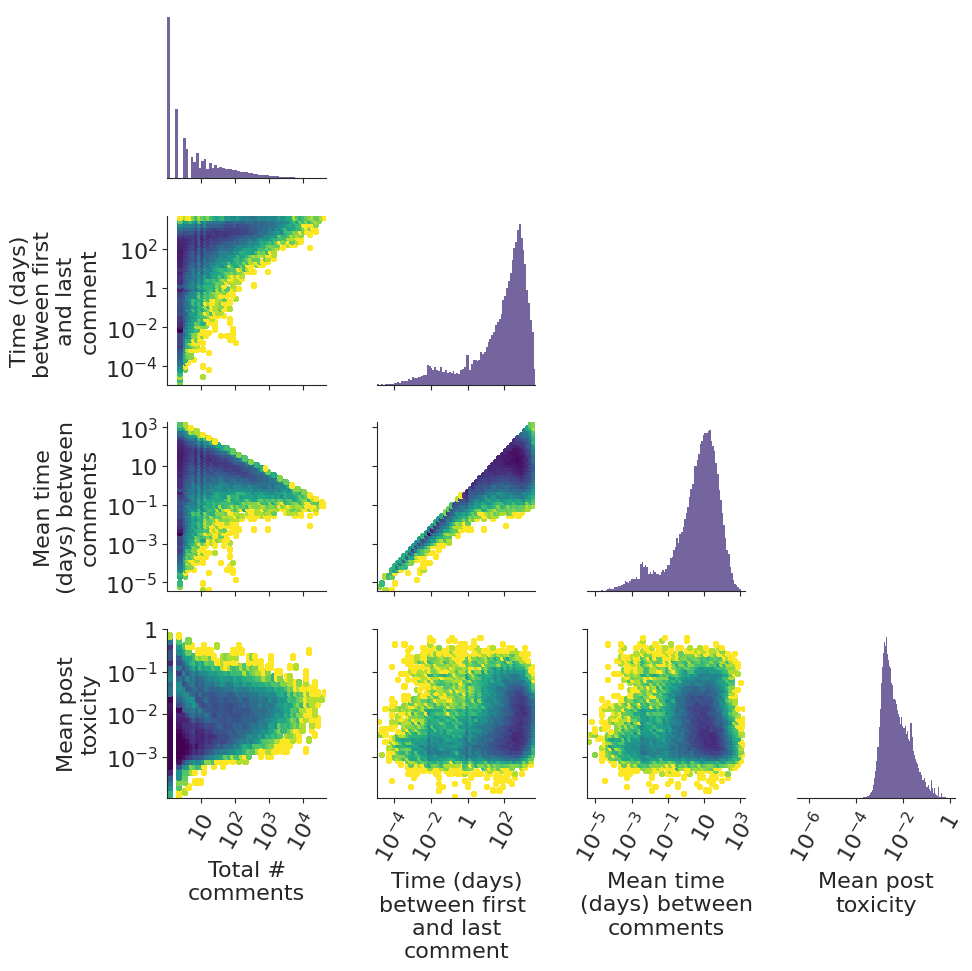} 
    \caption{Reddit user distribution plots for 3 different metrics of \enquote{activity}, as well as users' mean comment toxicity.  Plots were generated using a sample of 50,000 users.
    \cbrk
    Note that users with only one comment are omitted from most subplots, as they have no valid (log) value for \enquote{time between first and last comment} and \enquote{time between comments}.}
    \label{fig_supp:SEa_pairplot_buckets}
\end{figure}

\clearpage
\subsection{Mean score figures with alternate buckets}

\begin{figure}[h!]
    \centering
    \includegraphics[width=0.67\columnwidth]{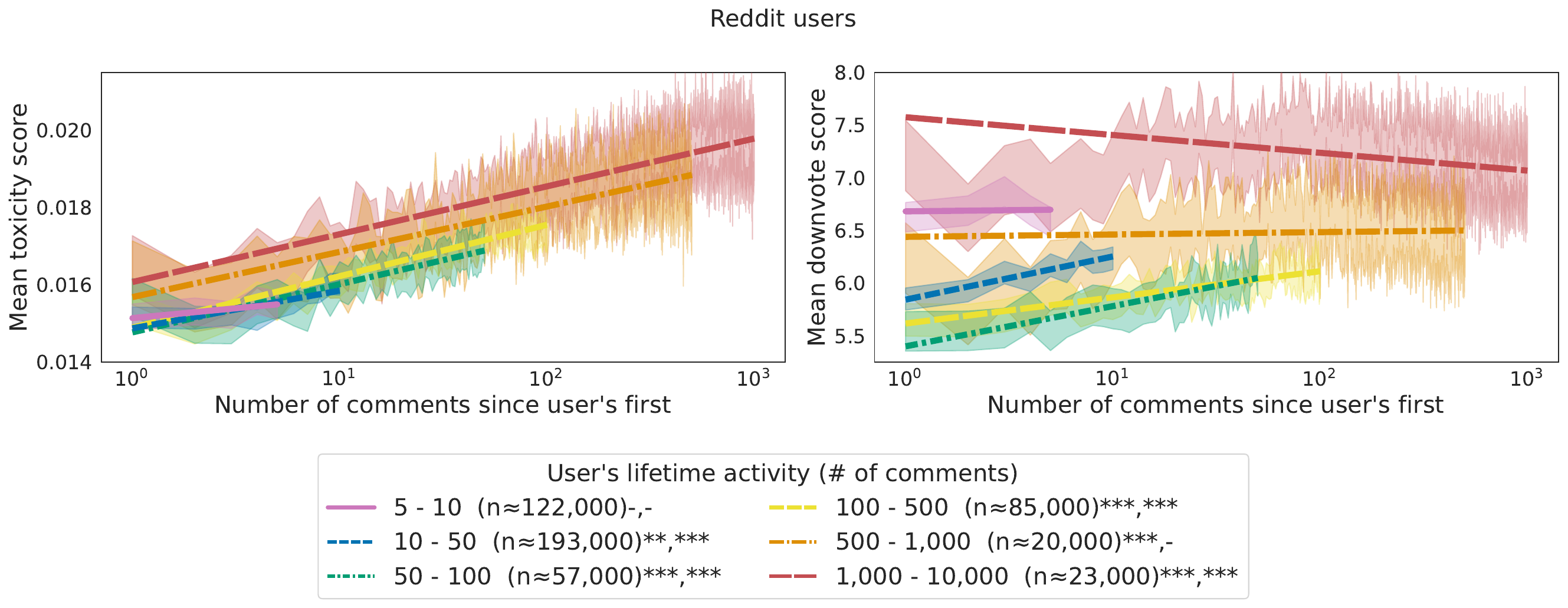} 
    \caption{Figure \ref{fig:B_rainbow} from main text, reproduced for comparison.}
    \label{fig_supp:B_repeat_SEb}
\end{figure}

\begin{figure}[h!]
    \centering
    \includegraphics[width=0.75\columnwidth]{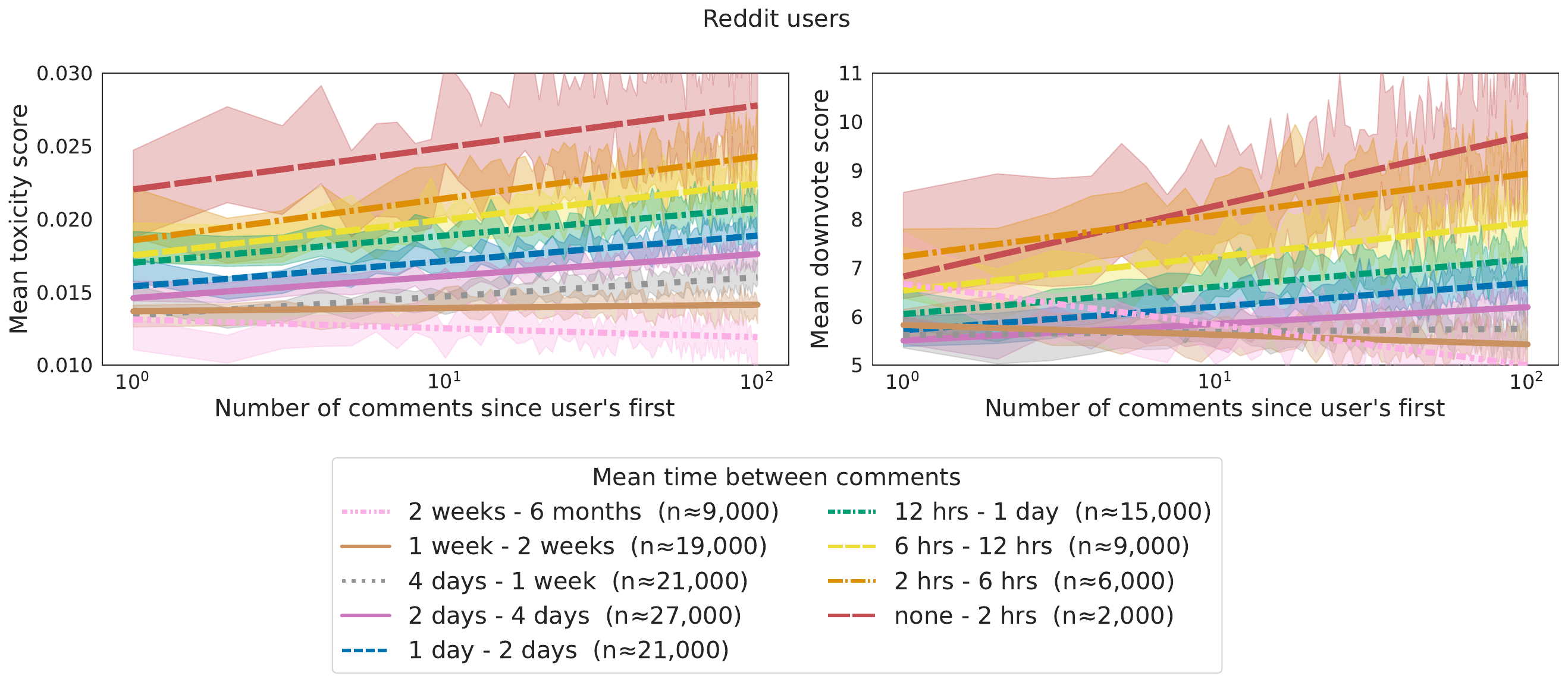} 
    \caption{Trends over Reddit users' lifetimes, from their first post to their hundredth. Limited to users with at least 100 comments. Here users are bucketed by their commenting speed instead of total volume, measured as mean time between comments.}
    \label{fig_supp:SEb1_rainbow_bucket_secs_per_post}
\end{figure}

\begin{figure}[h!]
    \centering
    \includegraphics[width=0.75\columnwidth]{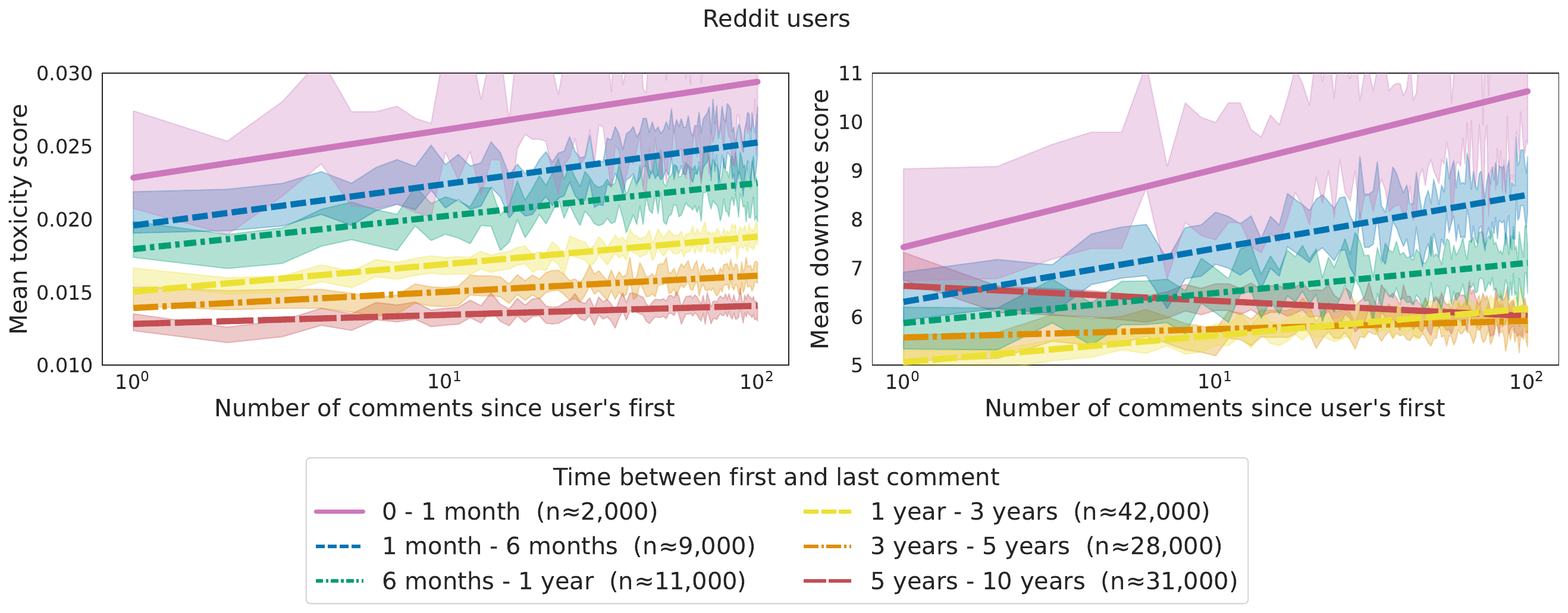} 
    \caption{Trends over Reddit users' lifetimes, from their first post to their hundredth. Limited to users with at least 100 comments. Here users are bucketed by the length of time they were active, rather than total number of comments. 
    \cbrk
    Note that, since we're limiting to users who made at least 100 comments, short-lived users are in some ways more likely to be \enquote{active} than longer-lived users, since they must have been commenting rapidly in order to make 100 comments in a shorter time span.}
    \label{fig_supp:SEb0_rainbow_bucket_timespan_secs}
\end{figure}

\clearpage
\subsection{Slope figures with alternate buckets}

\begin{figure}[h!]
    \centering
    \includegraphics[width=0.7\columnwidth]{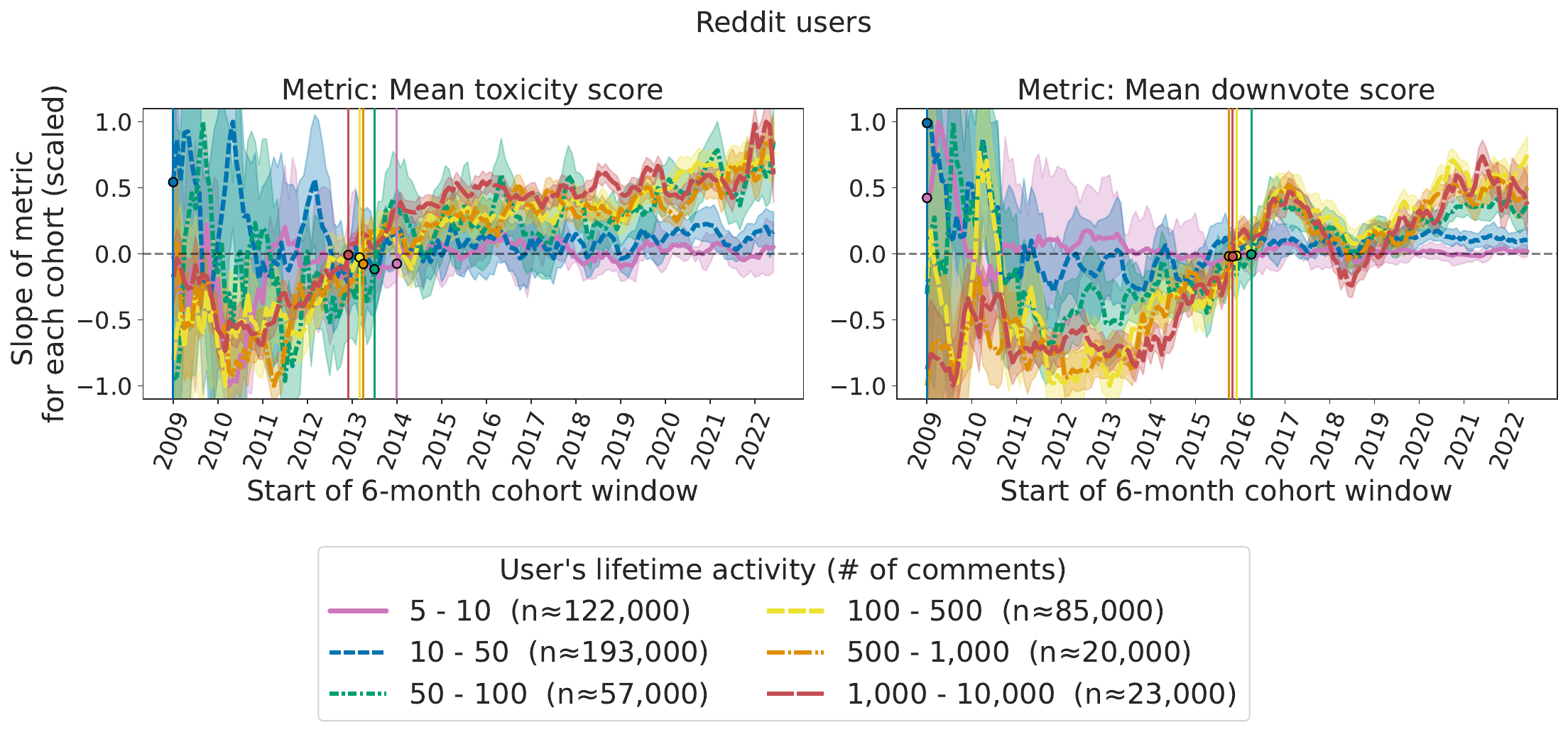} 
    \caption{Figure \ref{fig:D_slope} from main text, reproduced for comparison.}
    \label{fig_supp:D_repeat_SEc}
\end{figure}

\begin{figure}[h!]
    \centering
    \includegraphics[width=0.95\columnwidth]{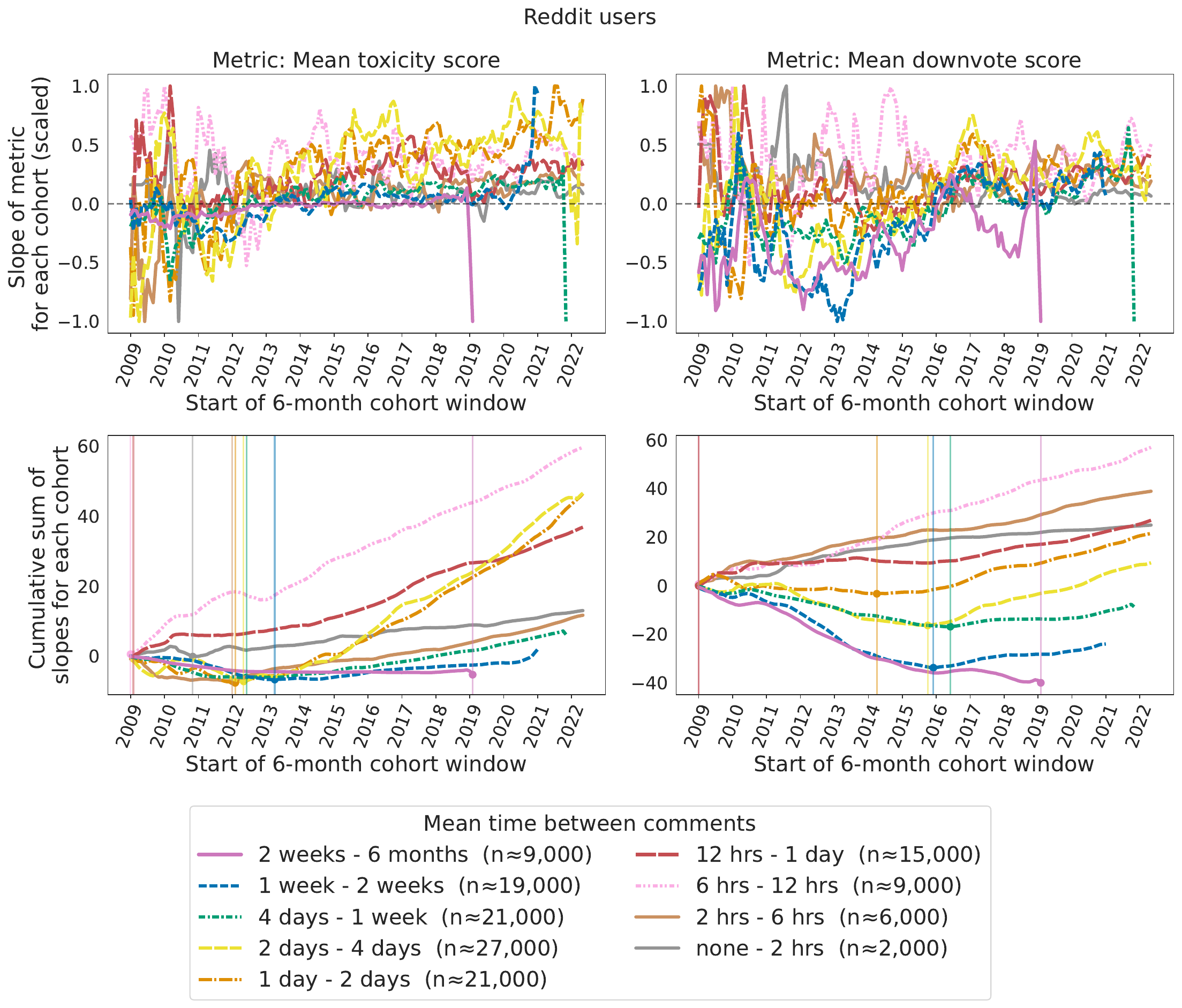} 
    \caption{Trends over Reddit users' lifetimes, where users are bucketed by their commenting speed instead of total volume, measured as mean time between comments. Limited to users with at least 100 comments.}
    \label{fig_supp:SEc1_slopes_bucket_secs_per_post}
\end{figure}

\begin{figure}[h!]
    \centering
    \includegraphics[width=0.95\columnwidth]{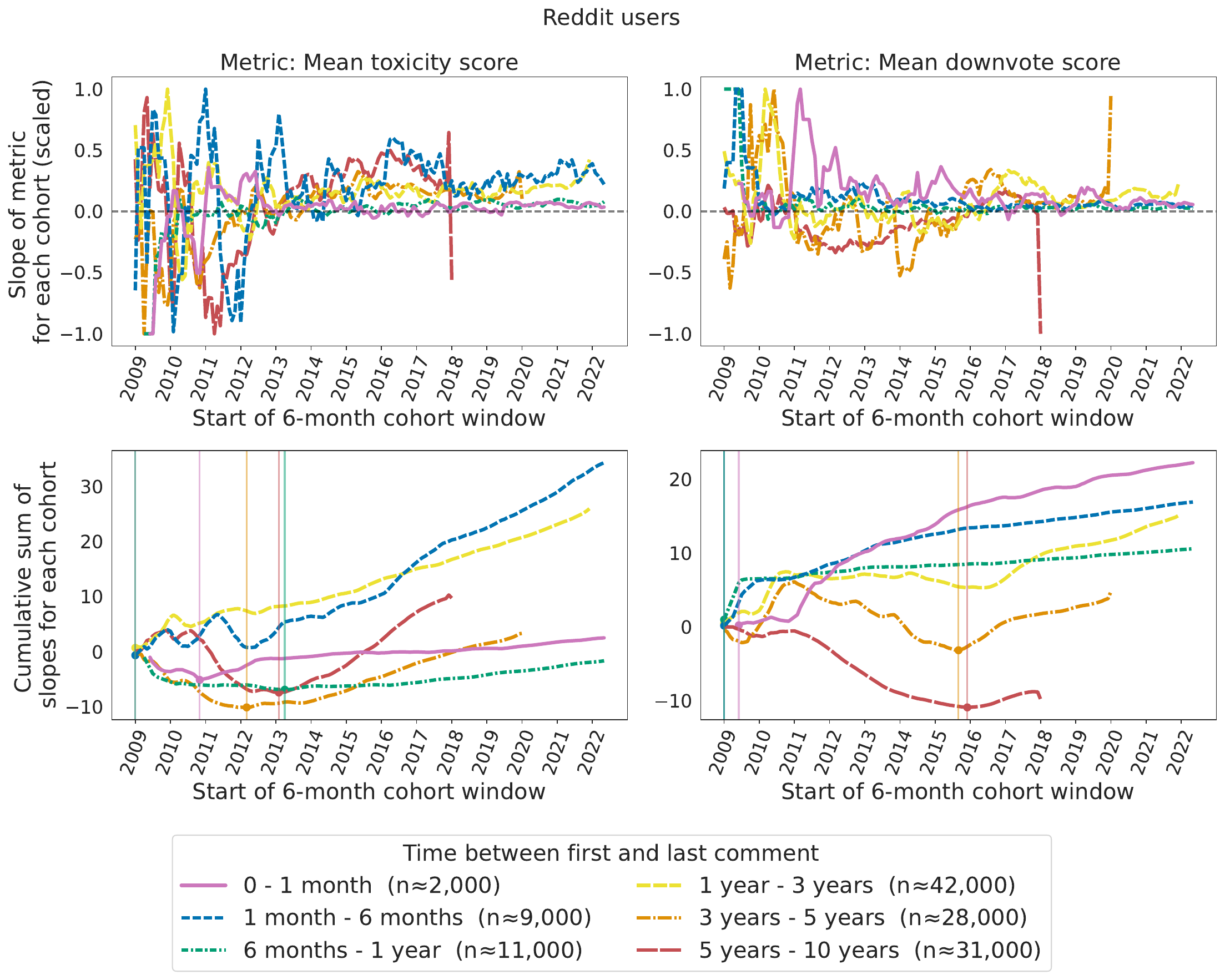} 
    \caption{Trends over Reddit users' lifetimes, where users are bucketed by the length of time they were active, rather than total number of comments. Limited to users with at least 100 comments.}
    \label{fig_supp:SEc0_slopes_bucket_timespan_secs}
\end{figure}

\clearpage
\section{Datasets: Sizes of datasets over time}
\label{section_supp:G_size_plots}

Note that cohorts include all users who began commenting in the 6-month period after the cohort window start date. Thus, cohorts with a start date in the last 6 months of the dataset are smaller due to truncation, and are excluded from most cohort analyses outside of this section.

\begin{figure}[h!]
    \centering
    \includegraphics[width=0.95\columnwidth]{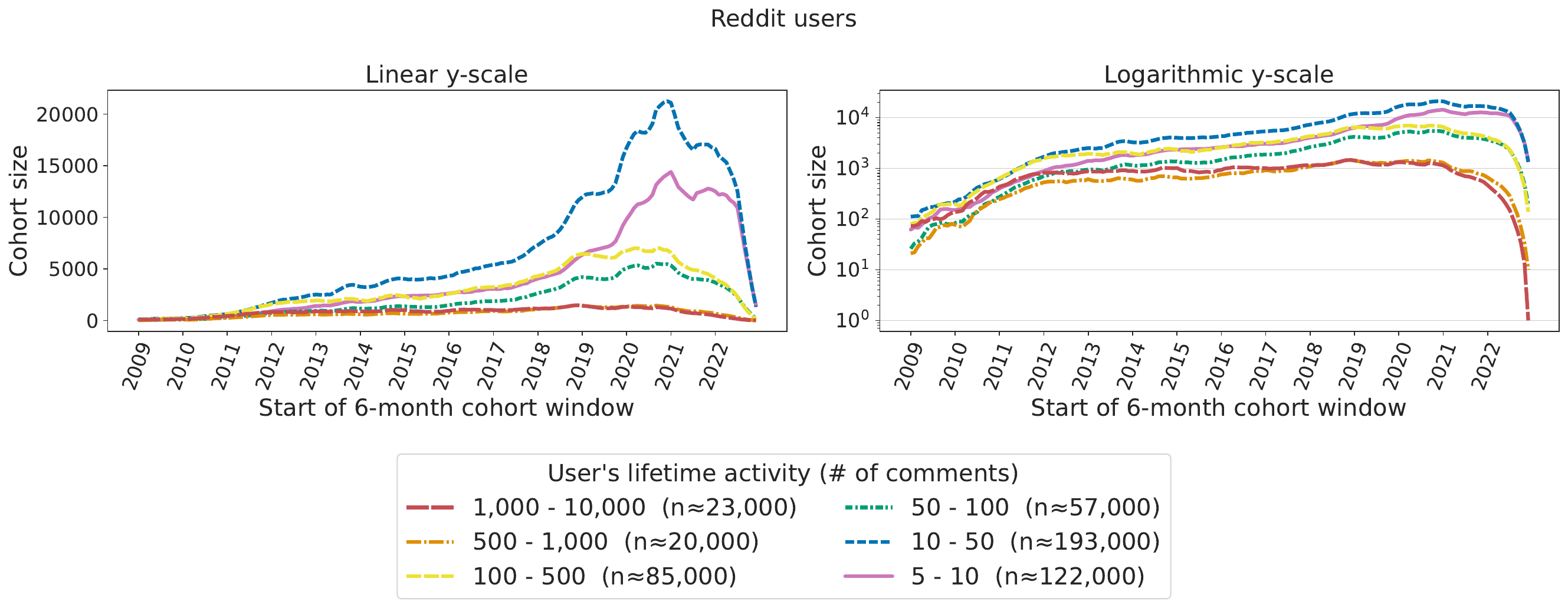} 
    \caption{Number of users in each cohort for the Reddit user dataset.}
    \label{fig_supp:SG0_sizes_auth_newdata}
\end{figure}

\begin{figure}[h!]
    \centering
    \includegraphics[width=0.95\columnwidth]{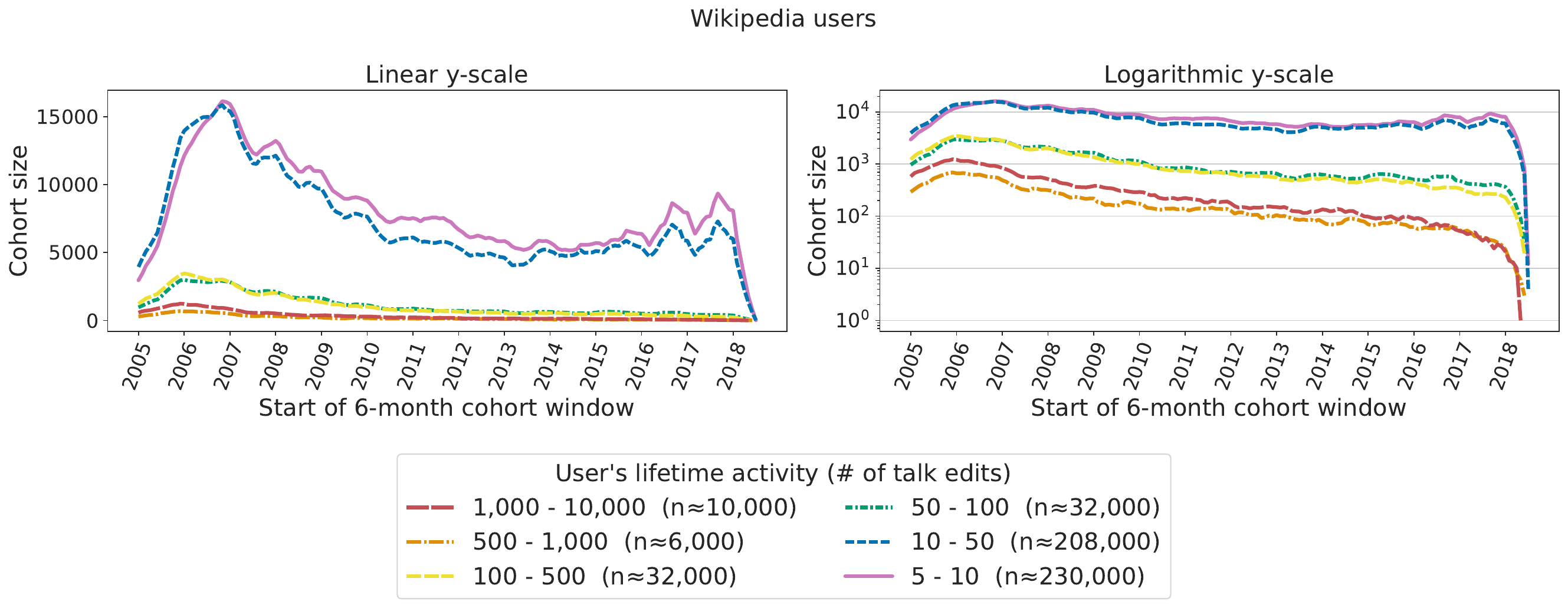} 
    \caption{Number of users in each cohort for the Wikipedia user dataset.}
    \label{fig_supp:SG1_sizes_wiki_auth}
\end{figure}

\begin{figure}[h!]
    \centering
    \includegraphics[width=0.95\columnwidth]{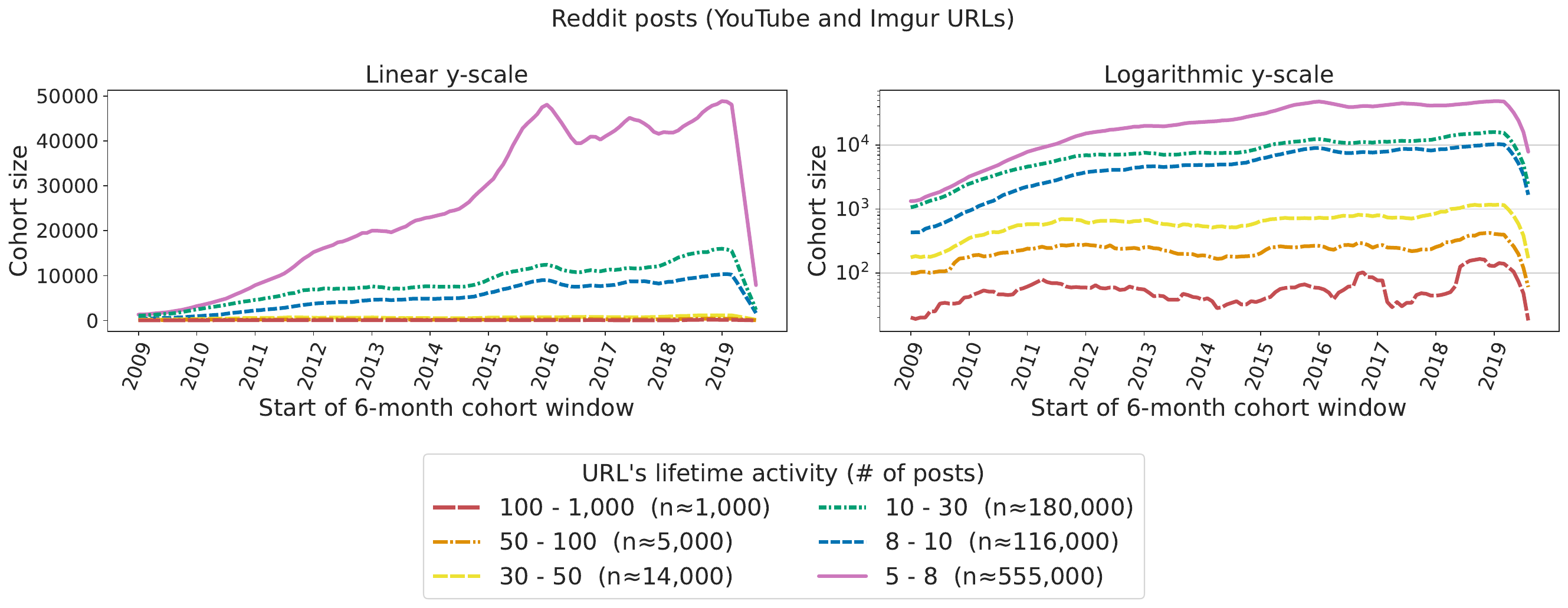} 
    \caption{Number of URLs in each cohort for the Reddit URL dataset.}
    \label{fig_supp:SG2_sizes_yt_img_combined_newscores}
\end{figure}

\begin{figure}[h!]
    \centering
    \includegraphics[width=0.95\columnwidth]{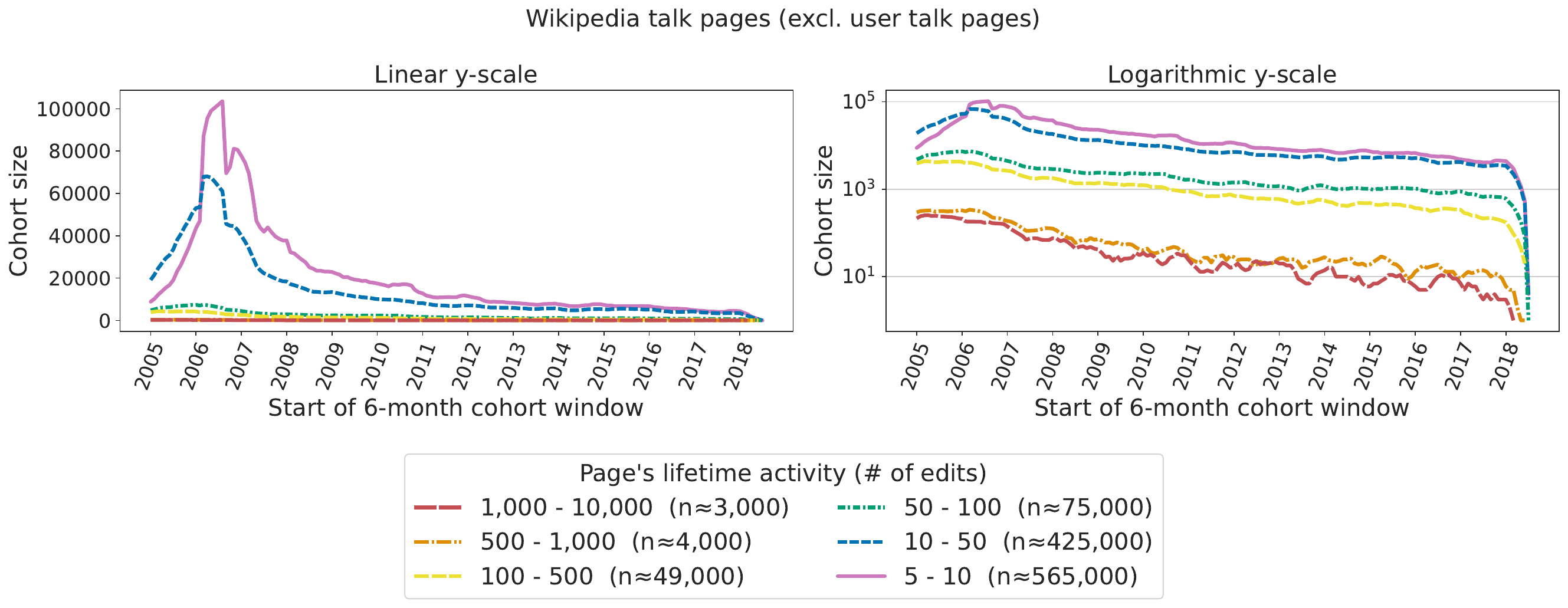} 
    \caption{Number of pages in each cohort for the Wikipedia talk page dataset (not including user talk pages).}
    \label{fig_supp:SG3_sizes_wiki_page_othertalk}
\end{figure}

\begin{figure}[h!]
    \centering
    \includegraphics[width=0.95\columnwidth]{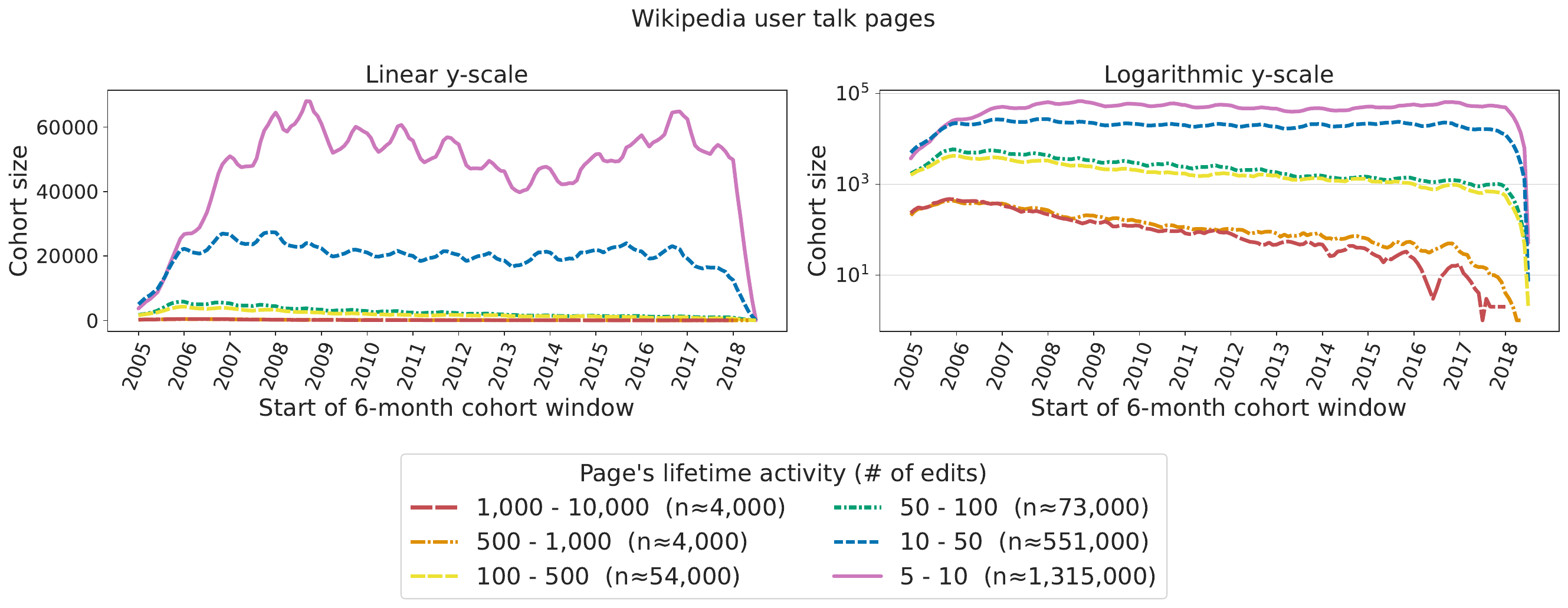} 
    \caption{Number of pages in each cohort for the Wikipedia talk page dataset, for user talk pages only.}
    \label{fig_supp:SG4_sizes_wiki_page_usertalk}
\end{figure}


\clearpage
\section{Datasets: Languages present in dataset}
\label{section_supp:SJ_languages}

For the Wikipedia dataset, only English-language Wikipedia was used, since in WikiConv SEVERE\_TOXICITY scores were only available for the English Wikipedia corpus.\cite{wikiconv_documentation}

\begin{figure}[h!]
    \centering
    \includegraphics[width=0.45\columnwidth]{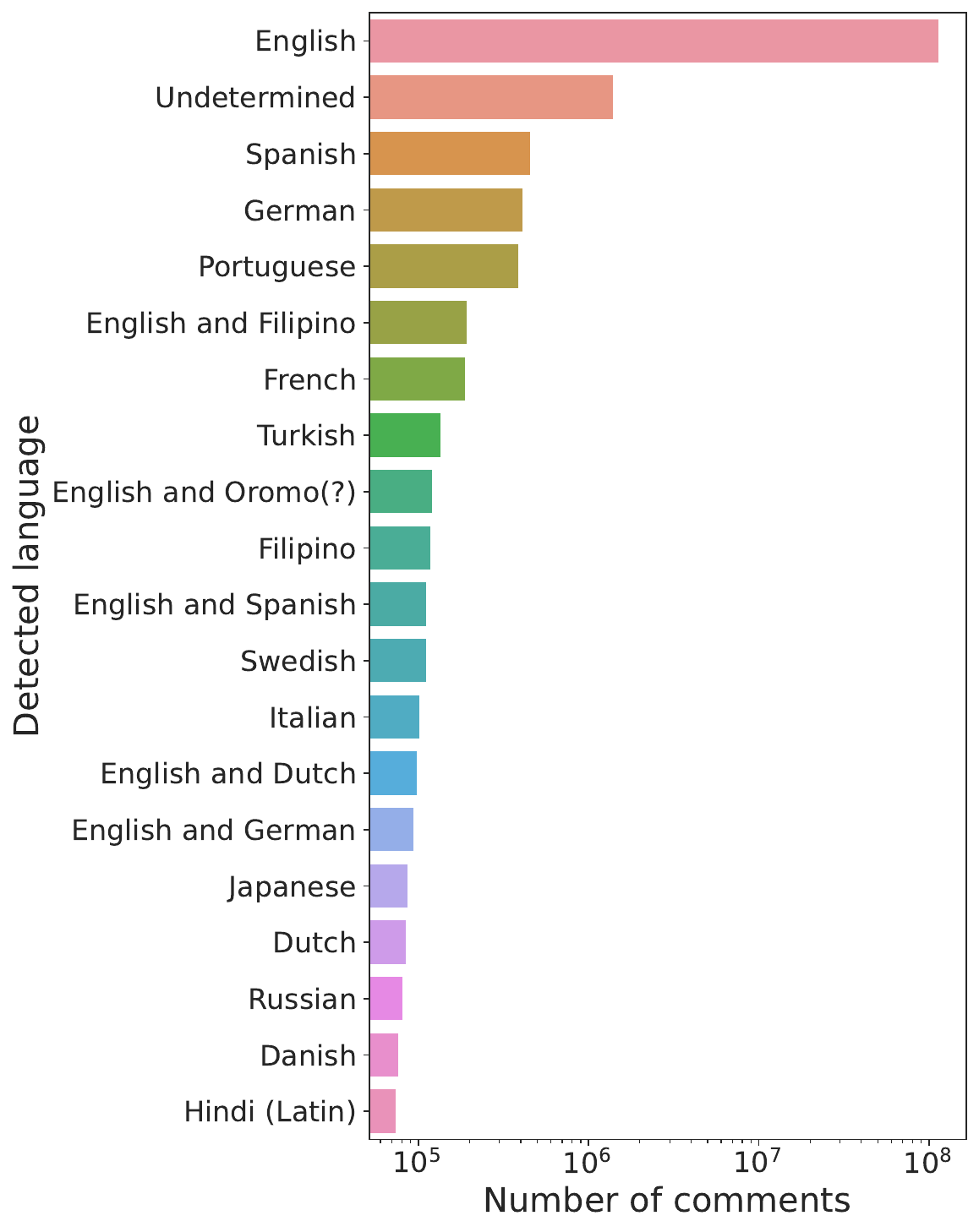} 
    \caption{Number of comments for each of the 20 most commonly detected languages or language combinations.
    \cbrk
    (Oromo has a question mark because unlike the other \enquote{English and} categories, Oromo isn't commonly detected alone. Spot-checking showed that these comments appeared to be in English only.)
    }
    \label{fig_supp:SJ_languages_detected}
\end{figure}

For the Reddit dataset, things are more complicated. Reddit is a majority English website, but is widely used around the world. We use the "detected language" from Perspective API to determine the language a comment was in. We find that 92.7\% of comments are detected as English only, and 95.5\% contain English as one of the detected languages. See Figure \ref{fig_supp:SJ_languages_detected} for the most commonly detected languages.

Further, Perspective API makes SEVERE\_TOXICITY scores available in multiple languages: ar, zh, cs, nl, en, fr, hi, hi-Latn, id, it, ja, ko, pl, pt, ru, sv.\cite{perspective_api_attribute_defs} 94.1\% of comments contain only languages scoreable by the toxicity model, and 96.8\% contain at least one scoreable language.

\begin{figure}[h!]
    \centering
    \includegraphics[width=0.45\columnwidth]{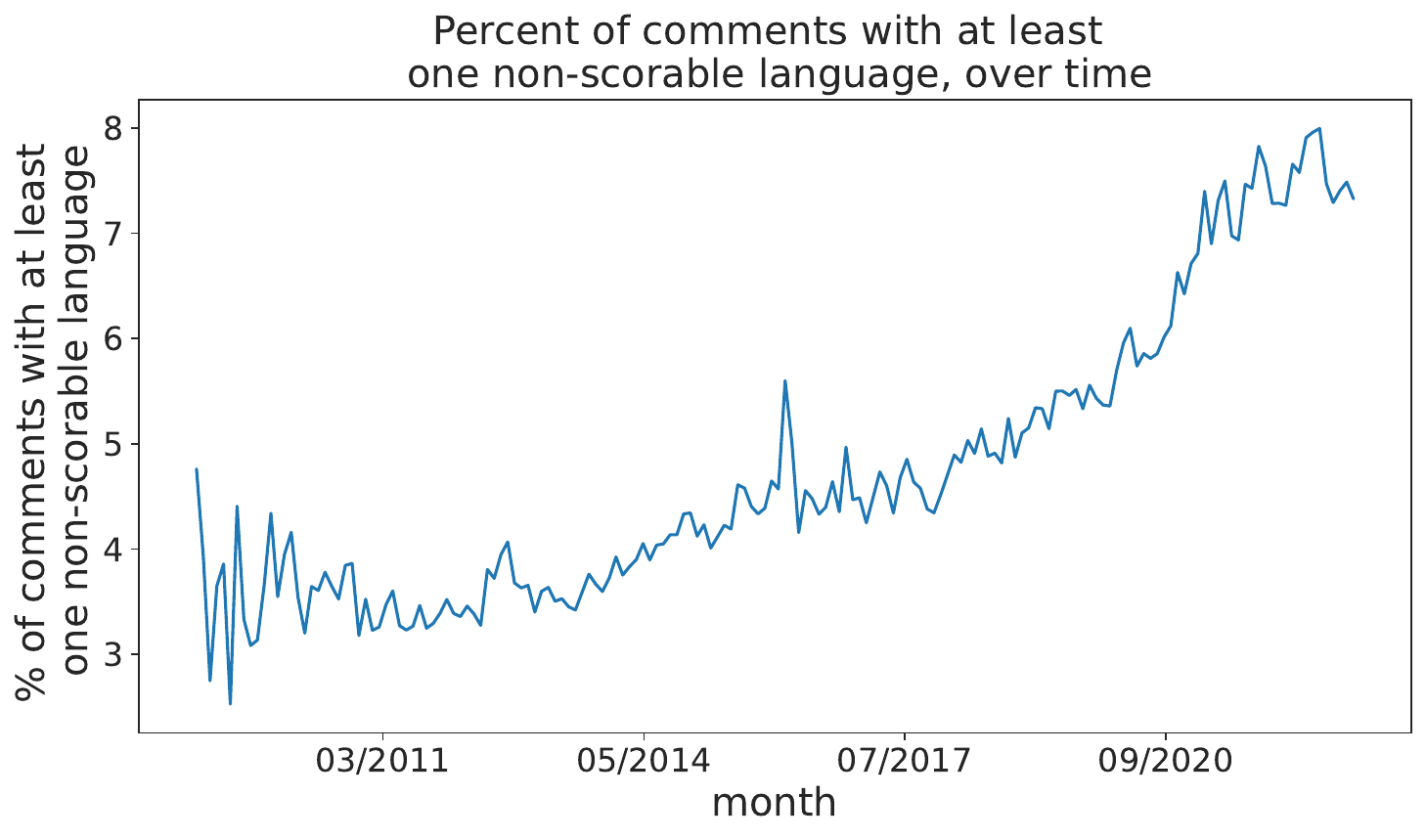} 
    \caption{Percentage of comments that aren't completely scorable by Perspective API, over time. Percentages of non-English comments and comments without any scorable languages show similar trends.
    }
    \label{fig_supp:SJ_languages_noneng}
\end{figure}

\section{Datasets: Deleted content}
\label{section_supp:SK_deleted_data}

 Wikipedia aims to avoid deleting edit history, and deletions are relatively rare.\cite{west2011what,2024wikipedia}

Reddit allows user deletion of comments, which are then removed from databases. Most commonly, both the author and text of the comment show up as \enquote{[deleted]}. Sometimes, the author remains, but the text of the comment becomes \enquote{deleted}, \enquote{redacted}, \enquote{removed}, or some variation. This depends on who deleted it (the author, a subreddit moderator, or a site moderator), when it was deleted, and other factors.

Unfortunately, we don't currently have access to the Reddit corpus, so we can only report older analyses. 

Within our 1\% sample, less than 0.003\% were comments that were variations on \enquote{deleted}, \enquote{redacted}, or\enquote{removed}. This was defined as any comment less than 14 characters in length that contained the strings \enquote{deleted}, \enquote{redacted}, or \enquote{removed}, case-insensitive. Spot checks showed that roughly half appeared to be actually deleted comments; the others were real comments like \enquote{It's deleted?}, etc.

For the entire 2005-2022 Reddit corpus (not our 1\% sample),  1,491,259,398 of 14,313,529,439 comments, or 10.4\%, had \enquote{[deleted]} as the author. From 2005-2019, 9\% were deleted, and from 2020-2022, 11.2\% were deleted. Since our 1\% sample is a sample of authors, not comments, none of these comments are included in our dataset. Some might have originally been made by authors in our sample, but since author information is deleted, we have no way to know. 

So, we don't have evidence that deletion percentages changed significantly over time, though we also don't have evidence to the contrary, since the only data point we have is deletion percentages for two long time periods. We also have no way to know if deletion percentages vary across users, or across a user's lifetime. This is a problem because if, for example, 10 out of a user's first 100 comments were deleted, we would mistakenly label their 100th comment as their 90th. Differences between users would mean that some users are misaligned when we're trying to get average slopes over many users. Differences over a user's lifetime would distort the shape of their trajectory. We can hope that the large size of the dataset smooths out these issues, but don't know of any way to confirm that.


\end{document}